\documentclass{article} 
\usepackage{iclr2027/iclr2027_conference,times}
\iclrfinalcopy

\usepackage{amsmath,amsfonts,bm}

\def\eqref#1{equation~\ref{#1}}

\def\1{\bm{1}}

\DeclareMathAlphabet{\mathsfit}{\encodingdefault}{\sfdefault}{m}{sl}
\SetMathAlphabet{\mathsfit}{bold}{\encodingdefault}{\sfdefault}{bx}{n}

\usepackage{hyperref}
\usepackage{url}

\usepackage{graphicx}
\usepackage{subcaption}
\usepackage{booktabs}
\usepackage{multirow}
\usepackage{colortbl}
\usepackage{tabularx}
\usepackage{array}

\title{\textit{Chaos in the Text}: Revealing the Modality Preference in Mixed-Modality Retrievers}

\author{%
\parbox{\textwidth}{\centering
  \vspace{0.8em}
  {\bfseries
  Yubo Sun\textsuperscript{1*}\hspace{1.5em}
  Chunyi Peng\textsuperscript{2*}\hspace{1.5em}
  Yukun Yan\textsuperscript{3\dag}\hspace{1.5em}
  Zhenghao Liu\textsuperscript{2\dag}\hspace{1.5em}
  Zhipeng Xu\textsuperscript{2}\\[0.4em]
  Sen Mei\textsuperscript{3}\hspace{1.5em}
  Linlin Xin\textsuperscript{1}\hspace{1.5em}
  Zheni Zeng\textsuperscript{4}\hspace{1.5em}
  Maosong Sun\textsuperscript{3}}
  \\[0.9em]
  {\normalfont\small
  \textsuperscript{1}University of the Chinese Academy of Sciences\hspace{1.5em}
  \textsuperscript{2}Northeastern University\\
  \textsuperscript{3}Tsinghua University\hspace{1.5em}
  \textsuperscript{4}Nanjing University}
  \\[0.5em]
  {\normalfont\footnotesize
  \textsuperscript{*}Equal contribution\hspace{1.5em}
  \textsuperscript{\dag}Corresponding author}
  \\[0.5em]
  {\normalfont\small
  Code: \href{https://github.com/OpenBMB/Trident}{\texttt{https://github.com/OpenBMB/Trident}}}
}}

\begin{document}

\maketitle

\begin{abstract}
Dense retrievers have made significant progress on text and image corpora, but whether these capabilities extend reliably to mixed corpora containing text, image, and fused text-image documents remains unclear. In this paper, we systematically examine retrievers across architectures and find that their performance is highly sensitive to modality composition. As image documents are progressively replaced with semantically corresponding text representations, retrieval performance follows a pronounced V-shaped curve, remaining strong on single-modality corpora but degrading substantially when modalities coexist. In particular, irrelevant text causes more severe degradation than an equal number of irrelevant images, a phenomenon we term \textbf{\textit{``Chaos in the Text''}}. Further analysis reveals modality preference, whereby text representations receive systematically higher similarity scores, allowing irrelevant text to outrank relevant images. To mitigate this bias, we introduce \textbf{Trident}, which constructs text, image, and fused text-image views of each document as co-equal positives and jointly optimizes relevance discrimination and positive-view balance through Multi-Positive View InfoNCE. Experiments across visual document and natural image benchmarks show that trident improves mixed-modality retrieval on both CLIP-based and VLM-based architectures, reduces sensitivity to modality composition and text distractors, and increases average single-modality retrieval performance.
\end{abstract}


\section{Introduction}
\label{sec:intro}


In recent years, image retrieval has made significant progress in natural-image~\citep{jia2021scaling, li2025unirag} and visual-document retrieval~\citep{faysse2025colpali, yu2025visrag} by directly encoding images and retrieving relevant content from image corpora. This approach preserves visual information often lost in text conversion and provides relevant evidence~\citep{peng2026conceptformer, tanaka2025vdocrag} to support better information gathering in multimodal RAG~\citep{sun2025visrag, peng2025mixture, xiong2026lang2act, yang2026memorycard}. However, real-world corpora often contain a mixture of text-only, image-only, and fused text-image documents, with relevant evidence potentially residing in any modality~\citep{wei2024uniir, sun2026hievidr}, which requires retrievers to compare query–document relevance across modalities and rank all candidates in a unified list.

\begin{figure}[t]
    \centering
    \begin{subfigure}[b]{0.43\textwidth}
        \centering
        \includegraphics[width=\textwidth]{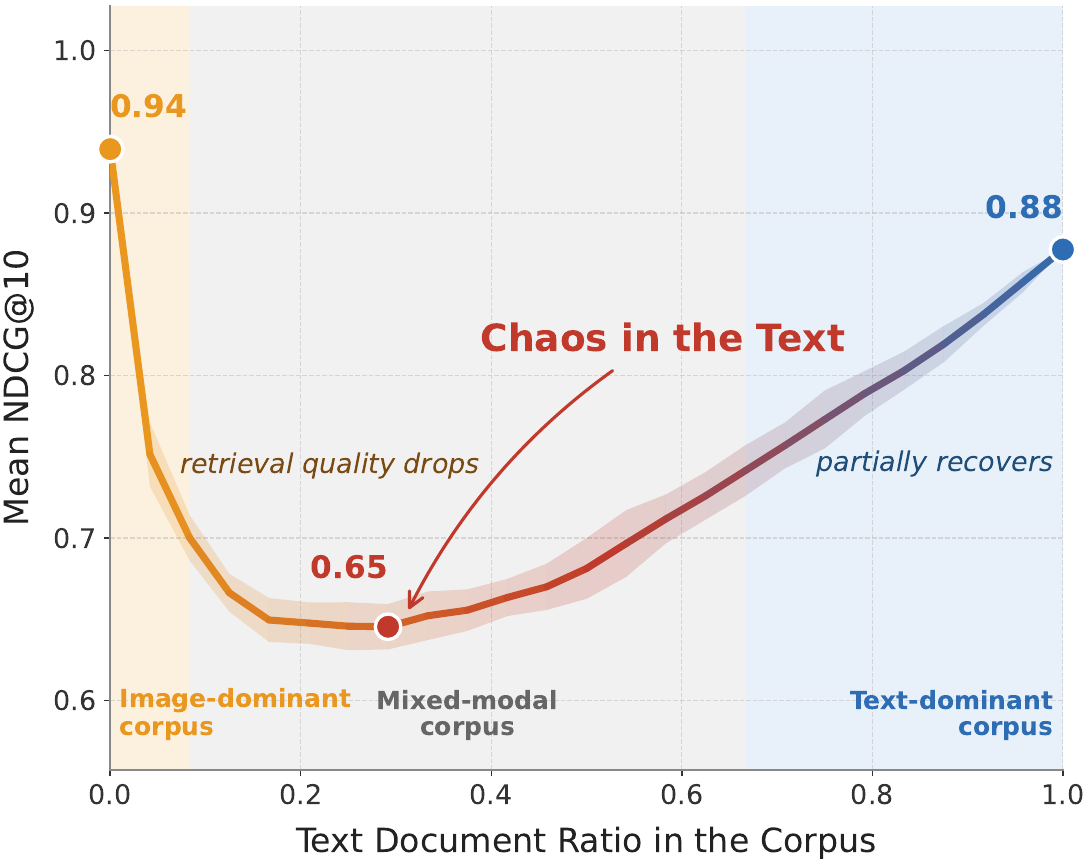}
        \caption{}
        \label{fig:intro_teaser:curve}
    \end{subfigure}
    \hfill
    \begin{subfigure}[b]{0.56\textwidth}
        \centering
        \includegraphics[width=\textwidth]{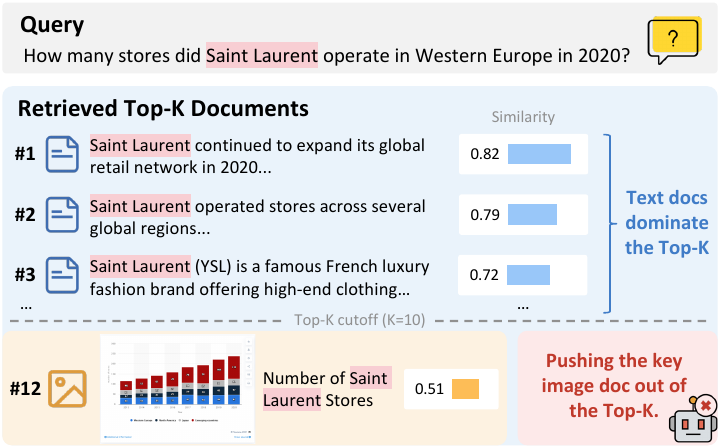}
        \caption{}
        \label{fig:intro_teaser:mechanism}
    \end{subfigure}
    \caption{(a) \textbf{Retrieval robustness degrades sharply in mixed-modality corpora.} As image documents are progressively replaced by text documents, NDCG@10 follows an asymmetric V-shaped curve. (b) \textbf{Chaos in the Text:} text documents are systematically over-ranked regardless of semantic relevance, pushing key documents out of the top-$K$ when the answer resides in the image modality.}
    \label{fig:intro_teaser}
\end{figure}

Despite its importance, we find that existing multimodal retrievers are highly sensitive to the modality composition of mixed-modality corpora. As illustrated in Figure~\ref{fig:intro_teaser}, progressively replacing documents in an image corpus with their semantically corresponding text representations yields a pronounced ``V-shaped'' performance curve. Models perform well on purely image or purely text corpora, while degrading substantially when the two modalities coexist. Further inspection reveals that models tend to rank irrelevant text documents above highly relevant image documents, pushing visual evidence containing the answer out of the top-K results. We term this phenomenon \textbf{\textit{``Chaos in the Text''}}. These findings indicate that strong retrieval performance on single-modality corpora does not guarantee reliable access to relevant evidence in mixed-modality settings.

To investigate the cause of this phenomenon, we further analyze similarity distributions across document modalities. Specifically, we construct three semantically corresponding representations for each document: text, image, and fused text-image, and compare their similarities to the same set of queries. We find that multiple retrievers systematically assign higher similarity scores to text representations than to image representations, even when considering only positive query–document pairs. This indicates that retrieval scores reflect not only semantic relevance but also a systematic influence of document modality, revealing a Modality Preference. When documents of different modalities compete in the same candidate pool, this bias translates into a ranking advantage, allowing irrelevant text documents to outrank relevant image documents.

To alleviate this phenomenon, we introduce \textbf{Trident}, a contrastive learning framework that balances modalities through multiple positive views. Specifically, Trident constructs text, image, and fused text-image representations of each document as three co-equal positives for the same query and jointly optimizes them with a Multi-Positive InfoNCE objective under shared normalization. While separating relevant documents from irrelevant ones, this objective discourages positive probability mass from concentrating on a single modality, encouraging the model to account for relevant evidence across modalities and thereby mitigating modality preference in mixed-modality retrieval.


Extensive experiments show that trident substantially mitigates the modality preference while improving overall retrieval performance. After trident training, models exhibit more stable retrieval performance on mixed-modality corpora, with reduced sensitivity to modality proportions and text distractors, while also achieving higher average retrieval performance on the original single-modality corpora. These improvements hold across both CLIP-based and VLM-based retrieval architectures, demonstrating the effectiveness and broad applicability of our training approach.

\section{How Retrieval Models Perform on Mixed-Modality Corpora?}
\label{sec:how}

Although multimodal retrieval models perform strongly on single-modality corpora (text-only or image-only), real-world corpora, as discussed in Section~\ref{sec:intro}, typically mix text, image, and text-image fused documents. How reliably models retrieve from such mixed-modality corpora, however, has not been systematically quantified. We therefore examine retrieval performance in this setting.

\begin{figure}[t]
    \centering
    \begin{subfigure}[b]{0.32\textwidth}
        \centering
        \includegraphics[width=\textwidth]{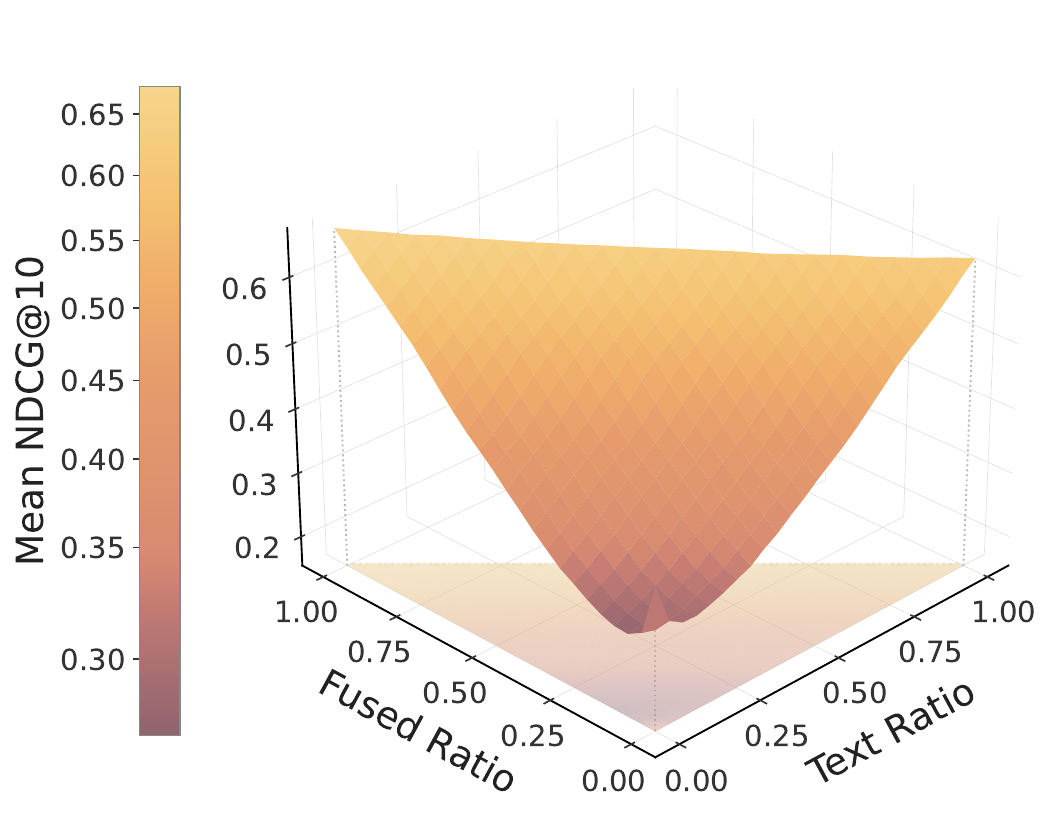}
        \caption{Jina-CLIP-v2 on ChartQA}
        \label{fig:ndcg_modelA_datasetX}
    \end{subfigure}
    \hfill
    \begin{subfigure}[b]{0.32\textwidth}
        \centering
        \includegraphics[width=\textwidth]{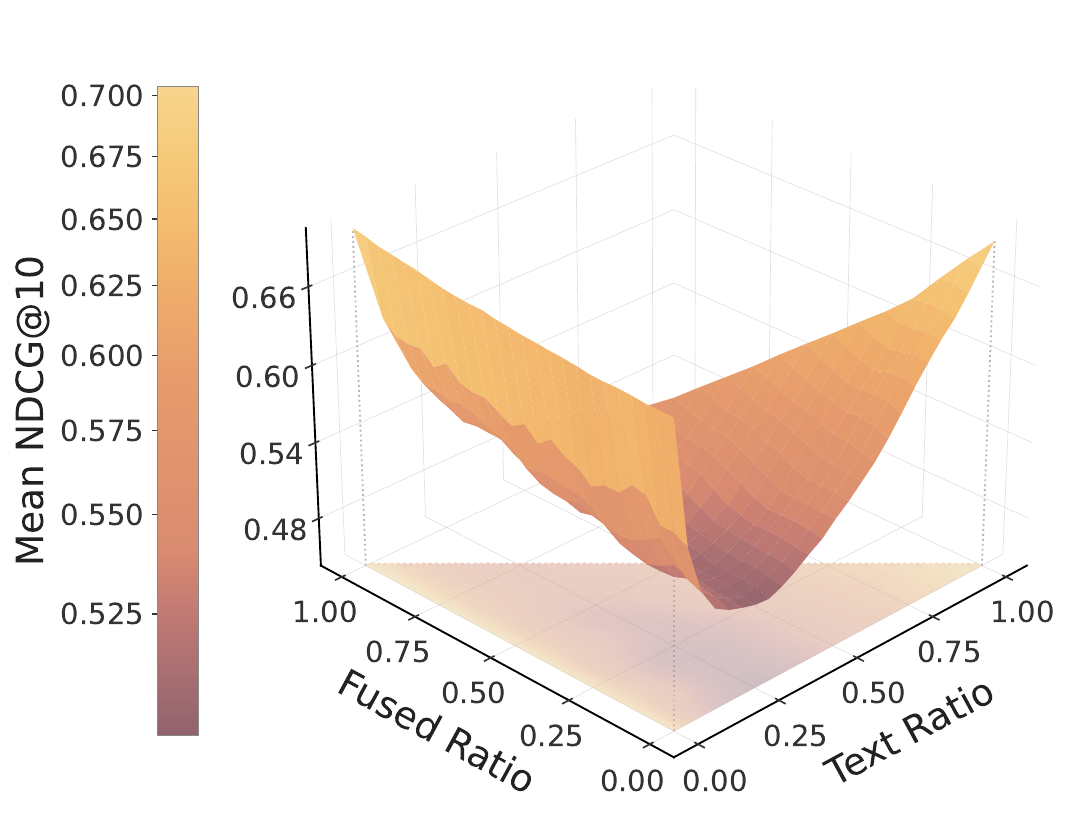}
        \caption{Qwen3-VL-E-2B on ChartQA}
        \label{fig:ndcg_modelA_datasetY}
    \end{subfigure}
    \hfill
    \begin{subfigure}[b]{0.32\textwidth}
        \centering
        \includegraphics[width=\textwidth]{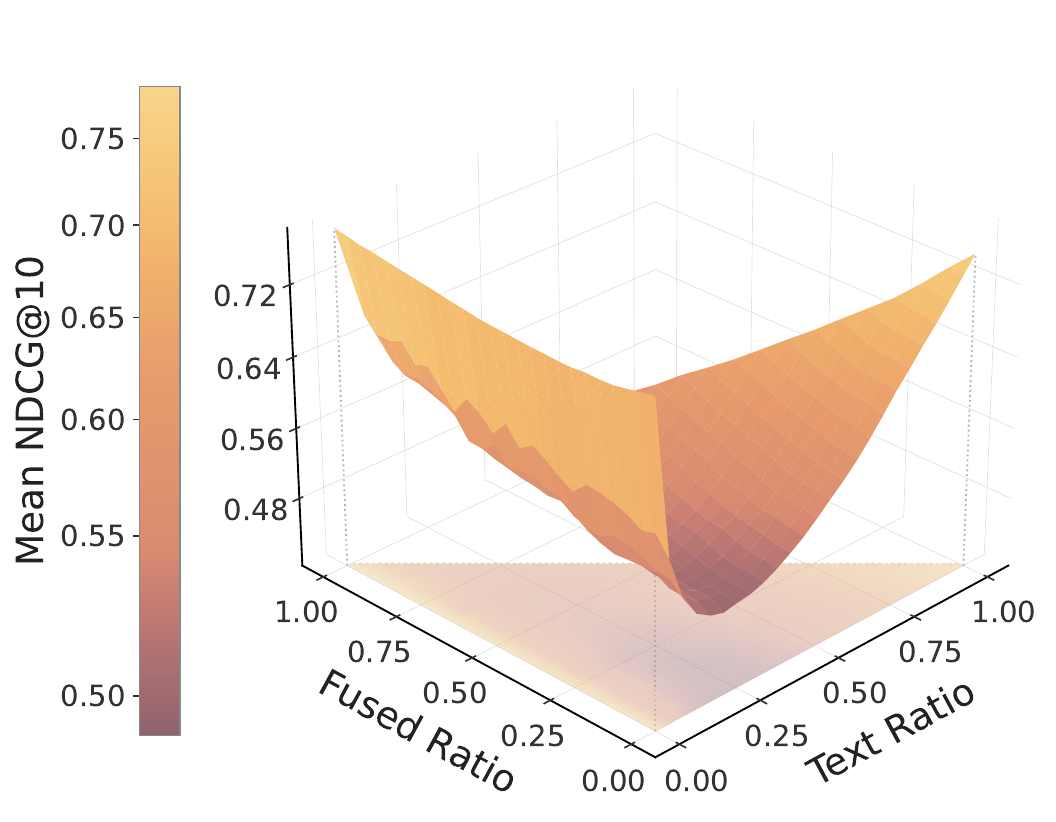}
        \caption{Qwen3-VL-E-8B on ChartQA}
        \label{fig:ndcg_modelA_datasetZ}
    \end{subfigure}

    \begin{subfigure}[b]{0.32\textwidth}
        \centering
        \includegraphics[width=\textwidth]{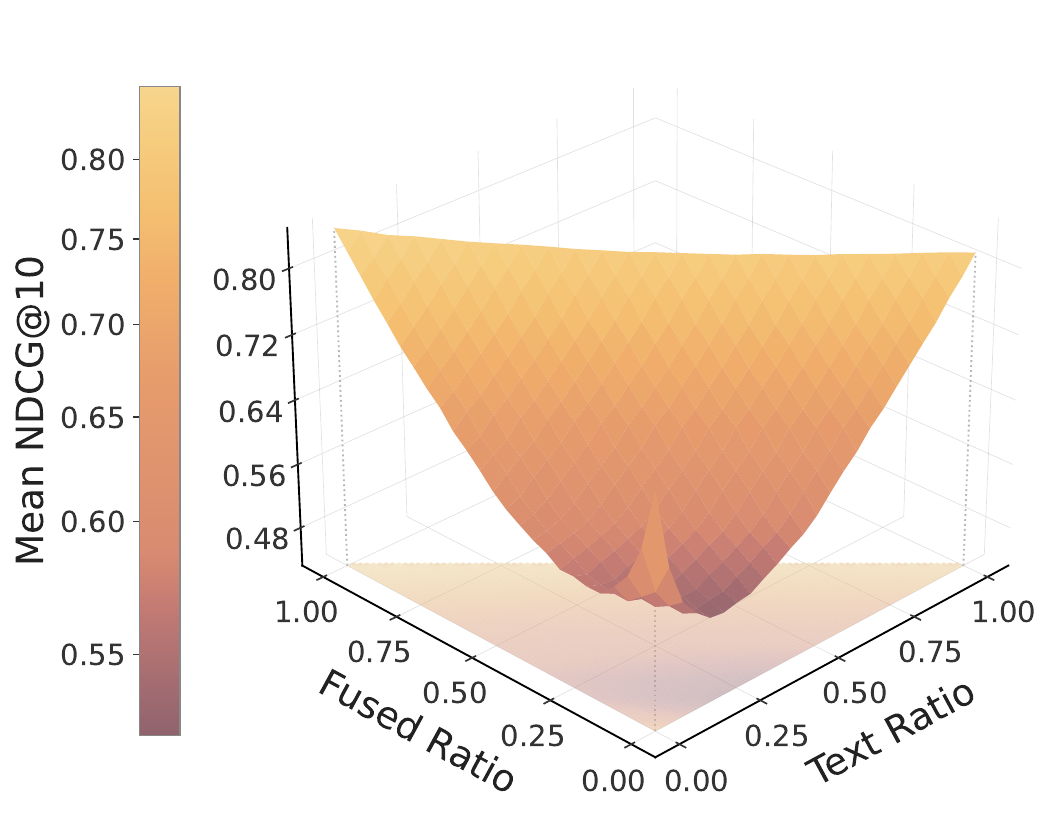}
        \caption{Jina-CLIP-v2 on SlideVQA}
        \label{fig:ndcg_modelB_datasetX}
    \end{subfigure}
    \hfill
    \begin{subfigure}[b]{0.32\textwidth}
        \centering
        \includegraphics[width=\textwidth]{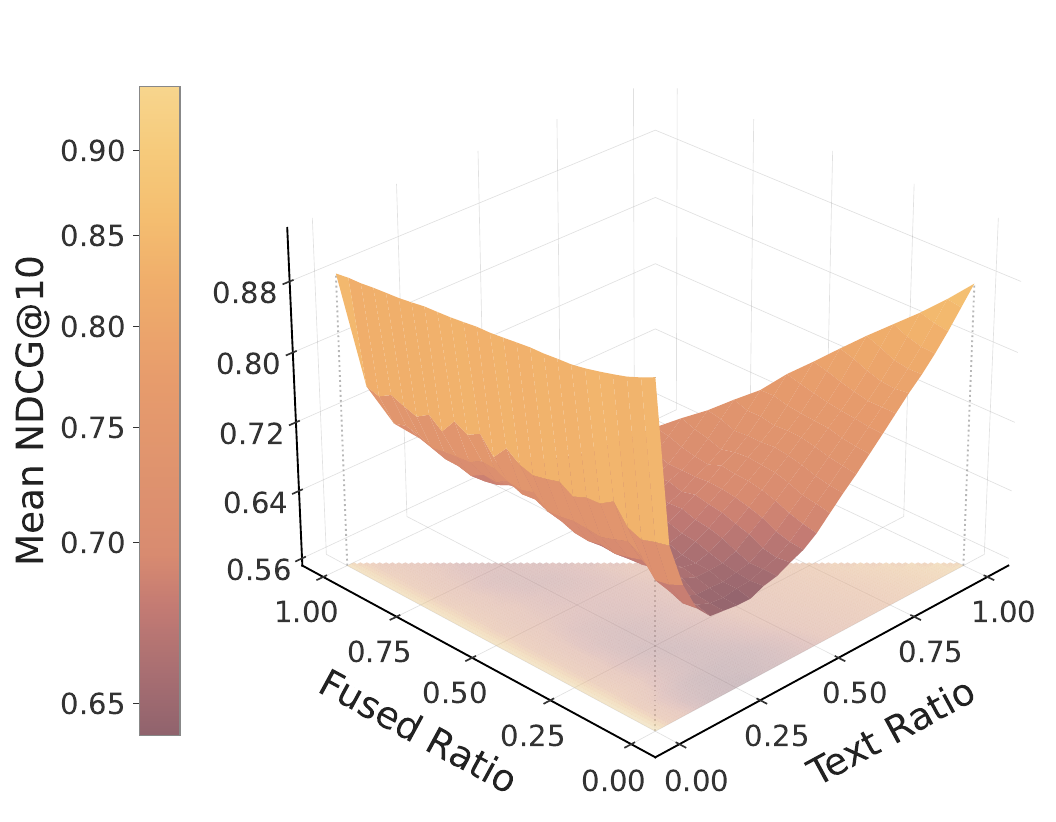}
        \caption{Qwen3-VL-E-2B on SlideVQA}
        \label{fig:ndcg_modelB_datasetY}
    \end{subfigure}
    \hfill
    \begin{subfigure}[b]{0.32\textwidth}
        \centering
        \includegraphics[width=\textwidth]{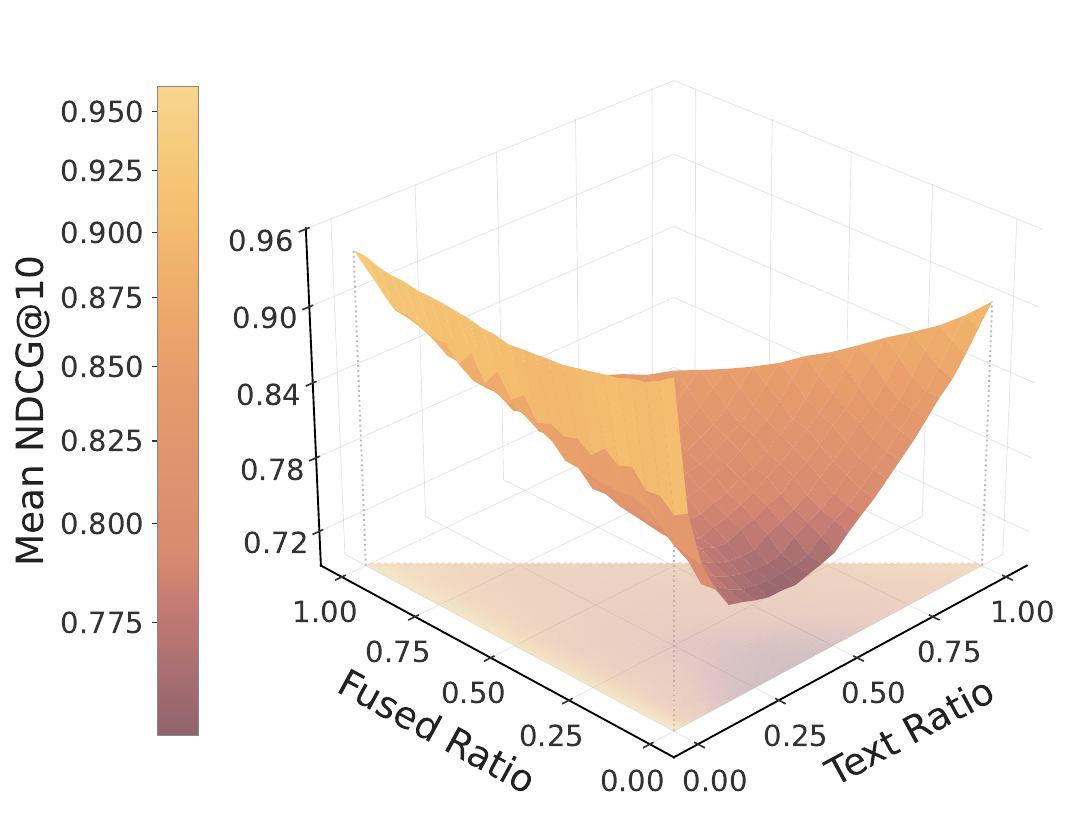}
        \caption{Qwen3-VL-E-8B on SlideVQA}
        \label{fig:ndcg_modelB_datasetZ}
    \end{subfigure}

    \caption{Retrieval performance is highly sensitive to the proportion of text, image, and fused documents in the corpus. Each surface shows mean NDCG@10 as the corpus's text and fused document ratios vary, across three models and two benchmarks.}
    \label{fig:ndcg_surfaces_grid}
\end{figure}

\subsection{Experimental Setup and Models}
\label{sec:how:setup}

\paragraph{Task formulation and metric.}
Given a text query $q$, retrieval aims to identify documents containing
the information needed to answer it.
Let $\mathcal{C}_{\mathrm{T}}$, $\mathcal{C}_{\mathrm{I}}$, and
$\mathcal{C}_{\mathrm{F}}$ denote the text, image, and fused text-image
candidate sets, respectively.
Text-to-image retrieval ranks candidates from an image-only corpus:
\begin{equation}
\operatorname{TopK}_{d\in\mathcal{C}_{\mathrm{I}}}\operatorname{sim}(q,d).
\end{equation}
Mixed-modality retrieval instead ranks candidates jointly across the three modalities:
\begin{equation}
\operatorname{TopK}_{d\in\mathcal{C}_{\mathrm{T}}\cup\mathcal{C}_{\mathrm{I}}\cup\mathcal{C}_{\mathrm{F}}}\operatorname{sim}(q,d),
\end{equation}
where $\operatorname{sim}(q,d)$ denotes query--document similarity. This requires relevance scores to be comparable across modalities within a unified ranking. Each query may have one or more relevant documents. Following common practice~\citep{muennighoff2023mteb,meng2025vlm2vec}, we adopt NDCG@10 as our primary metric.

\paragraph{Benchmarks and mixed-modality corpus construction.}
We evaluate on ChartQA~\citep{masry2022chartqa} and SlideVQA~\citep{tanaka2023slidevqa}, following the modality-replacement protocol of~\citet{li2025closing}. For each image document, we use Qwen3-VL-235B-A22B-Instruct~\citep{qwen3technicalreport} to generate a detailed description and a brief summary. The description forms the Text view, the original image forms the Image view, and the summary paired with the image forms the Fused view. Given target proportions of the three modalities, we independently sample one view per document according to these proportions. This preserves document identities and corpus size while varying the modality composition.

\paragraph{Models.}
\label{sec:how:models}
We evaluate several state-of-the-art CLIP-based and VLM-based retrieval models under their default settings; for models without native joint image-text input, we construct a combined embedding by averaging the unimodal image and text embeddings with equal weight (0.5 each). CLIP-based: CLIP-ViT-L/14~\citep{radford2021learning}, SigLIP2-L/16-384~\citep{tschannen2025siglip}, AltCLIP~\citep{chen2023altclip}, and Jina-CLIP-v2~\citep{koukounas2024jina}, each additionally reported with the post-hoc modality-gap calibration of GR-CLIP~\citep{li2025closing}. VLM-based: Qwen3-VL-Embedding-2B/8B~\citep{qwen3vlembedding}, VisRAG-Ret~\citep{yu2025visrag}, UniME-Phi3.5-V-4.2B~\citep{gu2025breaking}, and Jina-embeddings-v5-omni-small~\citep{honicke2026jina}.

\subsection{Results and discussion}
\label{sec:2.2}

To systematically evaluate the robustness of retrieval models in a mixed-modality corpus, we independently scan over compositions of text, image, and fused document ratios in the corpus, and compute the mean NDCG@10 at each composition, yielding a complete performance surface. Figure~\ref{fig:ndcg_surfaces_grid} presents results for three representative models, Jina-CLIP-v2, Qwen3-VL-Embedding-2B, and Qwen3-VL-Embedding-8B, on two benchmarks, ChartQA and SlideVQA\footnote{Results for the remaining models are shown in Appendix~\ref{sec:appd:ndcg_surface}.}.



\textbf{V-shaped performance under modality mixing.} 
As shown in Figure~\ref{fig:ndcg_surfaces_grid}, both CLIP-based and VLM-based visual retrievers exhibit similar V-shaped performance curves on the two benchmarks. Starting from an image-only corpus, retrieval performance drops markedly as image documents are progressively replaced with semantically corresponding text representations, reaches a trough when the two modalities are mixed, and then recovers as the text proportion increases and the corpus approaches a text-only composition. This indicates that models perform well on purely image or purely text corpora but struggle to maintain stable retrieval performance when the two coexist. In contrast, increasing the proportion of fused text-image documents does not induce a similar V-shaped degradation and instead alleviates the performance drop caused by mixing text and image documents. This contrast highlights the degradation arising from the coexistence of text-only and image-only documents as a prominent challenge in mixed-modality retrieval.

\begin{figure*}[t!]
  \centering
  \includegraphics[width=\textwidth]{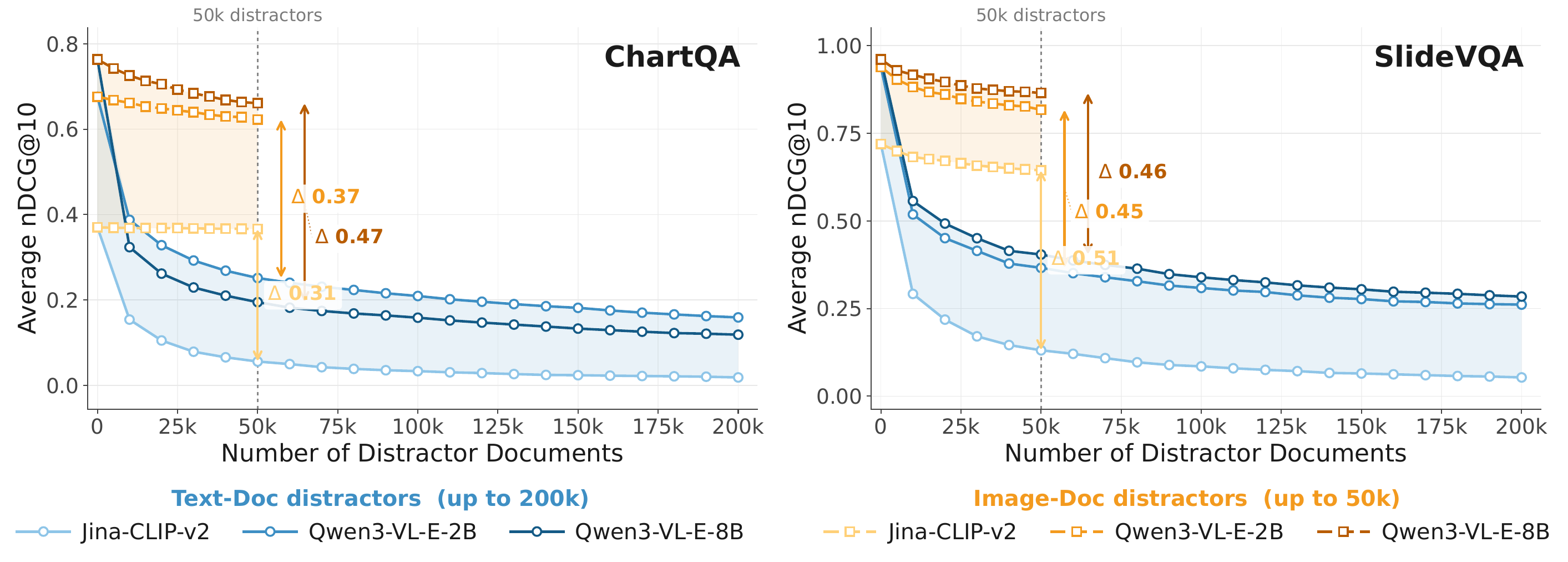}  
   \caption{Text distractors degrade retrieval performance far more severely than image distractors, despite both being equally irrelevant to every query. We report average NDCG@10 for three retrieval models on ChartQA (left) and SlideVQA (right) as the number of distractor documents increases.}
  \label{fig:distractor}
\end{figure*}

\textbf{``Chaos in the Text'' from text distractors.}
To further investigate this issue, we compare the effects of irrelevant documents of different modalities on retrieval performance in Figure~\ref{fig:distractor}. Specifically, we start with each benchmark’s original image-only corpus, separately add irrelevant text passages and images sampled from the Wikipedia corpus~\citep{sun2026hievidr}, and track retrieval performance as the number of distractors increases.

We find that, with the same number of distractors, irrelevant images have a relatively modest effect on retrieval performance, whereas irrelevant text causes a sharp decline. This difference persists across models and benchmarks, indicating that candidate-pool expansion alone cannot explain the degradation. We term this substantial deterioration in retrieval performance caused by introducing irrelevant text into an image corpus ``Chaos in the Text''. This phenomenon further suggests that models may exhibit systematic scoring biases across document modalities, allowing modality to influence rankings that should be determined by semantic relevance.

\section{Why do Retrieval Models Exhibit Chaos in the Text?}
\label{sec:why}

Given that Chaos in the Text is consistently observed across different models and benchmarks, its underlying cause remains unclear. We hypothesize that retrieval models may not rank documents purely by semantic relevance, but instead exhibit an inherent bias toward text documents. To test this, we construct three semantically identical, modality-distinct corpora (pure text, pure image, and text-image fused) and issue the same queries against each, comparing the resulting similarity distributions to examine whether modality itself systematically influences ranking.


\begin{figure*}[t!]
  \centering
  \includegraphics[width=\textwidth]{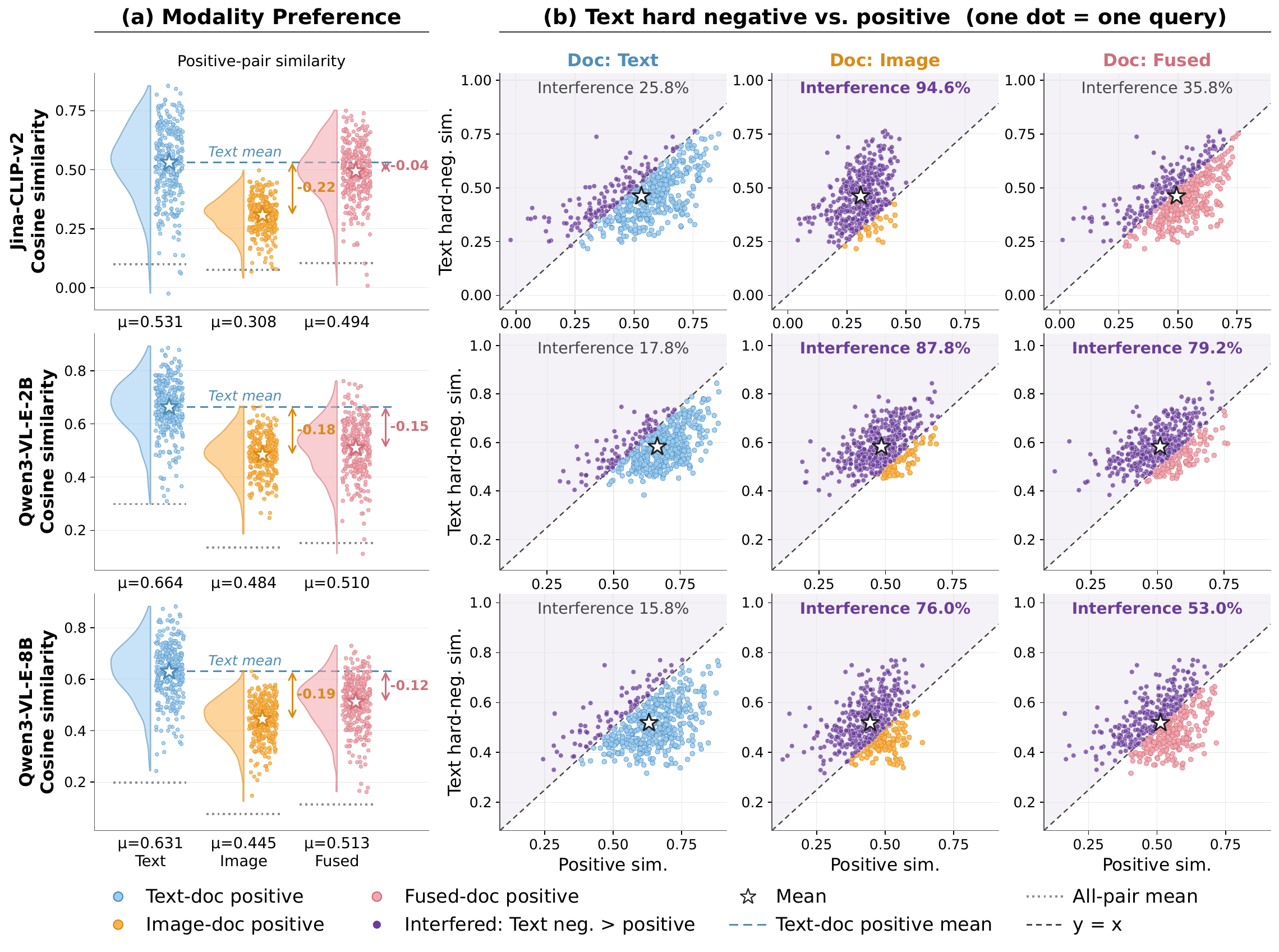}  
   \caption{(a) Violin plots show the similarity distributions across document modalities, revealing systematically higher similarity for text documents. (b) Each dot compares a query’s positive similarity with its text hard-negative similarity; points above y=x indicate cases where an irrelevant text document outranks the positive document, with the interference rate reporting their fraction.}
  \label{fig:preference}
\end{figure*}

\subsection{Experimental Setup}

We reuse the SlideVQA benchmark, retrieval models, and the three semantically aligned, modality-distinct parallel corpora (text, image, fused) from Section~\ref{sec:how:setup}. For each corpus, we compute cosine similarities between the same query set and its documents, yielding a similarity distribution per (query, doc modality) combination. We report two statistics: the all-pair mean, averaged over all query-document pairs, characterizing the overall similarity level of that modality; and the positive-pair mean, averaged only over ground-truth relevant pairs annotated in the qrels, characterizing the similarity score when relevance genuinely holds. A systematic discrepancy in these means across the three corpora would indicate that retrieval ranking is not governed purely by semantic relevance, but is confounded by modality itself.



\subsection{Results and discussion}
We analyze modality preference from two perspectives, examining cross-modal similarity differences among relevant documents and ranking interference from irrelevant text.

\paragraph{Cross-Modal Score Differences among Relevant Documents.}
As shown in Figure~\ref{fig:preference}(a), all three models exhibit similar score disparities for the same queries and their semantically corresponding relevant documents. The mean similarity between text queries and text positives is higher than that for image and fused text-image positives, with image positives showing the largest gap. Although the fused view narrows this gap, its similarity remains below that of the text view. This indicates that even when documents are relevant to the query, models assign different scores depending on their modality. The consistent trend across CLIP-based retrievers and VLM-based retrievers of different scales suggests that this modality preference is not specific to a particular architecture or model size.

\paragraph{Ranking Interference from Irrelevant Text.}
To examine whether these score differences affect the ranking of relevant and irrelevant documents, Figure~\ref{fig:preference}(b) compares each query’s similarity to its relevant document and a text hard negative. Each point represents a query, with positive similarity on the horizontal axis and text hard-negative similarity on the vertical axis. Points above the diagonal therefore indicate that irrelevant text receives a higher score than the relevant document.

When the relevant document is presented as an image, most queries lie above the diagonal across all three models, indicating that irrelevant text frequently outscores genuinely relevant images. This interference is substantially reduced when the relevant document is presented as text. Fused views alleviate interference to some extent, but a considerable proportion of queries remain affected, with varying improvements across models. These comparisons show that modality preference extends beyond differences in mean similarity to directly disrupt relevance-based ranking. The scoring advantage of text can outweigh the semantic relevance of image documents.

\section{How can we Mitigate Modality Preference?}

To alleviate the modality preference and promote joint learning across modalities, we introduce \textbf{trident}, a contrastive learning framework based on multiple positive views. Trident constructs text, image, and fused text-image views as three co-equal positives for the same query and jointly optimizes them within a shared candidate space. This enables relevant evidence across modalities to contribute to retrieval learning while reducing the influence of modality preference on ranking.

\begin{figure*}[t!]
  \centering
  \includegraphics[width=\textwidth]{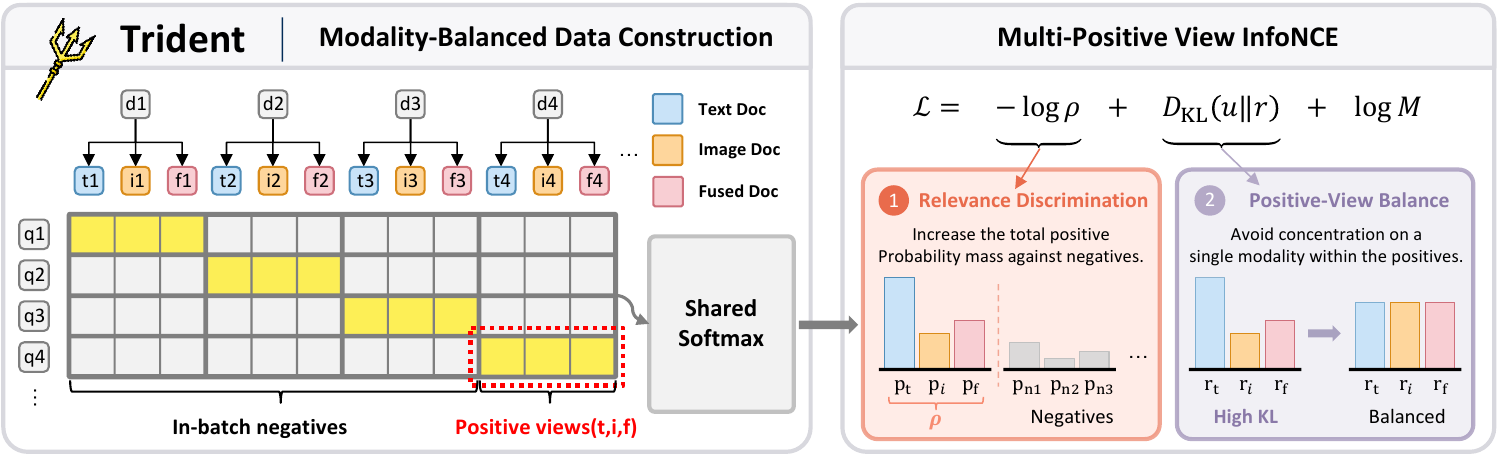}  
    \caption{\textbf{Overview of Trident.} Left: for each document, we construct tri-modal positive views (text, image, and fused), and all other views in the batch serve as negatives. The three positives and the negatives are jointly normalized by a shared softmax.
Right: Multi-Positive View InfoNCE decomposes into relevance discrimination, which increases the total positive probability mass, and positive-view balance, which prevents the mass from being dominated by a single modality.}
    \label{fig:method}
\end{figure*}

\subsection{trident}
\label{sec:trident}
We introduce trident through modality-balanced data construction and joint optimization of multiple positive views.
\paragraph{Modality-balanced data construction.}
To mitigate modality preference when matching queries with documents, we introduce Modality-Balanced Data Construction, which provides each query with multiple positive views across modalities. Existing training data typically pairs a query with a relevant document in a single modality, offering limited comparable supervision across modalities for the same query. We extend this single-positive pairing into a three-view pairing, enabling the model to jointly learn the relevance of different modality representations of the same document.

Specifically, we use Qwen3-VL-235B-A22B-Instruct to generate a detailed textual description and a concise summary for each image document. The detailed description forms the Text view, the original image forms the Image view, and the summary paired with the original image forms the Fused view. All three views serve as positives for the corresponding query, providing modality-balanced supervision for joint optimization.

\paragraph{Multi-positive view InfoNCE.}
After constructing modality-balanced multi-view data, training each modality with an independently normalized InfoNCE objective remains insufficient. Such objectives can learn within-modality rankings without constraining score baselines across modalities. For a query $q$ and modality $m\in\{t,i,f\}$, let $\mathcal{D}_m$ denote the candidate documents of modality $m$, including the positive document $d_m^{+}$ and the negatives. Let $s_m(q,d)$ denote the similarity score between $q$ and $d$, with $s_m^{+}=s_m(q,d_m^{+})$, and let $\tau>0$ denote the temperature. Adding the same modality-specific offset $c_m(q)$ to every candidate score leaves the within-modality softmax probability unchanged:
\begin{equation}
\frac{\exp\!\left((s_m^{+}+c_m(q))/\tau\right)}
{\sum_{d\in\mathcal{D}_m}
\exp\!\left((s_m(q,d)+c_m(q))/\tau\right)}
=
\frac{\exp(s_m^{+}/\tau)}
{\sum_{d\in\mathcal{D}_m}
\exp(s_m(q,d)/\tau)}.
\label{eq:modality_shift}
\end{equation}
Thus, even with balanced training samples across modalities, their relevance scores are not necessarily directly comparable.

To constrain cross-modal score baselines and mitigate modality preference, we introduce Multi-Positive View InfoNCE, which places all three positive views and the in-batch negatives in a shared softmax denominator and averages the log-probabilities of the positives. Given the positive-view set $\mathcal{P}=\{t,i,f\}$ for query $q$ and the negative set $N$, the training objective is
\begin{equation}
\mathcal{L}_\text{trident}(q)
=
-\frac{1}{|\mathcal{P}|}
\sum_{p\in\mathcal{P}}
\log
\frac{\exp(\mathrm{sim}(q,p)/\tau)}
{\displaystyle
\sum_{j\in\mathcal{P}\cup\mathcal{N}}
\exp(\mathrm{sim}(q,j)/\tau)}.
\label{eq:mp_infonce}
\end{equation}
Within this shared normalization space, each positive is compared against candidate documents across modalities, jointly constraining relative scores both within and across modalities.

\paragraph{Objective interpretation.}
To understand how this objective mitigates modality preference, we decompose Eq.~\eqref{eq:mp_infonce} into relevance discrimination and positive-view balance. Let $s_j=\operatorname{sim}(q,j)$, and define
\begin{equation}
p_j =
\frac{\exp(s_j/\tau)}
{\sum_{k\in\mathcal{P}\cup\mathcal{N}}\exp(s_k/\tau)},
\qquad
\rho=\sum_{m\in\mathcal{P}}p_m,
\qquad
r_m=\frac{p_m}{\rho},
\end{equation}
where $\rho$ is the total probability mass assigned to all positives, and $r_m$ is the relative share of the $m$-th view within the positive set. Let $M=|P|$ and $u_m=1/M$. The objective can then be rewritten as
\begin{equation}
\mathcal{L}_\text{trident}(q)
=
\underbrace{-\log\rho}_{\text{relevance discrimination}}
+
\underbrace{D_{\mathrm{KL}}(u\|r)}_{\text{positive-view balance}}
+
\log M.
\label{eq:loss_decomposition}
\end{equation}

The first term encourages the positive set as a whole to capture probability mass from the negatives, thereby distinguishing relevant from irrelevant documents. With this term alone, however, the model may rely primarily on the easiest-to-match text positive while neglecting the image or fused view. The second term penalizes probability imbalance within the positive set, preventing a single modality from monopolizing the positive probability mass and encouraging the model to learn relevant evidence across modalities.

The complete decomposition and gradient derivation are provided in the Appendix~\ref{app:objective_analysis}. For the current query, this objective increases the competitiveness of relevant images receiving insufficient probability mass and lowers the scores of high-scoring irrelevant text, while discouraging relevant text from dominating the positive probability mass. By jointly enforcing relevance discrimination and positive-view balance, trident translates modality-balanced supervision into cross-modal ranking constraints, thereby mitigating modality preference.

\subsection{Results and discussion}
We evaluate trident through overall retrieval performance, robustness to changes in modality composition, and relationships between modality representations. The detailed experimental setup is shown in the Appendix~\ref{sec:appd:training}.

\paragraph{Overall performance.}
Table~\ref{tab:main_results} reports NDCG@10 on six visual document retrieval benchmarks under Origin, each benchmark’s original image-only corpus, and Mix, a unified mixed-modality pool merged across the six benchmarks. We compare CLIP-based and VLM-based retrievers, GR-CLIP-calibrated models~\citep{li2025closing}, and trident variants.

Trident method achieves the highest Mix average within both architecture categories. Most baselines degrade substantially from Origin to Mix, while GR-CLIP partially improves mixed-modality retrieval but remains below trident, indicating additional benefits from modality-balanced training beyond post-hoc calibration. trident-Qwen3VL-2B also outperforms the larger Qwen3-VL-Embedding-8B on Mix, and both trident variants improve Origin averages over their corresponding baselines. Compared with w/o Multi-Positive View InfoNCE, the full objective further improves average performance on both architectures, particularly the VLM backbone, demonstrating the complementary benefits of balanced data and joint optimization.

\begin{table*}[t]
\centering
\caption{Main results on six visual document retrieval benchmarks under the \textbf{Origin} (each dataset's original, single-modality corpus) and \textbf{Mix} (mixed-modality corpora merged across all six datasets) settings. The best result in each column is in \textbf{bold}; the second best is \underline{underlined}. Our proposed models are highlighted in gray, with the ablation (w/o Multi-Positive View InfoNCE) shown below.}
\label{tab:main_results}
\renewcommand{\arraystretch}{1.15}  
\setlength{\tabcolsep}{4pt}          
\resizebox{\textwidth}{!}{%
\begin{tabular}{l *{14}{c}}
\toprule
\multirow{2.5}{*}{\textbf{Model}}
 & \multicolumn{2}{c}{ChartQA} & \multicolumn{2}{c}{DocVQA} & \multicolumn{2}{c}{InfoVQA}
 & \multicolumn{2}{c}{SlideVQA} & \multicolumn{2}{c}{ViDoSeek} & \multicolumn{2}{c}{Dude}
 & \multicolumn{2}{c}{\textbf{Avg}} \\
\cmidrule(lr){2-3} \cmidrule(lr){4-5} \cmidrule(lr){6-7}
\cmidrule(lr){8-9} \cmidrule(lr){10-11} \cmidrule(lr){12-13} \cmidrule(lr){14-15}
 & Origin & Mix & Origin & Mix & Origin & Mix & Origin & Mix & Origin & Mix & Origin & Mix & Origin & Mix \\
\midrule
\multicolumn{15}{c}{\textit{CLIP-based Models}} \\
\midrule
SigLIP2-L/16-384
  & \textbf{43.98} & 4.06 & 32.91 & 2.73 & 54.86 & 2.10 & 56.04 & 5.20 & 48.52 & 5.00 & 8.84 & 3.95 & 40.86 & 3.84 \\
\quad + GR-CLIP
  & \underline{40.09} & 19.59 & 31.74 & 11.07 & 54.95 & 16.88 & 56.80 & 28.00 & 53.62 & 22.62 & 11.60 & 4.45 & 41.47 & 17.10 \\
CLIP-ViT-L/14
  & 22.48 & 18.44 & 32.14 & 1.22 & 68.11 & 6.49 & 74.38 & 6.81 & 65.15 & 19.01 & 26.13 & 3.18 & 48.07 & 9.19 \\
\quad + GR-CLIP
  & 19.77 & 26.87 & 31.31 & 17.81 & 67.07 & 48.25 & 74.39 & 46.55 & \underline{68.10} & 50.34 & \underline{26.76} & 26.22 & 47.90 & 36.01 \\
AltCLIP
  & 23.30 & 14.64 & 33.84 & 14.13 & \textbf{69.18} & 25.01 & \textbf{76.24} & 21.66 & 67.99 & 29.85 & 26.61 & 12.73 & 49.53 & 19.67 \\
\quad + GR-CLIP
  & 20.20 & 26.39 & 32.94 & 24.84 & 68.72 & 51.00 & 75.77 & 47.82 & \textbf{69.48} & 46.24 & 25.86 & 24.88 & 48.83 & 36.86 \\
Jina-CLIP-v2
  & 37.00 & 45.53 & 46.23 & 31.32 & 64.10 & 45.84 & 71.93 & 52.08 & 64.32 & 53.01 & 24.94 & 28.32 & 51.42 & 42.68 \\
\quad + GR-CLIP
  & 34.94 & 45.75 & 45.56 & 34.39 & 62.96 & 54.03 & 73.11 & 60.64 & 64.92 & 59.77 & 25.92 & 32.17 & 51.24 & 47.79 \\
\rowcolor{gray!15}
\textbf{Trident-JinaCLIP} \textit{(Ours)}
  & 31.40 & \underline{50.25} & \textbf{54.10} & \textbf{46.47} & \underline{68.73} & \textbf{67.57} & \underline{75.84} & \textbf{67.06} & 61.78 & \textbf{61.67} & 26.36 & \textbf{40.06} & \textbf{53.04} & \textbf{55.51} \\
\rowcolor{gray!15}
\quad \textit{w/o Multi-Positive View InfoNCE} 
  & 31.08 & \textbf{51.06} & \underline{52.70} & \underline{45.21} & 68.29 & \underline{66.21} & 75.38 & \underline{67.22} & 61.51 & \underline{61.55} & \textbf{27.73} & \underline{39.42} & \underline{52.78} & \underline{55.11} \\
\midrule
\multicolumn{15}{c}{\textit{VLM-based Models}} \\
\midrule
UniME-Phi3.5-V-4.2B
  & 48.54 & 55.84 & 64.77 & 39.68 & 76.77 & 58.50 & 85.73 & 72.46 & 83.97 & \underline{74.93} & 44.02 & 42.71 & 67.30 & 57.35 \\
VisRAG-Ret
  & 55.57 & 26.91 & 79.27 & 45.49 & 88.88 & 55.32 & 92.46 & 57.58 & 80.32 & 55.63 & 57.80 & 37.28 & 75.72 & 46.37 \\
Jina-embeddings-v5-omni-small
  & 64.70 & 56.50 & 77.80 & 50.53 & 87.12 & 64.91 & 93.26 & 77.01 & 81.31 & 72.61 & 60.17 & \underline{49.90} & 77.39 & 61.91 \\
Qwen3-VL-Embedding-2B
  & 67.60 & 50.72 & 81.66 & 18.83 & 90.07 & 36.77 & \underline{93.94} & 45.75 & \underline{85.35} & 59.13 & 46.92 & 29.50 & 77.59 & 40.12 \\
Qwen3-VL-Embedding-8B
  & \textbf{76.35} & 45.13 & \textbf{86.64} & 27.56 & \textbf{93.62} & 44.56 & \textbf{96.09} & 55.30 & \textbf{86.26} & 73.57 & \textbf{61.85} & 36.87 & \textbf{83.47} & 47.17 \\
\rowcolor{gray!15}
\textbf{Trident-Qwen3VL-2B} \textit{(Ours)}
  & \underline{70.19} & \textbf{65.82} & \underline{82.76} & \textbf{67.68} & \underline{91.28} & \textbf{80.09} & 93.72 & \textbf{88.54} & 83.74 & \textbf{81.98} & \underline{61.36} & \textbf{59.31} & \underline{80.51} & \textbf{73.90} \\
\rowcolor{gray!15}
\quad \textit{w/o Multi-Positive View InfoNCE} 
    & 67.41 & \underline{63.12} & 80.57 & \underline{54.44} & 89.47 & \underline{72.26} & 93.06 & \underline{83.97} & 78.43 & 74.76 & 55.20 & 49.58 & 77.36 & \underline{66.36} \\
\bottomrule
\end{tabular}
}
\end{table*}

\begin{figure}[t]
    \centering
    \begin{subfigure}[b]{0.49\textwidth}
        \centering
        \includegraphics[width=\textwidth]{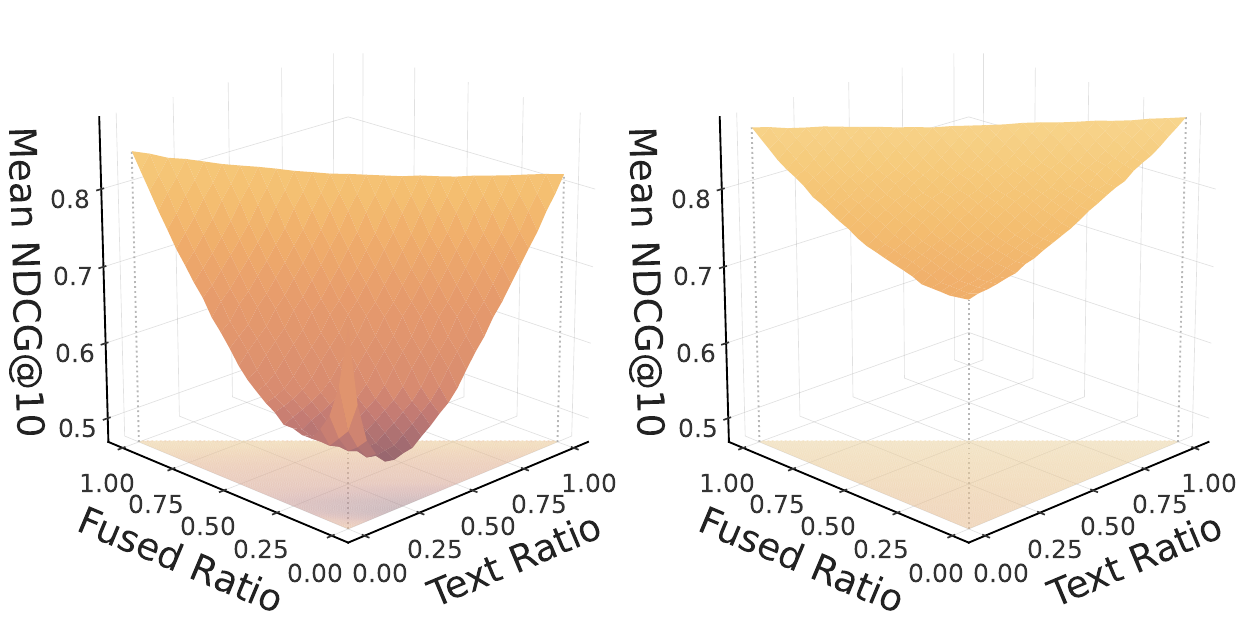}
        \caption{Training Comparison on Jina-Clip-v2}
        \label{fig:ndcg_compare_a}
    \end{subfigure}
    \hfill
    \begin{subfigure}[b]{0.49\textwidth}
        \centering
        \includegraphics[width=\textwidth]{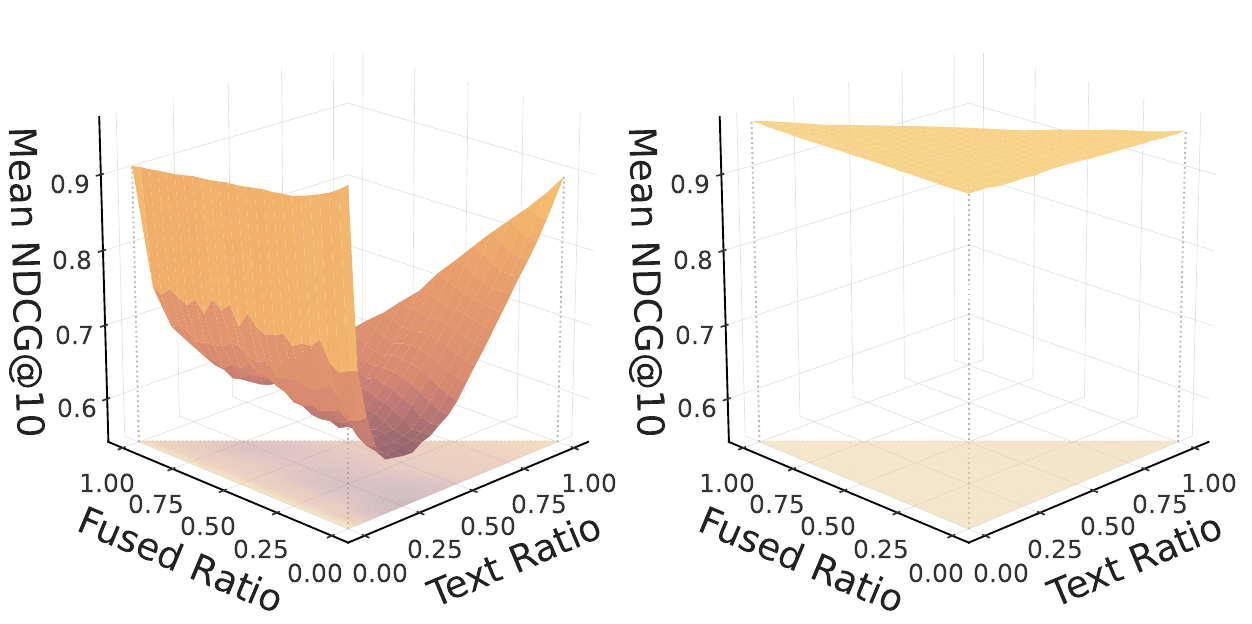}
        \caption{Training Comparison on Qwen3-VL-E-2B}
        \label{fig:ndcg_compare_b}
    \end{subfigure}

    \caption{Mean NDCG@10 on SlideVQA as a function of corpus modality composition (Fused Ratio, Text Ratio), before (left) and after (right) Trident training.}
    \vspace{-0.1in}
    \label{fig:ndcg_compare}
\end{figure}

\paragraph{Robustness to modality composition.}
Figure~\ref{fig:ndcg_compare} compares retrieval performance across modality proportions before and after training. Consistent with Section~\ref{sec:2.2}, both architectures initially exhibit pronounced valleys when different modalities coexist. After trident training, these valleys are substantially reduced, and the surfaces become higher and flatter. 

The VLM model achieves particularly stable performance, while the CLIP model retains only moderate variation. These consistent changes show that trident’s improvements extend across mixing ratios, reducing sensitivity to corpus modality composition.

\begin{figure*}[t!]
  \centering
  \includegraphics[width=\textwidth]{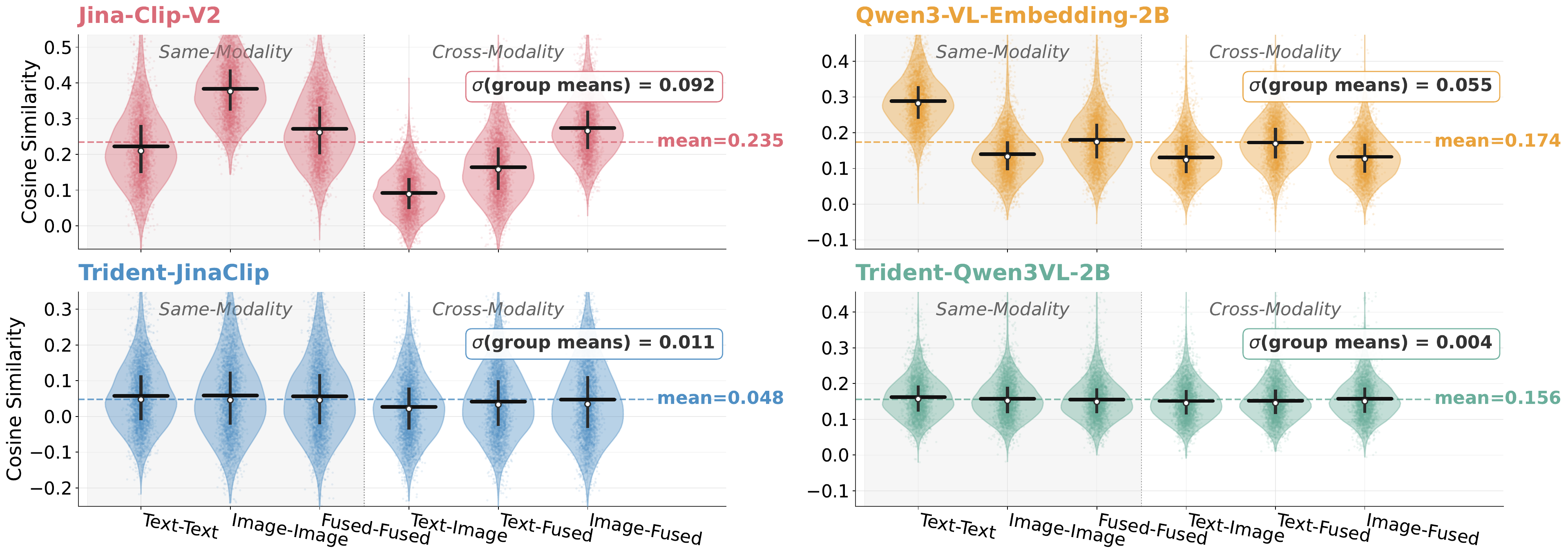}  
   \caption{Distribution of cosine similarities between same-modality and cross-modality positive pairs, before (top) and after (bottom) Trident training, on Jina-CLIP-v2 and Qwen3-VL-E-2B.}
  \label{fig:sim_change}
  \vspace{-0.2in}
\end{figure*}

\paragraph{Mitigation of modality preference.}
Figure~\ref{fig:sim_change} further examines similarity distributions between same-modality and cross-modality document representations. Before training, distribution centers differ markedly across modality pairs. Jina-CLIP-v2~\citep{koukounas2024jina} exhibits higher Image–Image similarity, whereas Qwen3-VL-Embedding-2B~\citep{qwen3vlembedding} favors Text–Text similarity, revealing architecture-dependent modality separation.

After trident training, these same-modality advantages diminish, and the distribution centers converge, narrowing the differences between same-modality and cross-modality similarities. Together with the flatter performance surfaces in Figure~\ref{fig:ndcg_compare}, these changes support the conclusion that trident mitigates modality preference and enables more stable mixed-modality retrieval.

\section{Related Work}

\textbf{Multimodal retrieval models.} Dense multimodal retrieval was established by dual-tower contrastive encoders such as CLIP~\citep{radford2021learning}, ALIGN~\citep{jia2021scaling}, and their successors SigLIP/SigLIP 2~\citep{zhai2023sigmoid,tschannen2025siglip}, which align text and images in a shared space but encode each modality through a separate tower. Research subsequently shifted toward vision-language model (VLM) backbones that ingest interleaved text and image tokens in a single transformer and emit one unified embedding, including VLM2Vec and VLM2Vec-V2~\citep{jiang2024vlm2vec,meng2025vlm2vec}, GME~\citep{zhang2024gme}, UniME~\citep{gu2025breaking}, and Qwen3-VL-Embedding~\citep{qwen3vlembedding}, which now dominate universal-embedding leaderboards. However, these systems, together with the standard benchmarks used to evaluate them, such as BEIR~\citep{thakur2021beir}, MTEB~\citep{muennighoff2023mteb}, and the document VQA datasets we adopt, are almost entirely built on single-modality corpora, and therefore cannot assess how multimodal retrieval models perform when retrieving over a mixed-modality corpus.

\textbf{Modality gap.} Prior work has shown that contrastive vision-language embeddings are not modality-agnostic.~\citet{liang2022mind} first noted the modality gap, in which image and text embeddings occupy disjoint cones of the hypersphere, an artifact of initialization and the contrastive temperature~\citep{shi2023towards}; later work attributes this to information imbalance across modalities~\citep{schrodi2025two,fahim2024s} and attempts to close it via geometric interventions such as manifold mixup~\citep{oh2023geodesic,eslami2024mitigate}. Yet these studies measure the gap purely geometrically and evaluate it on cross-modal alignment tasks, where a uniform offset between the two clusters is harmless, since all candidates share a modality and the offset cancels out in ranking. We find, however, that in mixed-modality retrieval, what corrupts the ranking is not the geometric modality gap, but a modality preference internal to the model.


\section{Conclusion}

In this paper, we reveal the sensitivity of multimodal retrieval models to corpus modality composition and identifies \textbf{\textit{``Chaos in the Text''}}, a phenomenon in which introducing irrelevant text into an image corpus substantially degrades retrieval performance. Further analysis reveals systematic biases in similarity scores across modalities, compromising the comparability of candidate documents. To mitigate this issue, we introduce trident, which combines modality-balanced data construction with Multi-Positive View InfoNCE to jointly promote relevance discrimination and positive-view balance. Experiments on CLIP-based and VLM-based architectures show that trident improves mixed-modality retrieval, reduces sensitivity to modality proportions, and increases average single-modality retrieval performance.

\subsection*{AI Use Statement}

We used generative AI tools to assist with writing and language polishing, including improving the clarity, readability, and presentation of the manuscript. All AI-assisted content was reviewed and revised by the authors. The authors take full responsibility for the final content of this work.

\bibliography{iclr2027/iclr2027_conference}
\bibliographystyle{iclr2027/iclr2027_conference}

\newpage
\appendix

\section{NDCG@10 Surfaces for the Remaining Retrievers}
\label{sec:appd:ndcg_surface}

\begin{figure}[t]
    \centering
    \begin{subfigure}[b]{0.32\textwidth}
        \centering
        \includegraphics[width=\textwidth]{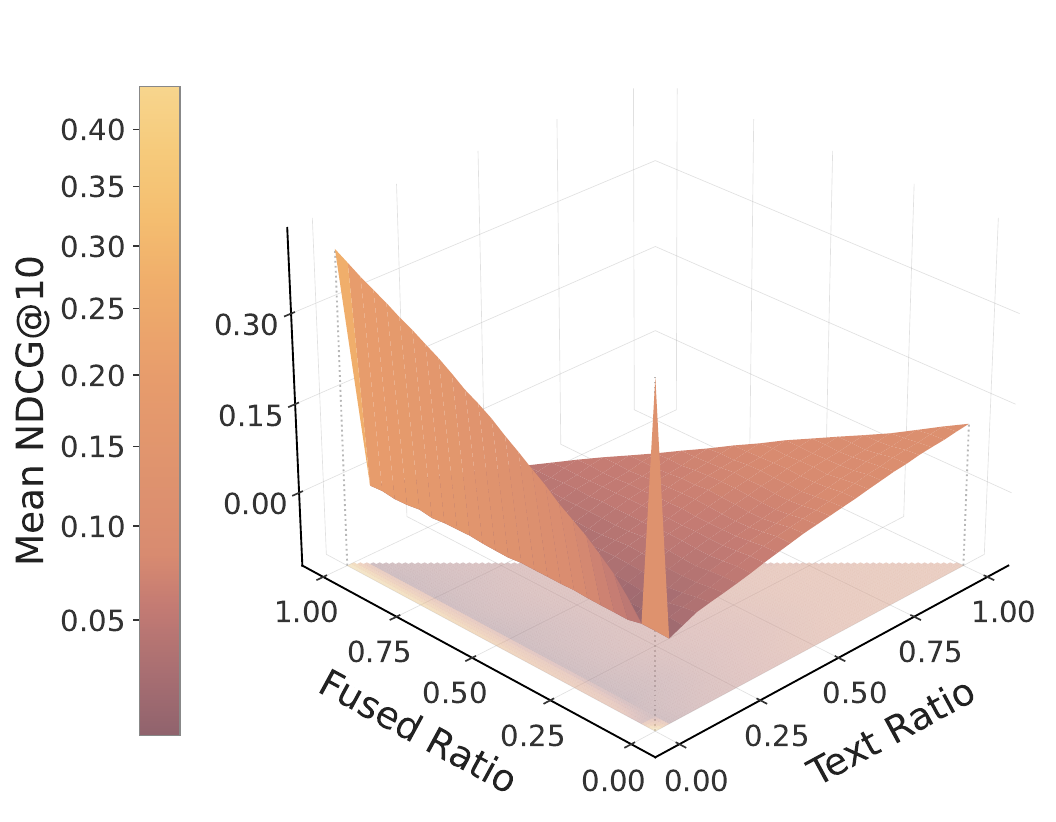}
        \caption{SigLIP2-L on ChartQA}
        \label{fig:ndcg_modelA_datasetX}
    \end{subfigure}
    \hfill
    \begin{subfigure}[b]{0.32\textwidth}
        \centering
        \includegraphics[width=\textwidth]{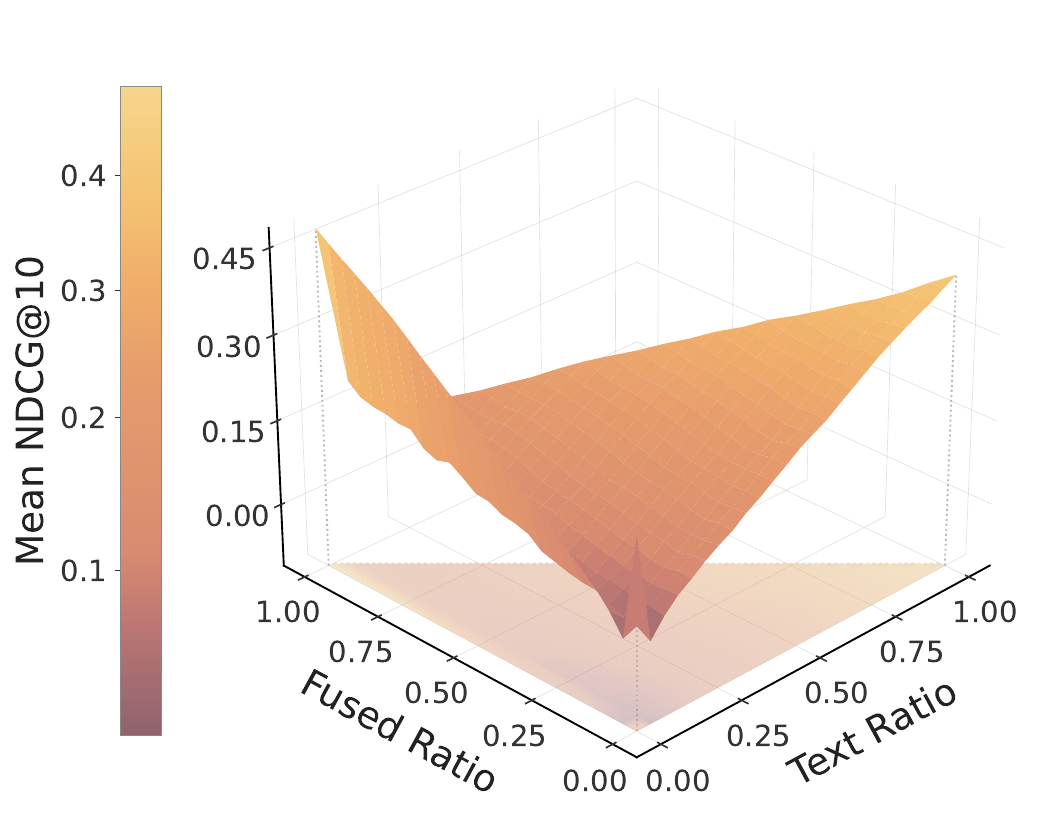}
        \caption{CLIP-ViT-L/14 on ChartQA}
        \label{fig:ndcg_modelA_datasetY}
    \end{subfigure}
    \hfill
    \begin{subfigure}[b]{0.32\textwidth}
        \centering
        \includegraphics[width=\textwidth]{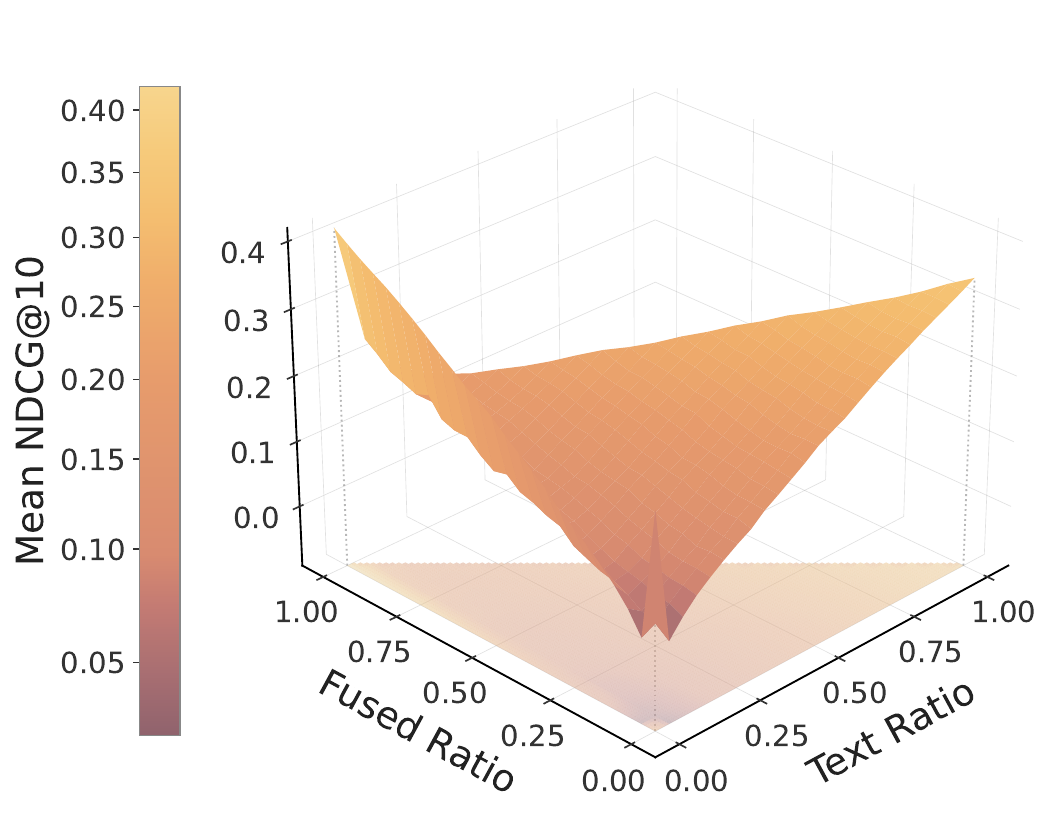}
        \caption{AltCLIP on ChartQA}
        \label{fig:ndcg_modelA_datasetZ}
    \end{subfigure}

    \vspace{0.5em}

    \begin{subfigure}[b]{0.32\textwidth}
    \centering
    \includegraphics[width=\textwidth]{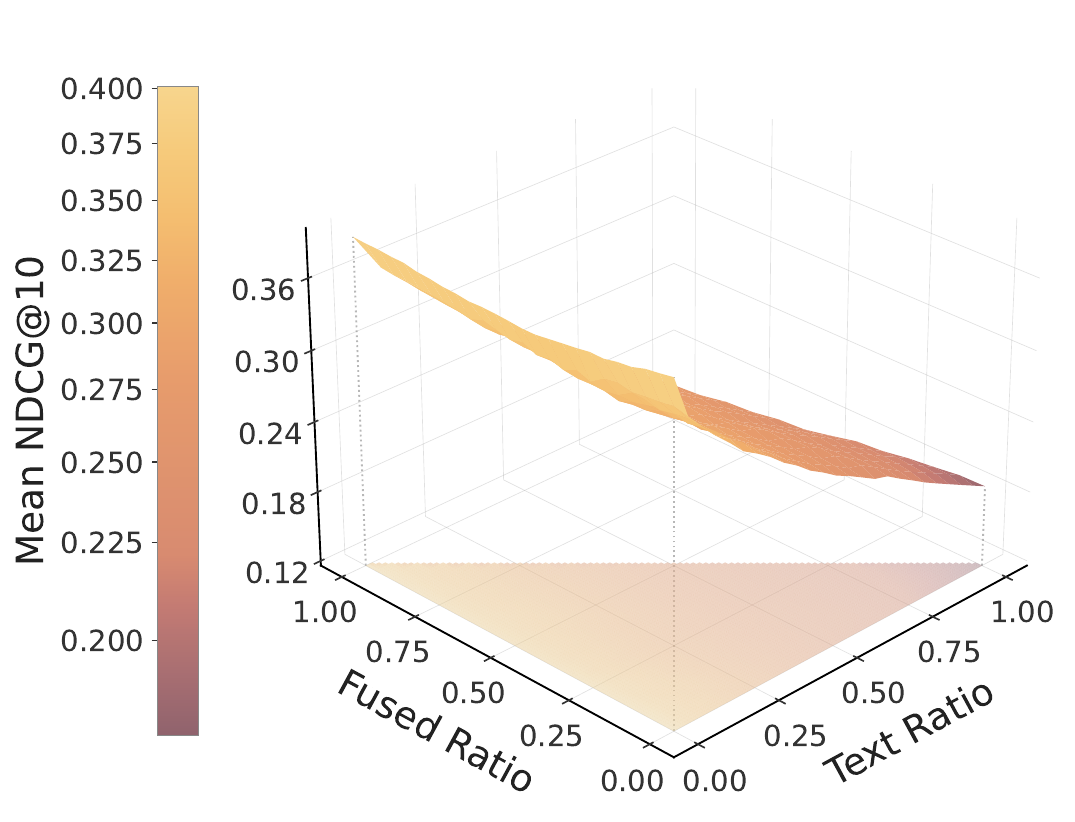}
    \caption{\quad+GR-CLIP on ChartQA}
    \label{fig:ndcg_modelA_datasetX}
    \end{subfigure}
    \hfill
    \begin{subfigure}[b]{0.32\textwidth}
        \centering
        \includegraphics[width=\textwidth]{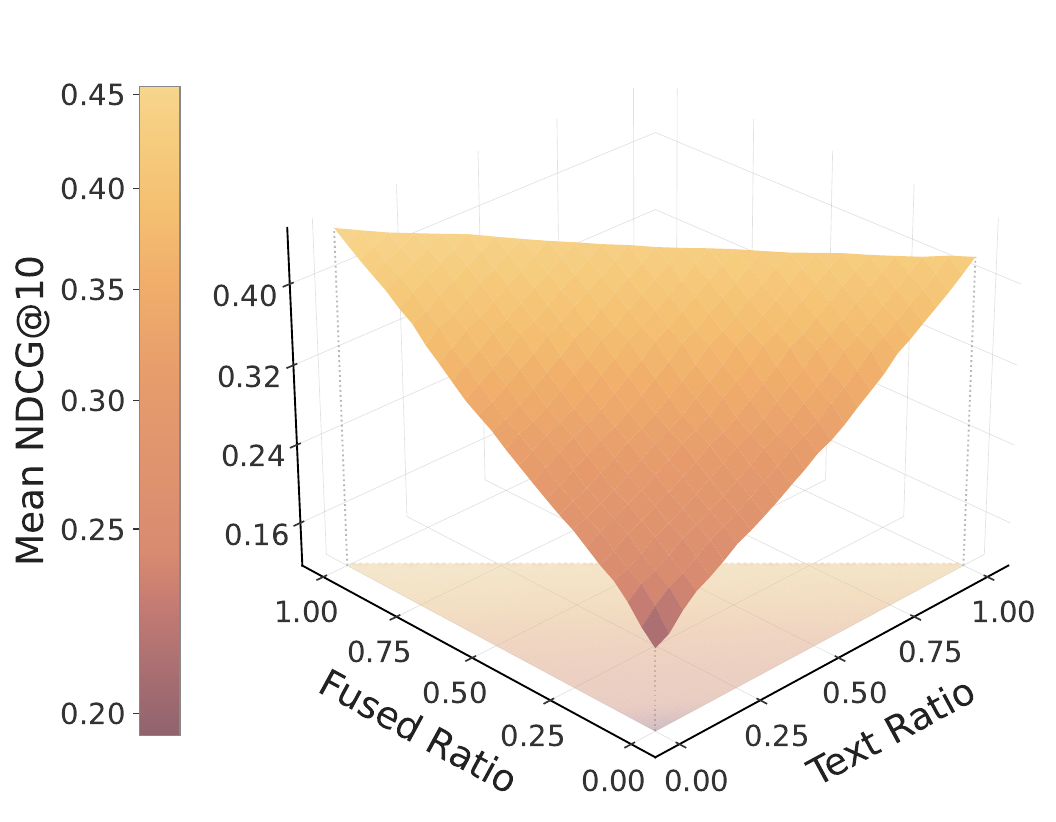}
        \caption{\quad+GR-CLIP on ChartQA}
        \label{fig:ndcg_modelA_datasetY}
    \end{subfigure}
    \hfill
    \begin{subfigure}[b]{0.32\textwidth}
        \centering
        \includegraphics[width=\textwidth]{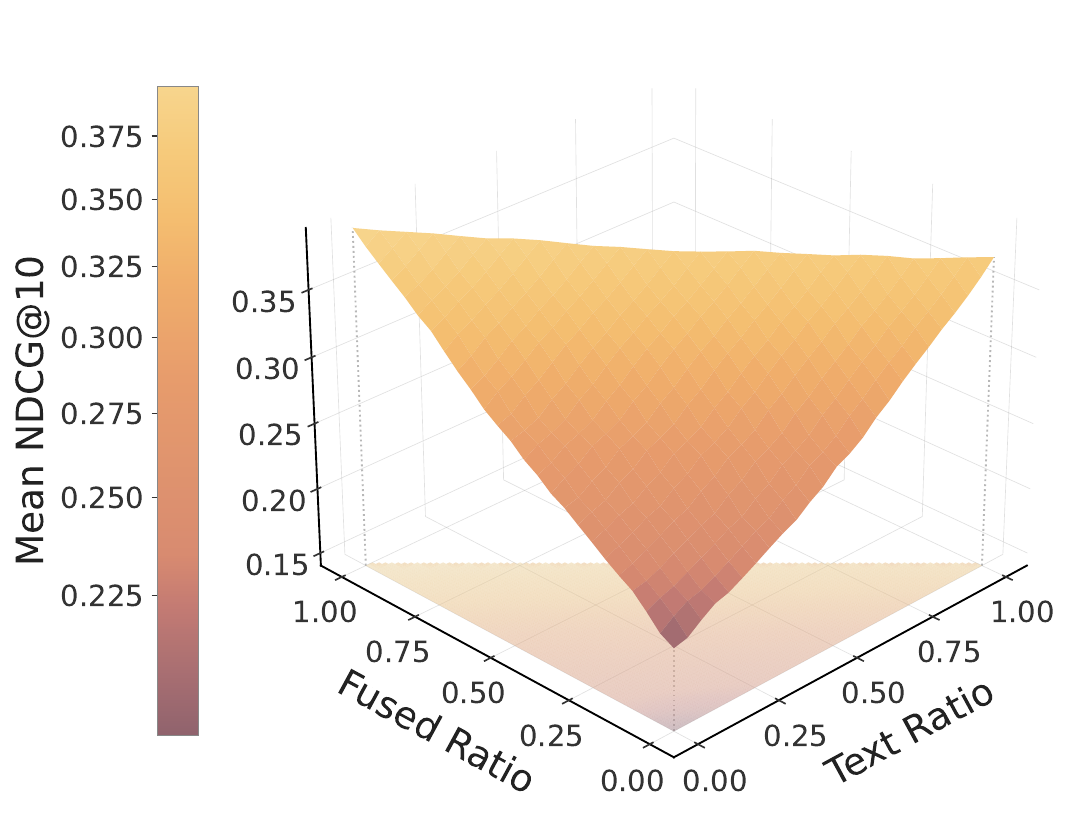}
        \caption{\quad+GR-CLIP on ChartQA}
        \label{fig:ndcg_modelA_datasetZ}
    \end{subfigure}

    \vspace{0.5em}

    \begin{subfigure}[b]{0.32\textwidth}
    \centering
    \includegraphics[width=\textwidth]{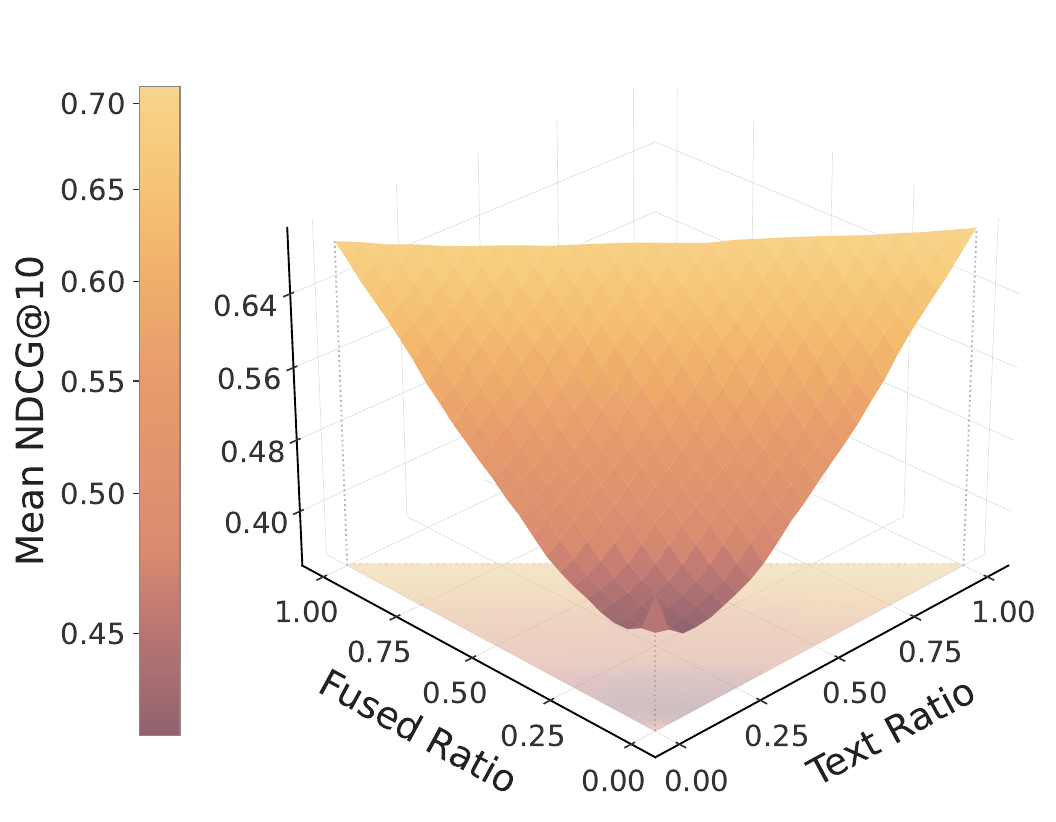}
    \caption{UniME-Phi3.5 on ChartQA}
    \label{fig:ndcg_modelA_datasetX}
    \end{subfigure}
    \hfill
    \begin{subfigure}[b]{0.32\textwidth}
        \centering
        \includegraphics[width=\textwidth]{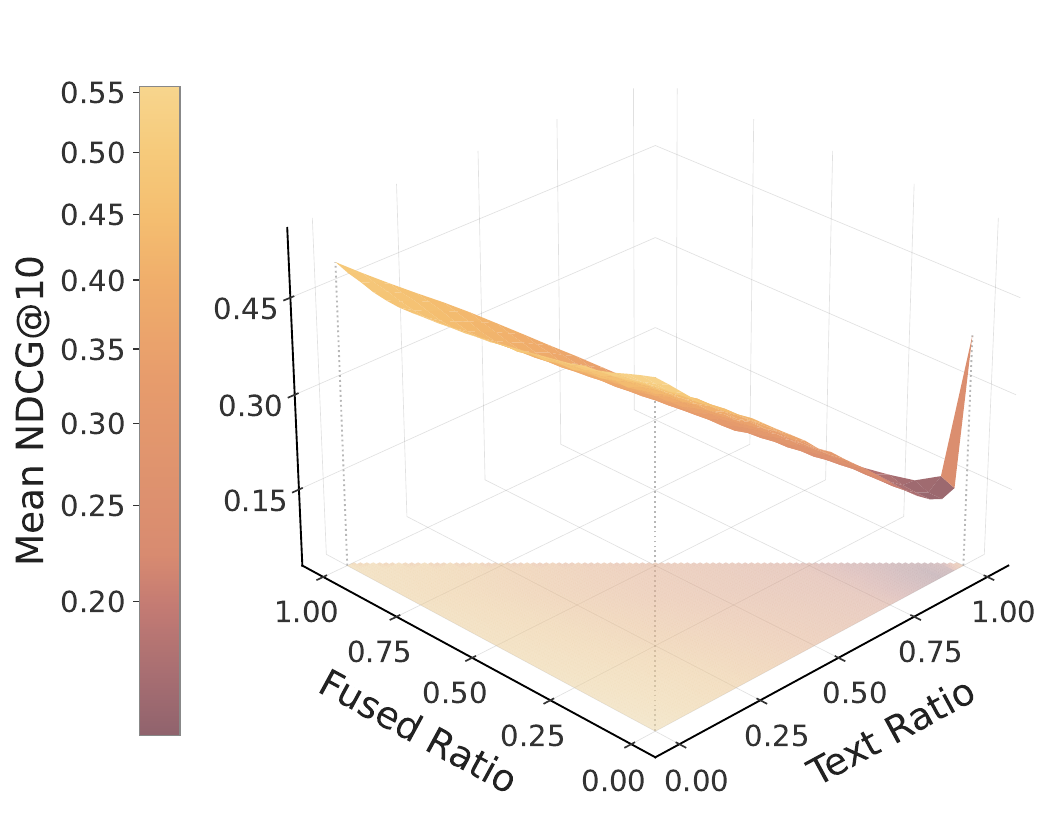}
        \caption{VisRAG-Ret on ChartQA}
        \label{fig:ndcg_modelA_datasetY}
    \end{subfigure}
    \hfill
    \begin{subfigure}[b]{0.32\textwidth}
        \centering
        \includegraphics[width=\textwidth]{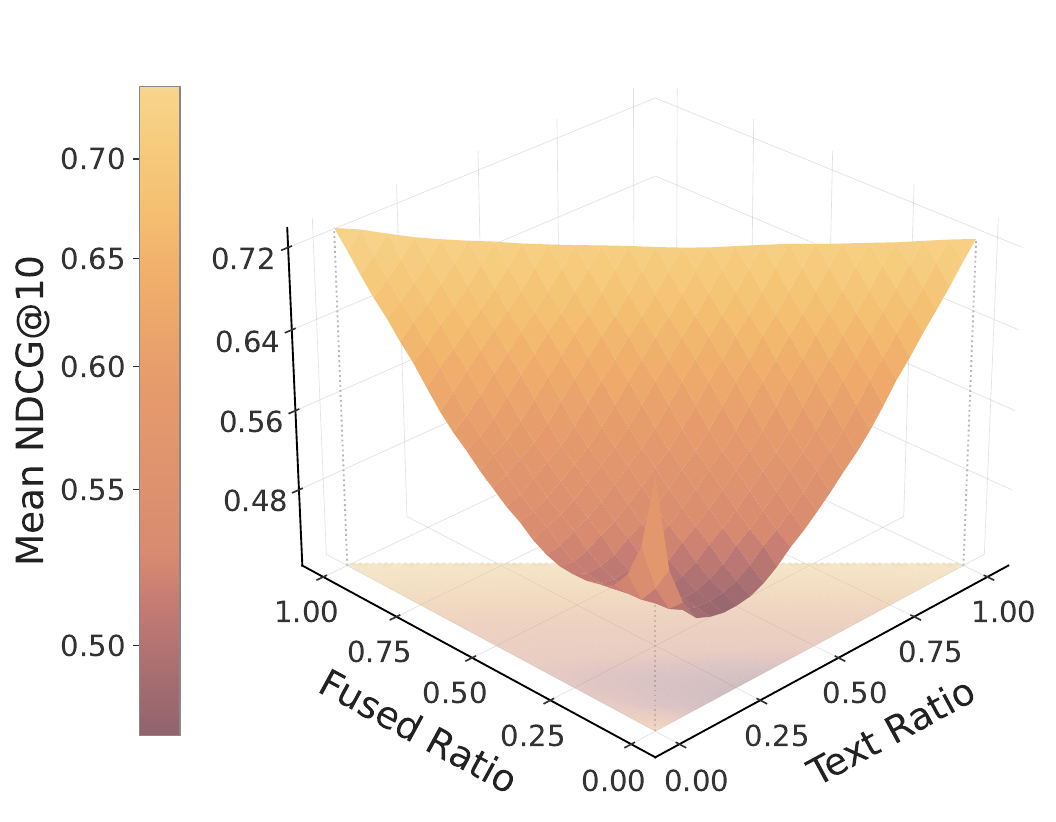}
        \caption{Jina-V5-small on ChartQA}
        \label{fig:ndcg_modelA_datasetZ}
    \end{subfigure}

    \vspace{0.5em}

    \caption{Mean NDCG@10 as a function of corpus modality composition (text ratio, fused ratio) across nine models on ChartQA. The three ratios sum to one, $\textit{image ratio} = 1 - \textit{text ratio} - \textit{fused ratio}$, so the surface in the figure is indexed by only two free variables, (text ratio, fused ratio).}
    \label{fig:ndcg_surfaces_grid_appd1}
\end{figure}

\begin{figure}[t]
    \centering
    \begin{subfigure}[b]{0.32\textwidth}
        \centering
        \includegraphics[width=\textwidth]{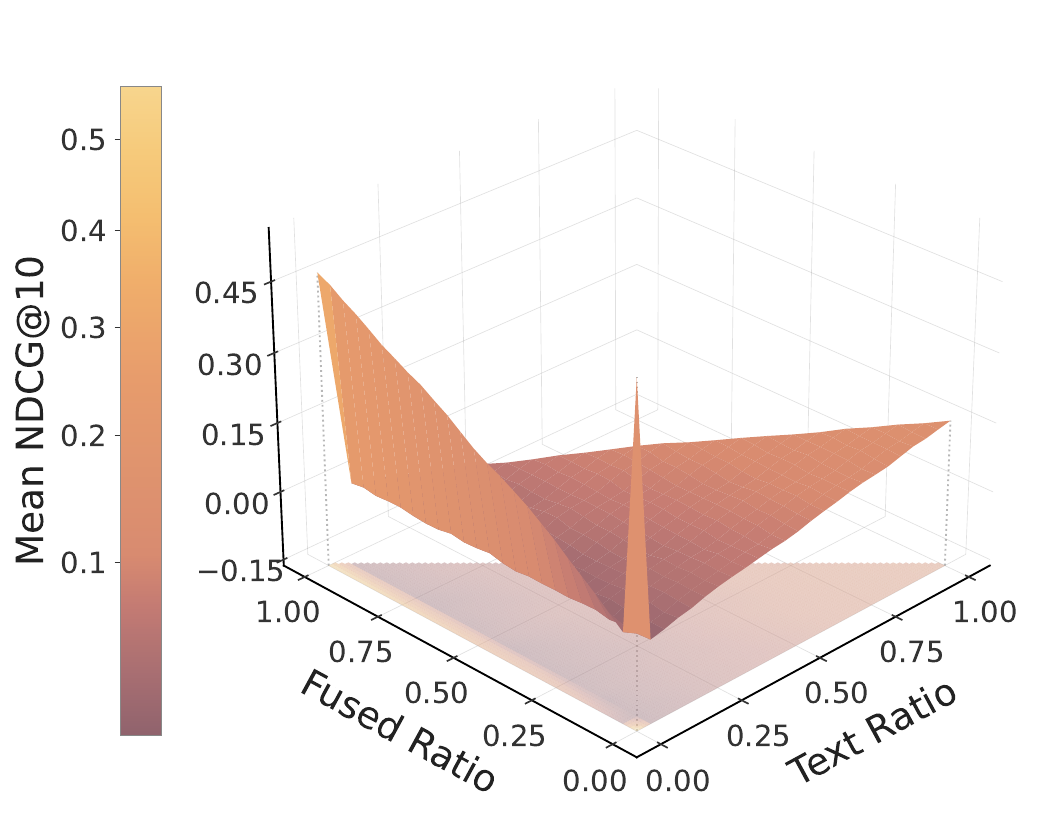}
        \caption{SigLIP2-L on SlideVQA}
        \label{fig:ndcg_modelA_datasetX}
    \end{subfigure}
    \hfill
    \begin{subfigure}[b]{0.32\textwidth}
        \centering
        \includegraphics[width=\textwidth]{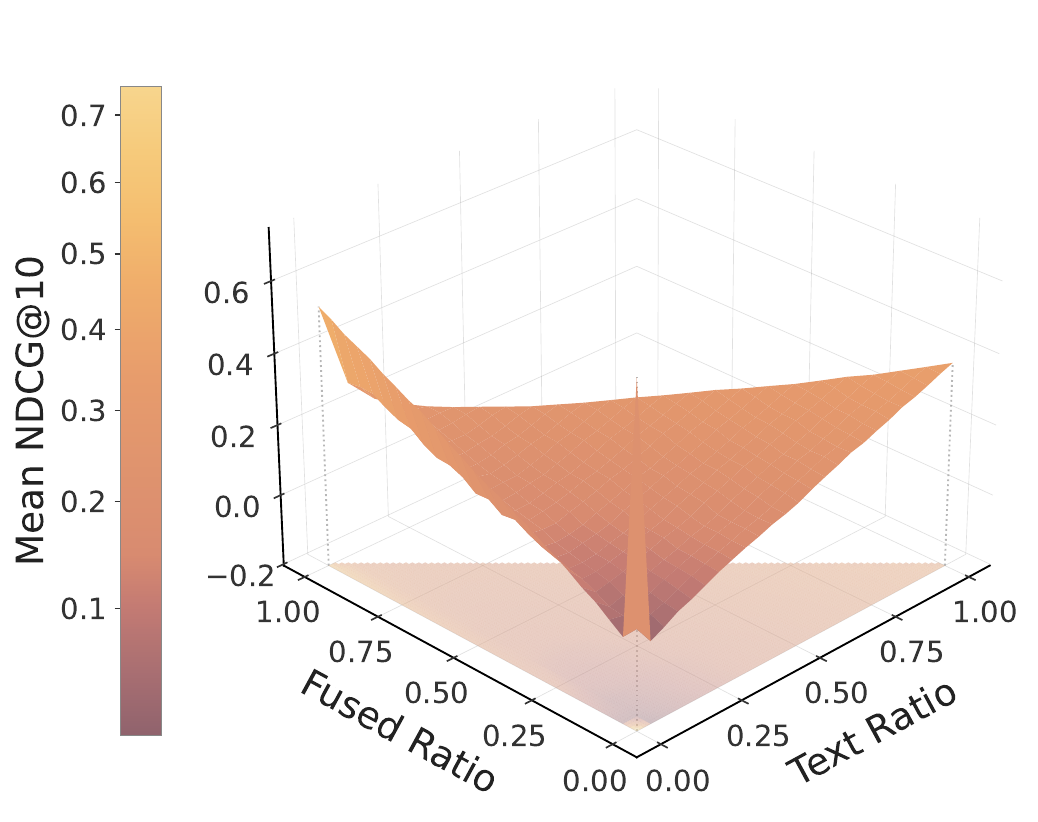}
        \caption{CLIP-ViT-L/14 on SlideVQA}
        \label{fig:ndcg_modelA_datasetY}
    \end{subfigure}
    \hfill
    \begin{subfigure}[b]{0.32\textwidth}
        \centering
        \includegraphics[width=\textwidth]{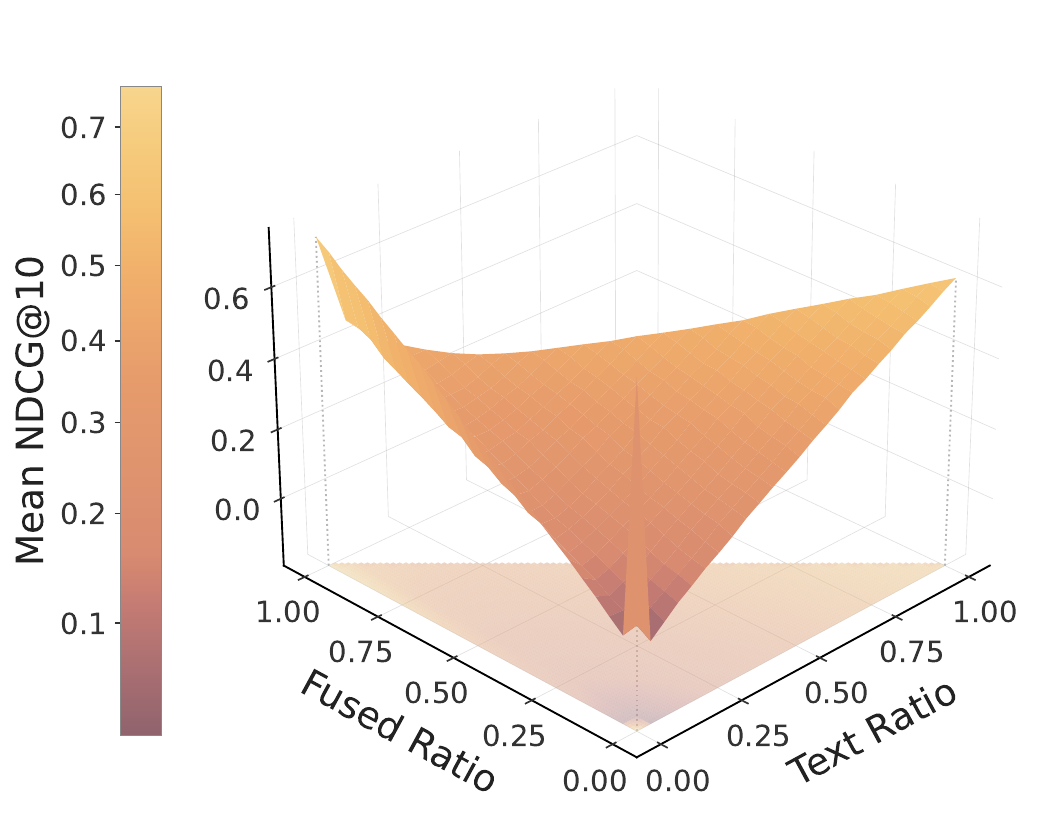}
        \caption{AltCLIP on SlideVQA}
        \label{fig:ndcg_modelA_datasetZ}
    \end{subfigure}

    \vspace{0.5em}

    \begin{subfigure}[b]{0.32\textwidth}
    \centering
    \includegraphics[width=\textwidth]{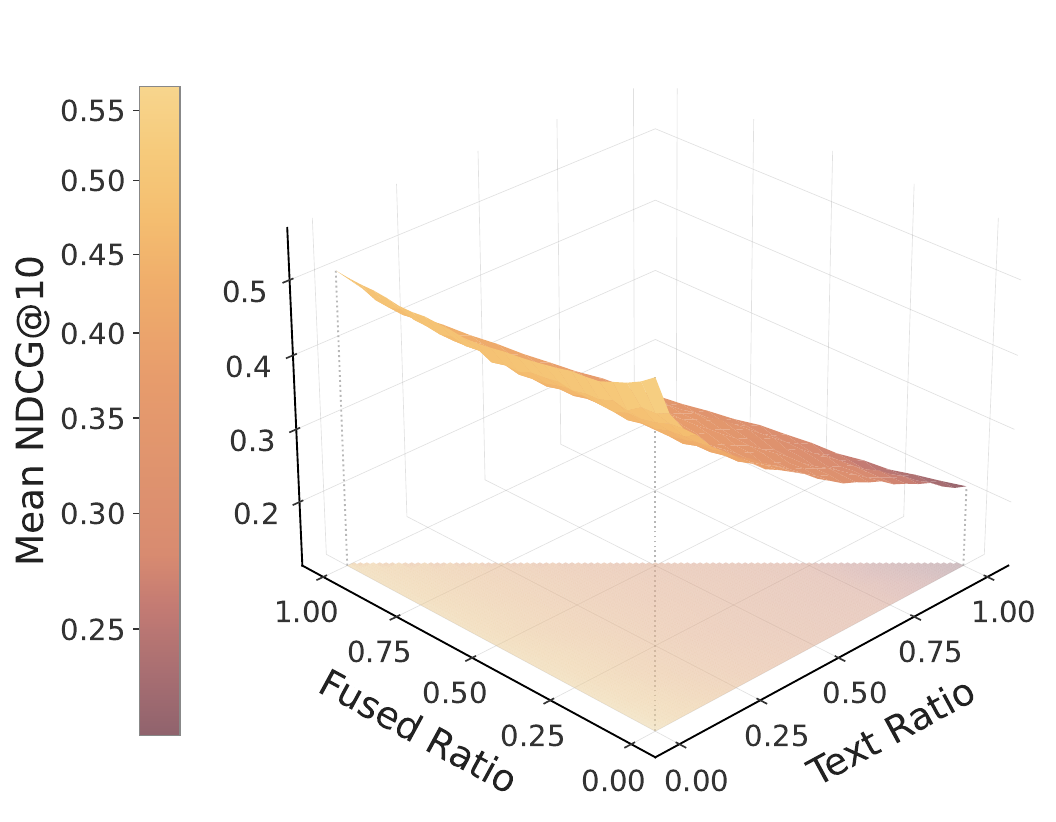}
    \caption{\quad+GR-CLIP on SlideVQA}
    \label{fig:ndcg_modelA_datasetX}
    \end{subfigure}
    \hfill
    \begin{subfigure}[b]{0.32\textwidth}
        \centering
        \includegraphics[width=\textwidth]{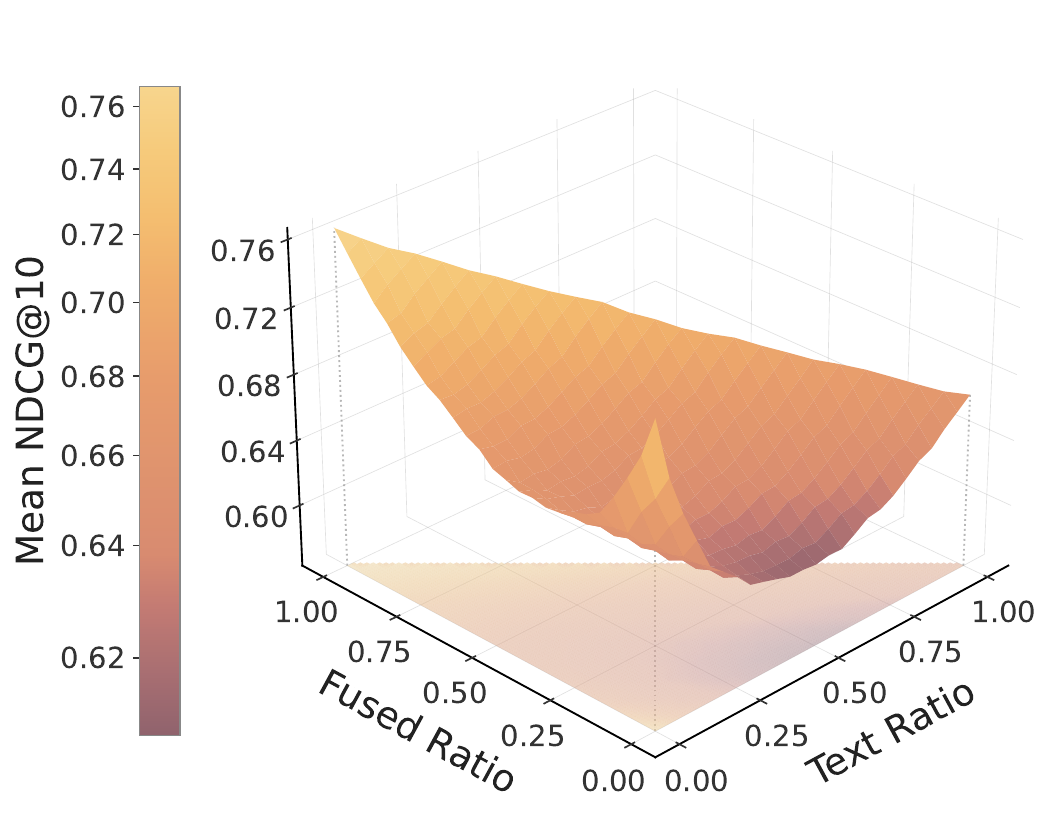}
        \caption{\quad+GR-CLIP on SlideVQA}
        \label{fig:ndcg_modelA_datasetY}
    \end{subfigure}
    \hfill
    \begin{subfigure}[b]{0.32\textwidth}
        \centering
        \includegraphics[width=\textwidth]{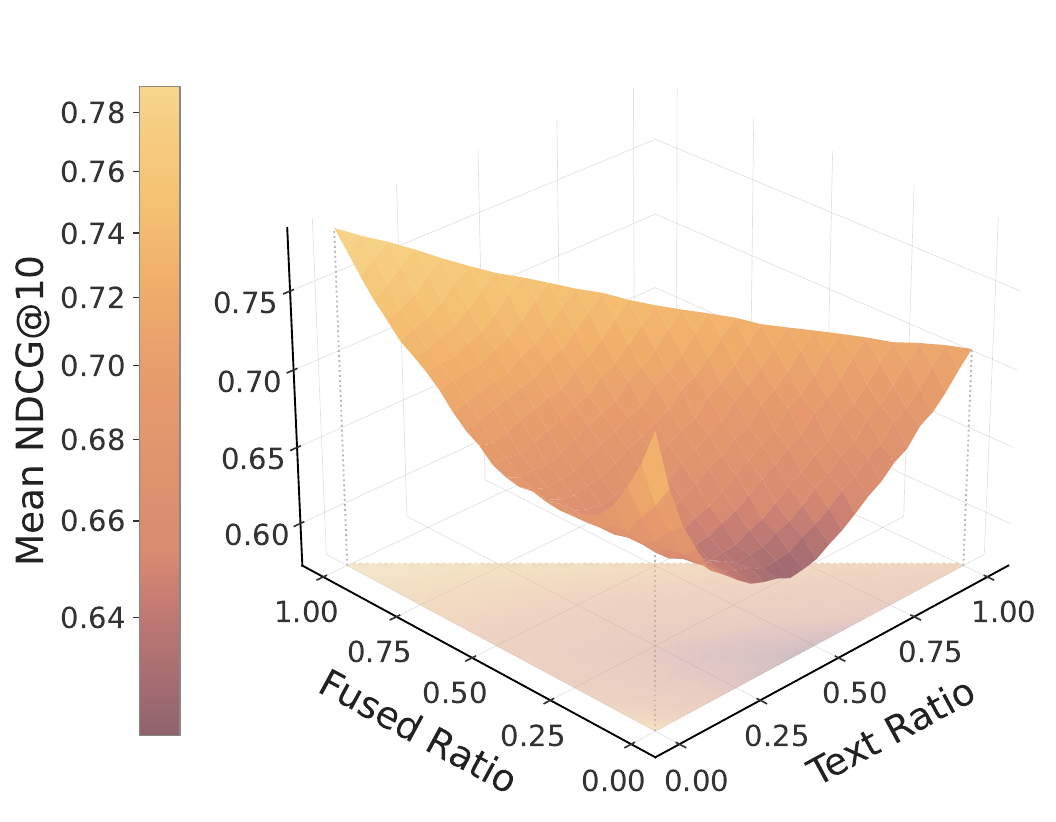}
        \caption{\quad+GR-CLIP on SlideVQA}
        \label{fig:ndcg_modelA_datasetZ}
    \end{subfigure}

    \vspace{0.5em}

    \begin{subfigure}[b]{0.32\textwidth}
    \centering
    \includegraphics[width=\textwidth]{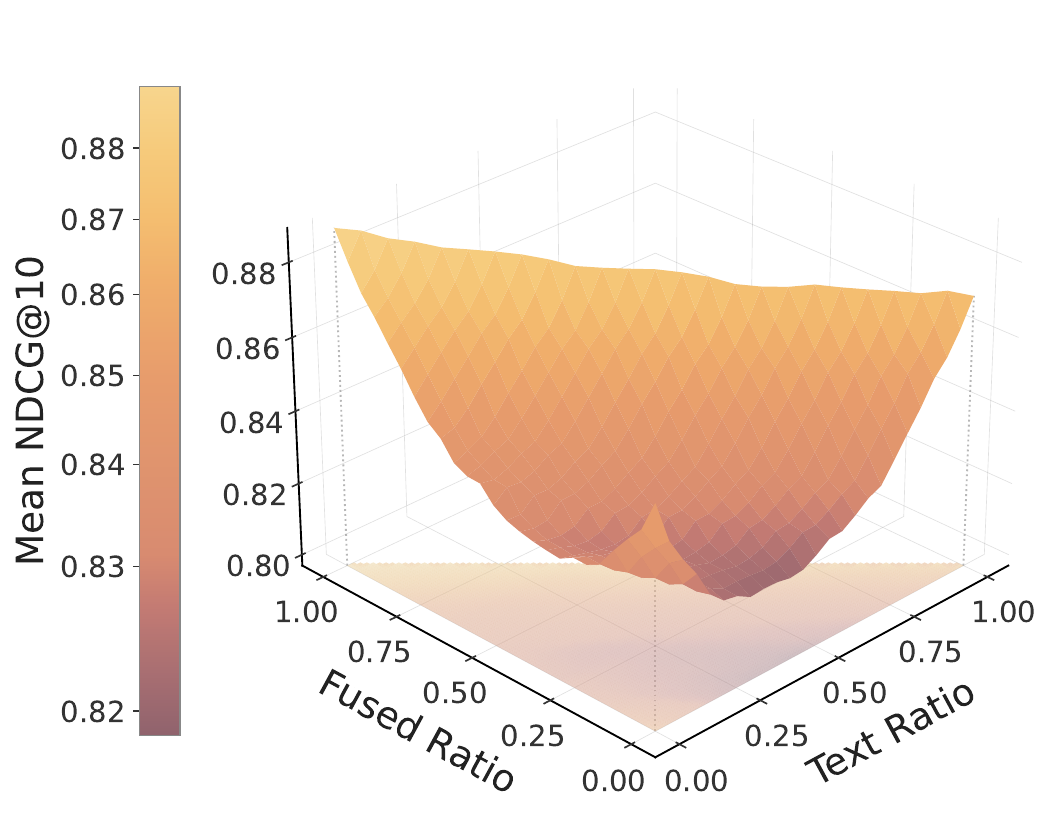}
    \caption{UniME-Phi3.5 on SlideVQA}
    \label{fig:ndcg_modelA_datasetX}
    \end{subfigure}
    \hfill
    \begin{subfigure}[b]{0.32\textwidth}
        \centering
        \includegraphics[width=\textwidth]{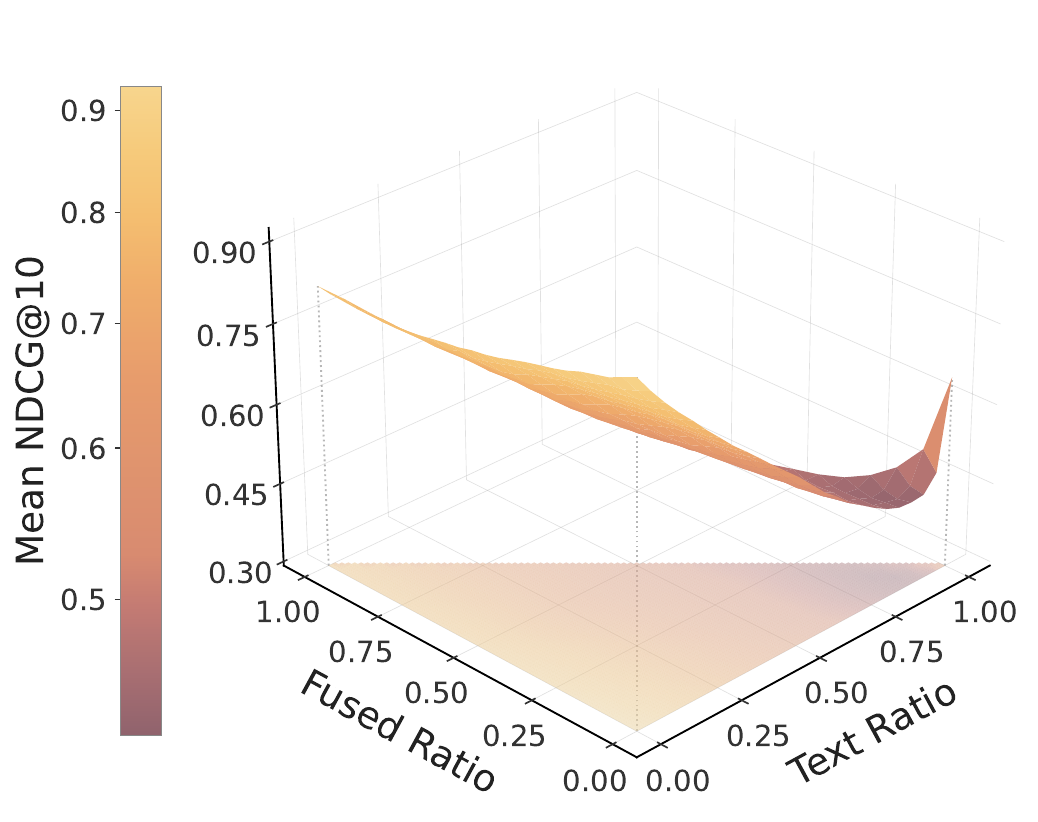}
        \caption{VisRAG-Ret on SlideVQA}
        \label{fig:ndcg_modelA_datasetY}
    \end{subfigure}
    \hfill
    \begin{subfigure}[b]{0.32\textwidth}
        \centering
        \includegraphics[width=\textwidth]{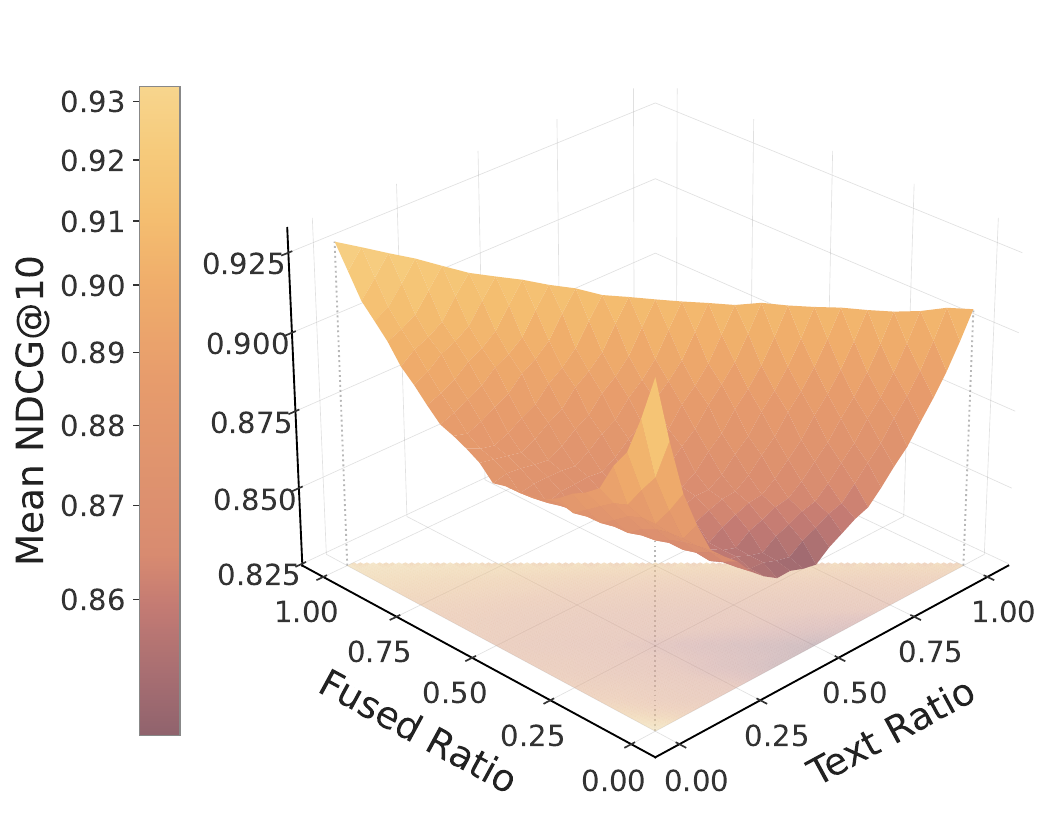}
        \caption{Jina-V5-small on SlideVQA}
        \label{fig:ndcg_modelA_datasetZ}
    \end{subfigure}

    \vspace{0.5em}

    \caption{Mean NDCG@10 as a function of corpus modality composition (text ratio, fused ratio) across nine models on SlideVQA. The three ratios sum to one, $\textit{image ratio} = 1 - \textit{text ratio} - \textit{fused ratio}$, so the surface in the figure is indexed by only two free variables, (text ratio, fused ratio).}
    \label{fig:ndcg_surfaces_grid_appd2}
\end{figure}

Figures~\ref{fig:ndcg_surfaces_grid_appd1} (ChartQA) and~\ref{fig:ndcg_surfaces_grid_appd2} (SlideVQA) present the complete performance surfaces for the remaining nine models on both benchmarks, further corroborating the conclusion drawn in the main text. The results show that the sensitivity of retrieval performance to the proportion of text, image, and fused documents in the corpus is not an idiosyncrasy of the representative models discussed above, but a deficiency shared by nearly all evaluated retrieval models. With the exception of the GR-CLIP-based subfigures, the remaining models exhibit the same asymmetric V-shaped surface observed in the main text on both benchmarks: as the corpus transitions from purely image documents toward an increasing proportion of text, NDCG@10 first drops markedly, reaching its minimum around some intermediate ratio, before recovering as the text ratio continues to increase toward a purely text corpus. Notably, SigLIP2-L/16-384 and VisRAG-Ret exhibit sharp, localized peaks near the boundaries of the surface (i.e., where one modality's proportion approaches one), indicating that these models achieve their best performance only when the corpus is composed of a single modality (purely text or purely image); performance drops sharply as soon as even a small proportion of another modality is introduced. This suggests that such models rely heavily on the modality "purity" of the corpus, rather than exhibiting genuine adaptability to mixed-modality settings.

In contrast, even GR-CLIP, a post-hoc correction method specifically designed to address mixed-modality retrieval for CLIP-based models, fails to fully resolve this robustness deficiency. As shown in subfigures (d)–(f), applying GR-CLIP correction narrows the overall range of performance fluctuation and reduces the absolute magnitude of degradation relative to the uncorrected models; however, the underlying V-shaped degradation pattern persists-performance still varies systematically with the proportion of text, image, and fused documents in the corpus, rather than being flattened into a stable surface. More importantly, GR-CLIP's correction depends on pre-computing modality-wise statistics over a large domain-specific reference corpus, a requirement that inherently limits its transferability and practicality in open-domain settings or under unknown corpus distributions. This indicates that even a targeted post-hoc correction method can only partially mitigate, rather than eliminate, the sensitivity of multimodal retrieval models to the modality proportions of the corpus.

Taken together, the complete results for all nine models on both benchmarks further reinforce the core conclusion of the main text: regardless of architecture, scale, training paradigm, or the use of targeted post-hoc correction, current mainstream multimodal retrieval models generally lack robustness to variation in the proportion of text, image, and fused documents in the corpus.

\section{Extended Chaos-in-the-Text Results Across Retrievers}
\label{sec:appd:distractor}

\begin{figure*}[t!]
  \begin{subfigure}[b]{1.0\textwidth}
    \centering
    \includegraphics[width=\textwidth]{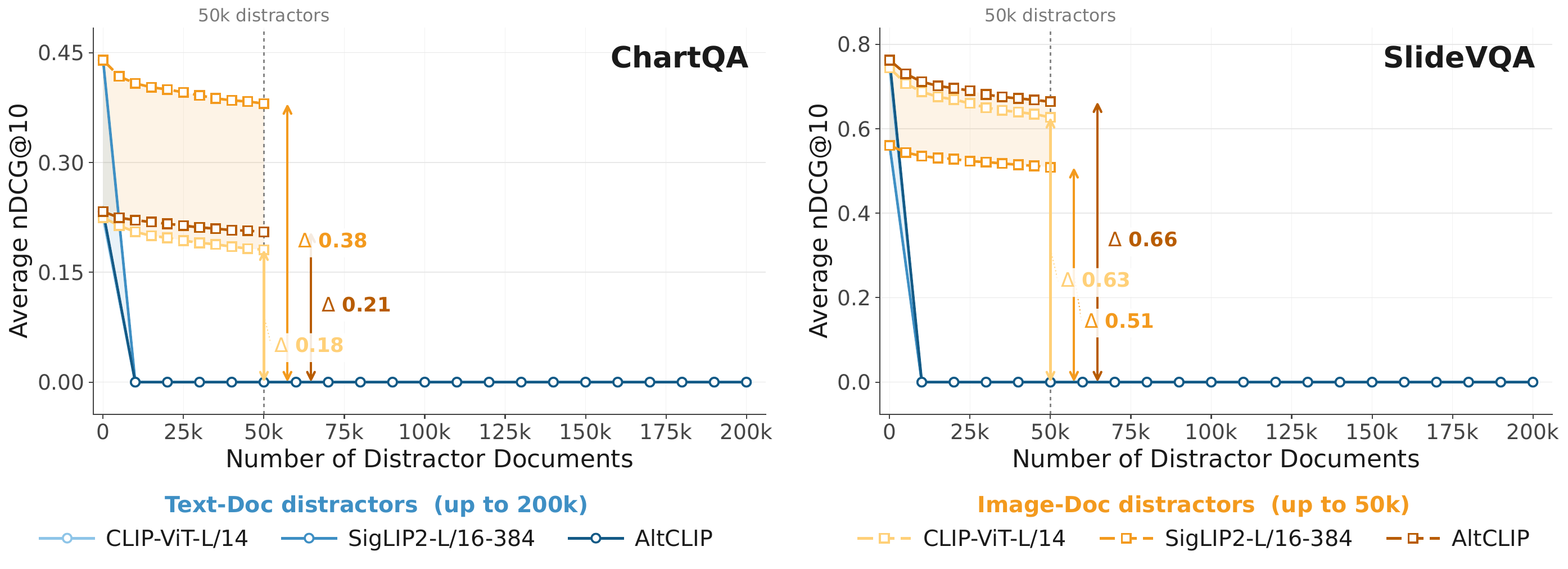}  
  \end{subfigure}
  
  \begin{subfigure}[b]{1.0\textwidth}
    \centering
    \includegraphics[width=\textwidth]{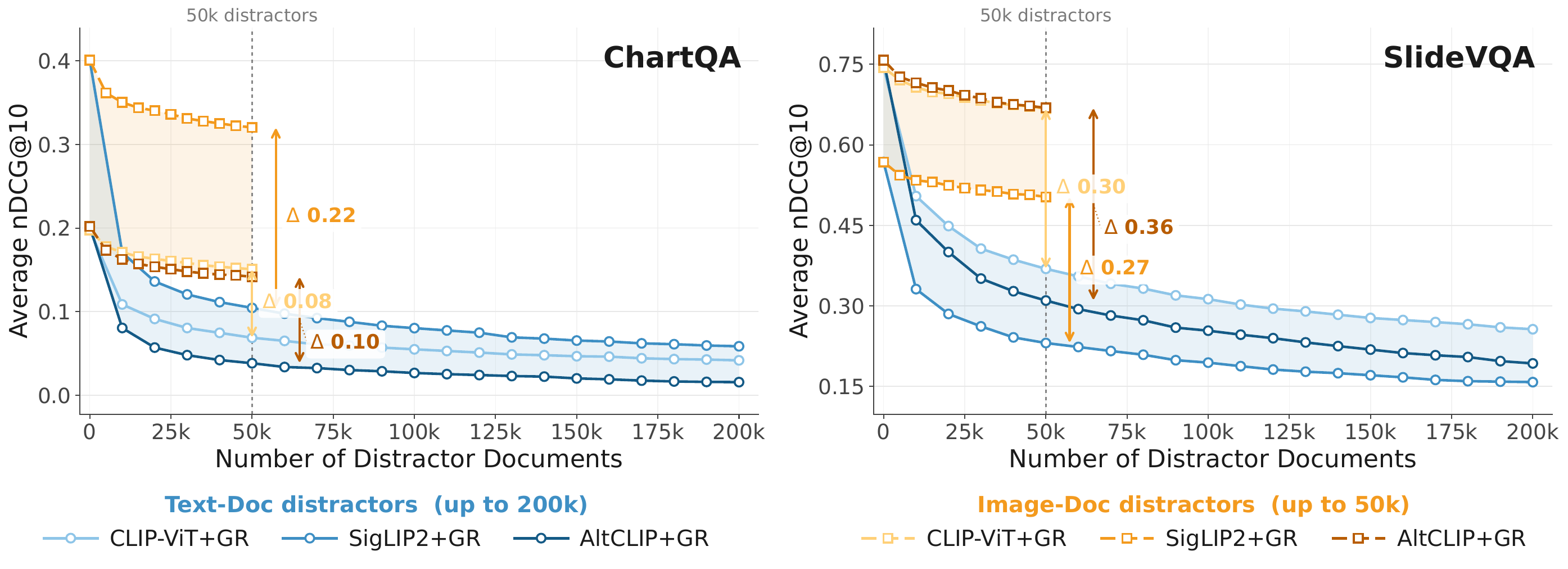}  
  \end{subfigure}

  \begin{subfigure}[b]{1.0\textwidth}
    \centering
    \includegraphics[width=\textwidth]{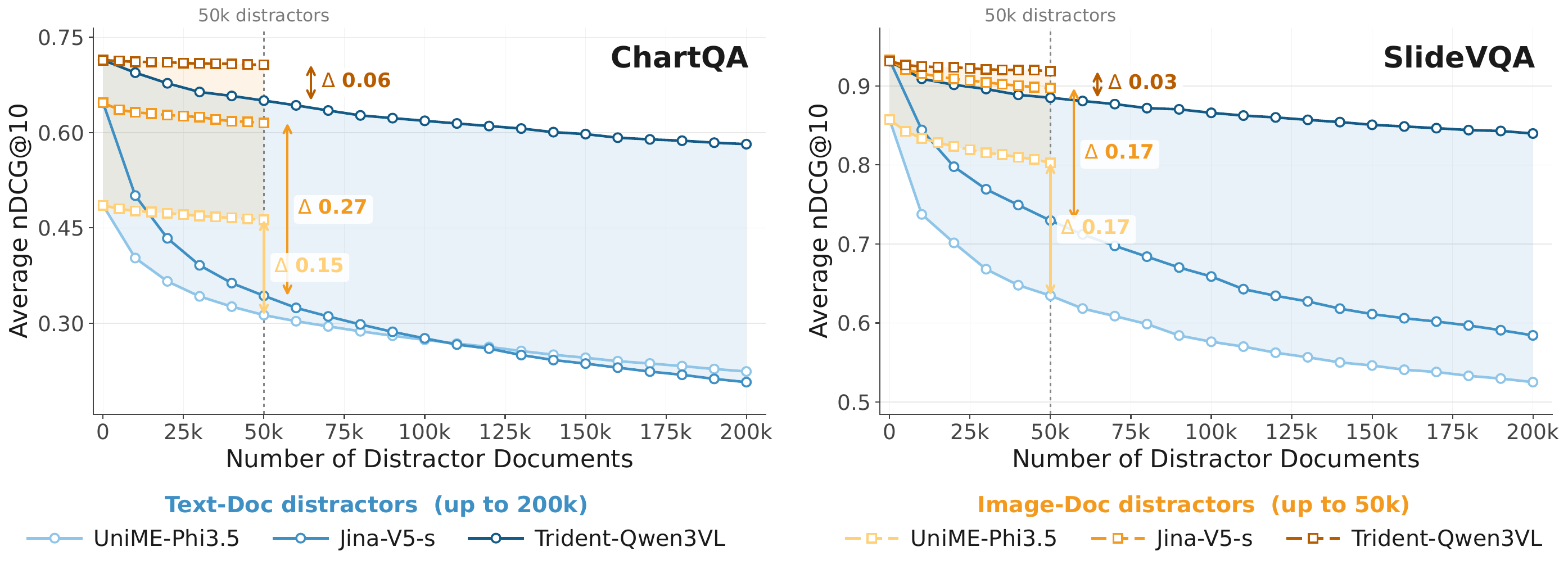}  
  \end{subfigure}
  
   \caption{Text distractors degrade retrieval performance far more severely than image distractors, despite both being equally irrelevant to every query. We report average NDCG@10 for nine retrieval models on ChartQA (left) and SlideVQA (right) as the number of distractor documents increases.}
  \label{fig:distractor_appd}
\end{figure*}

Figure~\ref{fig:distractor_appd} extends this analysis to the remaining retrieval models, following the same protocol as Figure~\ref{fig:distractor} in the main text. Across all three groups of models, CLIP-ViT-L/14, SigLIP2-L/16-384, and AltCLIP (top), their GR-CLIP-corrected counterparts (middle), and UniME-Phi3.5-V-4.2B, Jina-embeddings-v5-omni-small, and our Trident-Qwen3VL-2B-2B (bottom), the same asymmetry holds: text distractors depress nDCG@10 far more severely than an equal number of image distractors, and the gap $\Delta$ widens as the distractor pool grows. That this pattern reproduces consistently across a substantially broader and architecturally diverse set of encoders confirms that Chaos in the Text is not an idiosyncrasy of a specific model, but a systematic modality bias shared across dense multimodal retrievers.

A direct comparison between the top and middle rows, identical encoders before and after GR-CLIP correction, offers further evidence for the partial-mitigation conclusion drawn above. The gap $\Delta$ shrinks visibly once GR-CLIP is applied, indicating that this post-hoc correction does provide a measurable robustness gain even under the distractor setting. Yet the gap does not close: text distractors continue to outrank image distractors of equal irrelevance across the full range of corpus sizes, reaffirming that GR-CLIP alleviates rather than eliminates the underlying modality preference.

The bottom row contrasts UniME-Phi3.5-V-4.2B and Jina-embeddings-v5-omni-small with our Trident-Qwen3VL-2B-2B under the identical distractor protocol. While both baselines retain the characteristic large text-image gap, Trident-Qwen3VL-2B exhibits a markedly smaller $\Delta$ on both ChartQA and SlideVQA, with its text- and image-distractor curves remaining close throughout. This indicates that our method directly targets and substantially mitigates the model's modality preference, suppressing the Chaos in the Text effect and yielding retrieval rankings that are considerably more robust to the modality composition of the corpus.

\begin{figure*}[t!]
  \centering
  \includegraphics[width=\textwidth]{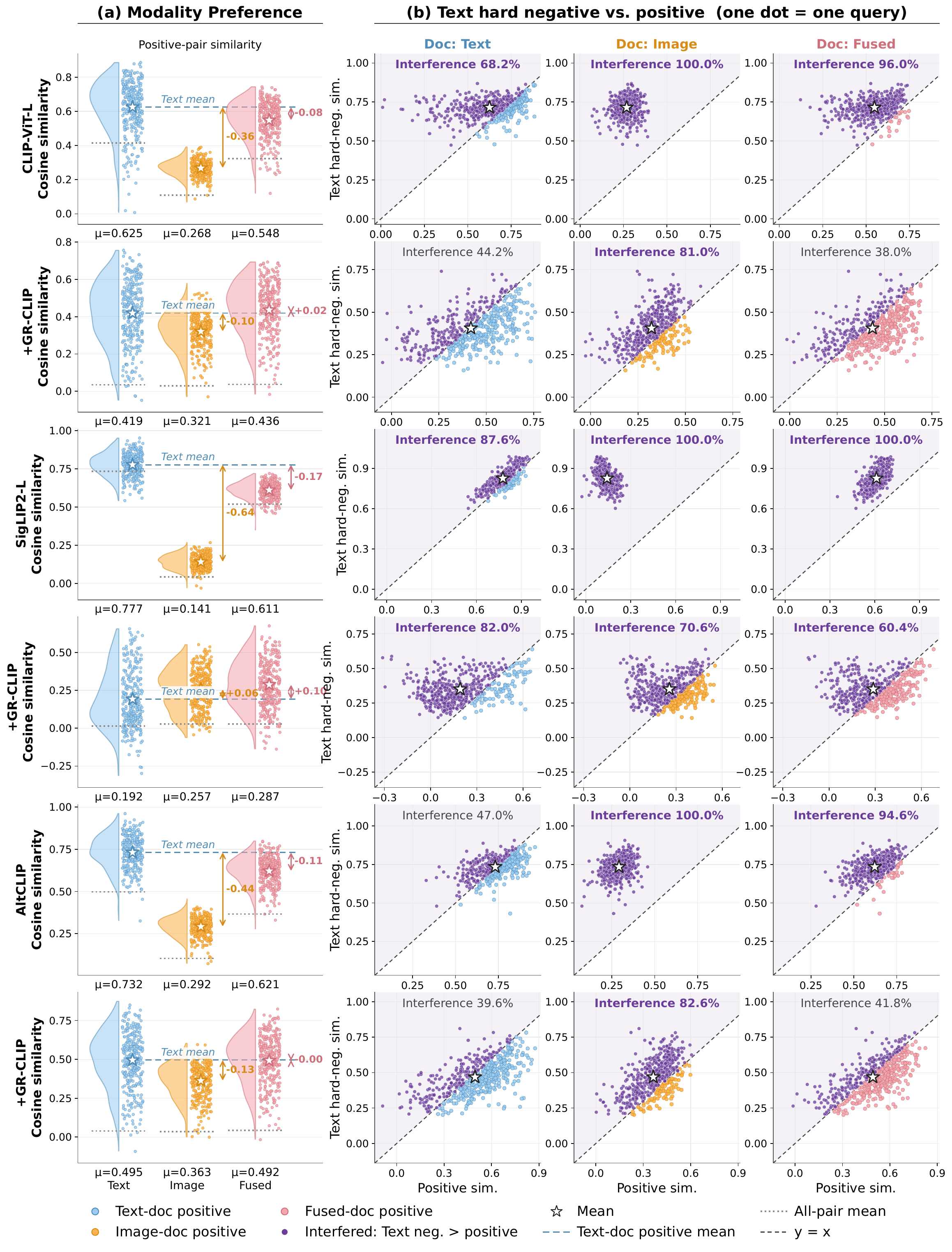}
   \caption{(a) Violin plots show the similarity distributions across document modalities, revealing systematically higher similarity for text documents. (b) Each dot compares a query’s positive similarity with its text hard-negative similarity; points above y=x indicate cases where an irrelevant text document outranks the positive document, with the interference rate reporting their fraction.}
  \label{fig:preference_appd1}
\end{figure*}

\begin{figure*}[t!]
  \centering
  \includegraphics[width=\textwidth]{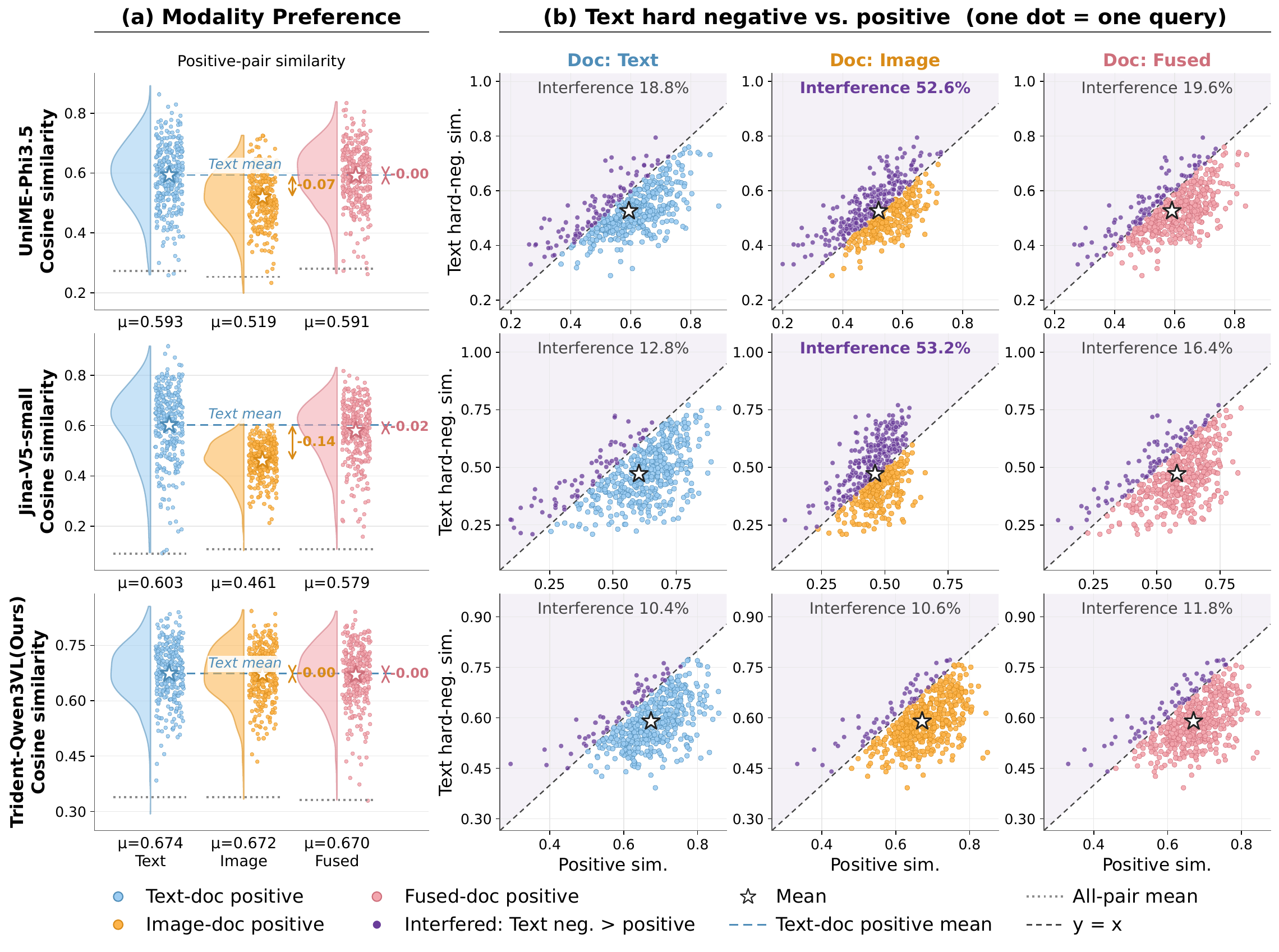}
   \caption{(a) Violin plots show the similarity distributions across document modalities, revealing systematically higher similarity for text documents. (b) Each dot compares a query’s positive similarity with its text hard-negative similarity; points above y=x indicate cases where an irrelevant text document outranks the positive document, with the interference rate reporting their fraction.}
  \label{fig:preference_appd2}
\end{figure*}

\section{Additional Evidence of Modality Preference Across Retrievers}
\label{sec:appd:preference}

We provide additional results complementing the analysis in Section~\ref{sec:why} and Figure~\ref{fig:preference}. Following the same protocol as in the main text, we examine modality preference from two perspectives: (i) score differences among semantically corresponding positive documents presented in different modalities, and (ii) ranking interference caused by irrelevant text documents. The additional results cover a broader set of CLIP-based retrievers and recent multimodal embedding models.

\paragraph{Additional CLIP-based retrievers.}
Figure~\ref{fig:preference_appd1} extends the analysis to CLIP-ViT-L/14, SigLIP2-L, and AltCLIP, together with their GR-CLIP-corrected counterparts. The positive-pair distributions in Figure~\ref{fig:preference_appd1}(a) reproduce the modality-dependent score differences observed in the main text. In particular, the similarity assigned to an image positive is substantially lower than that assigned to its semantically corresponding text positive for CLIP-ViT-L/14, SigLIP2-L, and AltCLIP. The fused view generally reduces this gap but does not consistently eliminate it.

More importantly, the same asymmetry appears at the ranking level in Figure~\ref{fig:preference_appd1}(b). When the relevant document is presented as an image, text hard negatives frequently receive higher similarity scores than the positive, producing high interference rates across the three retrievers. The effect is particularly pronounced for CLIP-ViT-L/14 and SigLIP2-L, where nearly all image-positive queries in our sample are affected. In contrast, interference is generally lower when the positive itself is presented as text. These results further support that modality preference is not merely a shift in average similarity, but can directly alter the ranking between relevant and irrelevant documents.

The GR-CLIP variants provide an additional comparison. Although post-hoc correction can reduce the positive-score gap for some models, the ranking-level behavior is not consistently resolved. For example, substantial text-negative interference remains for image positives after correction, and the effect varies considerably across backbone models and positive modalities. This suggests that calibrating the similarity space alone does not necessarily remove the underlying modality-dependent ranking behavior.

\paragraph{Other VLM-based retrievers and Trident.}
Figure~\ref{fig:preference_appd2} further evaluates UniME-Phi3.5-V-4.2B and Jina-embeddings-v5-omni-small, and compares them with our Trident-Qwen3VL-2B. The two recent baselines exhibit the same qualitative pattern. Their text positives receive higher similarity than image positives, and image positives are more susceptible to being outranked by irrelevant text. For example, the interference rate for image positives reaches 52.6\% for UniME-Phi3.5 and 53.2\% for Jina-V5-small, compared with 18.8\% and 12.8\%, respectively, when the positive is presented as text.

In contrast, Trident-Qwen3VL-2B produces closely aligned positive similarities across the three document views ($0.674$, $0.672$, and $0.670$ for text, image, and fused positives). Correspondingly, the strong modality-dependent ranking asymmetry is largely removed: the interference rates are 10.4\%, 11.8\%, and 11.8\% for text, image, and fused positives, respectively. Thus, Trident not only reduces the score discrepancy among semantically equivalent positive views, but also substantially reduces the tendency of irrelevant text documents to outrank relevant evidence solely because of its modality.

Taken together with the results in the main text, these additional experiments show that modality preference appears across retrievers with different architectures, training objectives, and model scales. They also highlight the distinction between globally adjusting similarity scores and learning a representation in which relevance ranking itself is less dependent on document modality.

\section{Additional Training Details}
\label{sec:appd:training}

\begin{table*}[t]
\centering
\caption{\textbf{Backbone configurations.} 
Model-specific configurations used for the two Trident variants.}
\label{tab:backbone_config}
\small
\setlength{\tabcolsep}{4pt}
\renewcommand{\arraystretch}{1.12}
\begin{tabularx}{\linewidth}{
    @{}
    >{\raggedright\arraybackslash}p{0.25\linewidth}
    >{\raggedright\arraybackslash}X
    >{\raggedright\arraybackslash}X
    @{}
}
\toprule
& \textbf{Qwen3-VL-2B} & \textbf{jina-clip-v2} \\
\midrule
Initialization
    & Qwen3-VL-2B-Instruct
    & jina-clip-v2 \\

Pooling
    & Last token
    & Modality-specific \\

Fine-tuning
    & LoRA ($r{=}32$, $\alpha{=}64$)
    & Full parameters \\

Vision tower
    & Frozen
    & Trainable \\

Image input
    & Dynamic resolution
    & $512{\times}512$ \\

Max. text length
    & 8192
    & 8192 \\

Instruction
    & Fig.~\ref{fig:template}
    & None \\

Text--image fusion
    & Joint encoding
    & Embedding fusion (0.5/0.5) \\
\bottomrule
\end{tabularx}
\end{table*}

\begin{figure}[t]
    \centering
    \begin{subfigure}[b]{\textwidth}
        \centering
        \includegraphics[width=\textwidth]{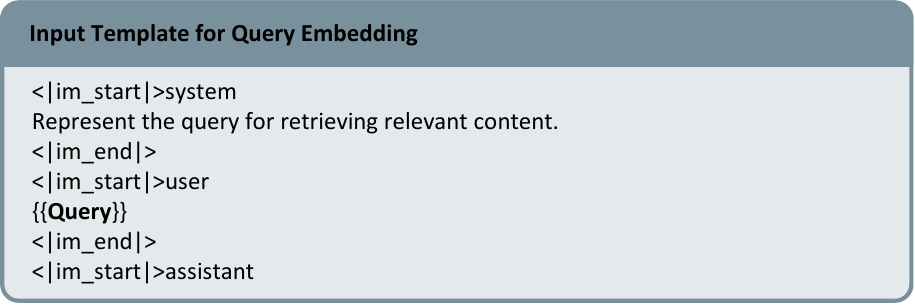}
        \label{fig:template1}
    \end{subfigure}
    \hfill
    \begin{subfigure}[b]{\textwidth}
        \centering
        \includegraphics[width=\textwidth]{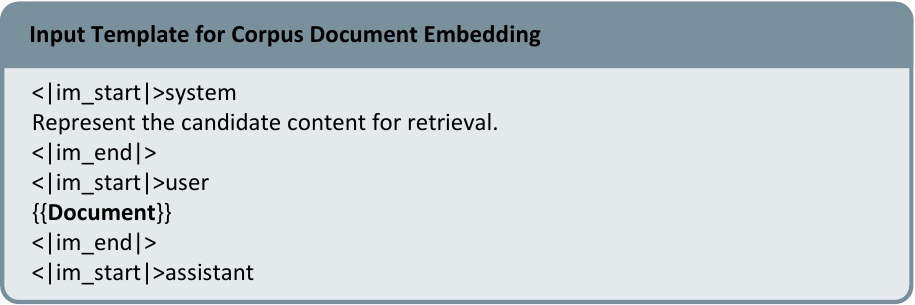}
        \label{fig:template2}
    \end{subfigure}

    \caption{Input template for trainning Trident-Qwen3VL-2B.}
    \label{fig:template}
\end{figure}
\begin{table*}[t]
\centering
\caption{\textbf{Optimization hyperparameters.}
The same optimization setup is used for both backbone variants.}
\label{tab:optim_hparams}
\small
\setlength{\tabcolsep}{8pt}
\renewcommand{\arraystretch}{1.12}
\begin{tabular}{@{}lc@{}}
\toprule
\textbf{Hyperparameter} & \textbf{Value} \\
\midrule
Epochs                  & 1 \\
Batch size / GPU        & 8 \\
Gradient accumulation   & 2 \\
Learning rate           & $1\times10^{-4}$ \\
LR schedule             & Cosine \\
Warmup ratio            & 5\% \\
Weight decay            & 0.01 \\
Gradient clipping       & 1.0 \\
Precision               & bfloat16 \\
Positive views / sample & 3 \\
\bottomrule
\end{tabular}
\end{table*}

\paragraph{Training data.}
We adopt OpenDocVQA~\citep{tanaka2025vdocrag}, a 41k-sample dataset, as training data, from which we construct tri-modal training instances following the procedure in Section~\ref{sec:how:setup}.

\paragraph{Backbones and baselines.}
We use the CLIP-based and VLM-based models of Section~\ref{sec:how:models} as baselines, and train one representative backbone per category to test the generality of our method: Jina-CLIP-v2 for CLIP-based models, and Qwen3-VL-2B-Instruct~\citep{qwen3technicalreport} for VLM-based models, chosen to enable a controlled comparison against Qwen3-VL-Embedding-2B.\footnote{Training hyperparameters and implementation details are in Appendix~\ref{sec:appd:training}.}

\paragraph{Evaluation benchmarks and settings.}
Since our training data comes from the Visual Document QA domain, we evaluate generalization on six benchmarks: ChartQA and SlideVQA (Section~\ref{sec:how:setup}), plus DocVQA~\citep{mathew2021docvqa}, InfoVQA~\citep{mathew2022infographicvqa}, ViDoSeek~\citep{wang2025vidorag}, and Dude~\citep{van2023document}.\footnote{We further validate trident on natural-image retrieval in Appendix~\ref{sec:appd:nature}, beyond the document domain.} We report two settings of increasing difficulty. In \textbf{Origin}, retrieval is performed over each dataset's original, single-modality (pure image) corpus. In \textbf{Mix}, we first convert each corpus into a mixed-modality one using the procedure of Section~\ref{sec:how:setup}, then merge the six resulting corpora into a single pool, so retrieval must additionally contend with a much larger set of visually and semantically similar distractors drawn from the other five datasets.

\paragraph{Training Setup.} We implement both backbone variants, Qwen3-VL-2B-Instruct and jina-clip-v2, and extend the training pipeline to support multi-view positive sampling, cross-GPU negative sharing, and false-negative masking. All experiments are conducted on 8 NVIDIA A100-80GB GPUs using Distributed Data Paralle. We use bfloat16 mixed-precision training and enable gradient checkpointing to reduce memory consumption. Unless otherwise specified, both models are trained for one epoch with three positive views per sample.

\paragraph{Backbone Configurations.} The backbone-specific configurations are summarized in Table~\ref{tab:backbone_config}. For Qwen3-VL-2B-Instruct, we follow the pooling strategy and instruction format of Qwen3-VL-Embedding, using last-token pooling and separate instruction templates for queries and documents, as illustrated in Figure~\ref{fig:template}. During training, we apply LoRA only to the language tower while keeping the vision tower frozen. This substantially reduces memory consumption while preserving the pretrained visual representations. For jina-clip-v2, which adopts a dual-tower architecture with pretrained vision--language alignment, we fine-tune both the text and vision towers end-to-end. For fused documents, the text and image embeddings are combined with equal weights to obtain the final representation.

\paragraph{Optimization.} We train both models for one epoch using a cosine learning-rate schedule with a 5\% warmup ratio. The learning rate is set to $1\times10^{-4}$ with a weight decay of $0.01$, and gradients are clipped to a maximum norm of $1.0$. Each GPU processes a batch of 8 samples with 2 gradient accumulation steps. The main optimization hyperparameters are summarized in Table~\ref{tab:optim_hparams}.

\section{Decomposition and Gradient Analysis of Multi-Positive View InfoNCE}
\label{app:objective_analysis}

\subsection{Objective Definition}
\label{app:objective_definition}

For a query $q$, let $P$ denote its set of semantically corresponding
positive views, $N$ its set of negative documents, and $C=P\cup N$
the complete candidate set. In Trident, $P$ contains the Text, Image,
and Fused views, giving $M=|P|=3$. The following derivation holds for
any $M\geq 1$.

Let $s_j=\operatorname{sim}(q,j)$ denote the similarity between the
query and candidate $j$, and let $\tau>0$ be the temperature.
The softmax probability over the shared candidate set is
\begin{equation}
    p_j =
    \frac{\exp(s_j/\tau)}{Z},
    \qquad
    Z = \sum_{k\in C}\exp(s_k/\tau).
    \label{eq:app_softmax}
\end{equation}

Multi-Positive View InfoNCE averages the negative log-probabilities
of all positive views,
\begin{equation}
    \mathcal{L}(q)
    =
    -\frac{1}{M}
    \sum_{m\in P}\log p_m.
    \label{eq:app_objective}
\end{equation}

We define the total probability mass assigned to the positive set
and the relative share of each positive view as
\begin{equation}
    \rho = \sum_{m\in P}p_m,
    \qquad
    r_m = \frac{p_m}{\rho}.
    \label{eq:app_positive_mass}
\end{equation}
Here, $\sum_{m\in P}r_m=1$.
We further define $u_m=1/M$, the uniform target distribution over
positive views.

\subsection{Relevance Discrimination and Positive-View Balance}
\label{app:objective_decomposition}

Since $p_m=\rho r_m$, Eq.~\eqref{eq:app_objective} can be rewritten as
\begin{equation}
\begin{aligned}
    \mathcal{L}(q)
    &=
    -\frac{1}{M}\sum_{m\in P}\log(\rho r_m) \\
    &=
    -\log\rho
    -\frac{1}{M}\sum_{m\in P}\log r_m.
\end{aligned}
\label{eq:app_loss_expansion}
\end{equation}

The KL divergence between the uniform distribution $u$ and the
relative positive-view distribution $r$ is
\begin{equation}
\begin{aligned}
    D_{\mathrm{KL}}(u\|r)
    &=
    \sum_{m\in P}\frac{1}{M}
    \log\frac{1/M}{r_m} \\
    &=
    -\log M
    -\frac{1}{M}\sum_{m\in P}\log r_m.
\end{aligned}
\label{eq:app_kl_expansion}
\end{equation}

Substituting Eq.~\eqref{eq:app_kl_expansion} into
Eq.~\eqref{eq:app_loss_expansion} gives
\begin{equation}
    \mathcal{L}(q)
    =
    \underbrace{-\log\rho}_{\text{relevance discrimination}}
    +
    \underbrace{D_{\mathrm{KL}}(u\|r)}_{\text{positive-view balance}}
    +
    \log M
.
\label{eq:app_loss_decomposition}
\end{equation}

The first term, $-\log\rho$, encourages the positive set to capture
probability mass from negative documents, making relevant documents
collectively more competitive. However, this term depends only on
the total positive probability mass, not on its distribution across
views. The model can therefore increase $\rho$ primarily through an
easy-to-match text positive while assigning little probability to
the image or fused positive.

The second term, $D_{\mathrm{KL}}(u\|r)$, constrains the probability
allocation within the positive set. It is minimized when
$r_m=1/M$ for every positive view, thereby penalizing excessive
concentration on a subset of views. Together, the two terms encourage
the model to distinguish relevant documents from irrelevant ones
while avoiding reliance on a single modality to match the positive
set. The constant $\log M$ does not affect optimization.

\subsection{Gradients with Respect to Positive and Negative Scores}
\label{app:score_gradients}

The derivative of the log-softmax is
\begin{equation}
    \frac{\partial\log p_m}{\partial s_j}
    =
    \frac{1}{\tau}
    \left(\mathbb{I}[j=m]-p_j\right),
    \label{eq:app_logsoftmax_gradient}
\end{equation}
where $\mathbb{I}[\cdot]$ is the indicator function.
Consequently, the gradient of the objective with respect to any
candidate score is
\begin{equation}
    \frac{\partial\mathcal{L}}{\partial s_j}
    =
    \frac{1}{\tau}
    \left(
        p_j-\frac{\mathbb{I}[j\in P]}{M}
    \right).
\label{eq:app_score_gradient}
\end{equation}

\paragraph{Positive views.}
For a positive view $m\in P$,
\begin{equation}
    \frac{\partial\mathcal{L}}{\partial s_m}
    =
    \frac{p_m-1/M}{\tau}.
    \label{eq:app_positive_gradient}
\end{equation}
When its probability is below $1/M$, the gradient is negative,
so gradient descent on the score increases it.
When its probability exceeds $1/M$, the gradient is positive,
and the update instead decreases its score.
Thus, a relevant image receiving insufficient probability is
encouraged to become more competitive, while a positive view
receiving excessive probability mass is suppressed.

Consider two positive views $a,b\in P$.
The difference between their score gradients is
\begin{equation}
    \frac{\partial\mathcal{L}}{\partial s_a}
    -
    \frac{\partial\mathcal{L}}{\partial s_b}
    =
    \frac{p_a-p_b}{\tau}.
    \label{eq:app_positive_gradient_difference}
\end{equation}
If the scores are treated as independent variables and updated
by one gradient descent step with learning rate $\eta$, then
\begin{equation}
    (s_a'-s_b')-(s_a-s_b)
    =
    -\frac{\eta}{\tau}(p_a-p_b).
    \label{eq:app_positive_gap_update}
\end{equation}
Whenever $s_a>s_b$, we also have $p_a>p_b$, so the update direction
tends to reduce their score difference.
Even when both positive probabilities are below $1/M$, the weaker
positive receives a larger upward score update.

\paragraph{Negative documents.}
For a negative document $n\in N$,
\begin{equation}
    \frac{\partial\mathcal{L}}{\partial s_n}
    =
    \frac{p_n}{\tau}
    >0.
    \label{eq:app_negative_gradient}
\end{equation}
Gradient descent on the score therefore decreases it.
For a fixed query, a higher-scoring negative receives a larger
softmax probability and consequently a stronger downward update.
An irrelevant text document assigned a high score is therefore
penalized more strongly. This behavior depends on relevance labels
and current scores, rather than on an assumption that text documents
are necessarily negative.

\subsection{Complementary Roles of the Two Components}
\label{app:component_gradients}

To distinguish relevance learning from view balancing, let
\begin{equation}
    \mathcal{L}_{\mathrm{rel}}=-\log\rho,
    \qquad
    \mathcal{L}_{\mathrm{bal}}=D_{\mathrm{KL}}(u\|r).
    \label{eq:app_component_definitions}
\end{equation}

For a positive view $m$, the relevance-discrimination gradient is
\begin{equation}
    \frac{\partial\mathcal{L}_{\mathrm{rel}}}{\partial s_m}
    =
    \frac{p_m-r_m}{\tau}
    =
    -\frac{(1-\rho)r_m}{\tau},
    \label{eq:app_relevance_positive_gradient}
\end{equation}
while the positive-view balance gradient is
\begin{equation}
    \frac{\partial\mathcal{L}_{\mathrm{bal}}}{\partial s_m}
    =
    \frac{r_m-1/M}{\tau}.
    \label{eq:app_balance_positive_gradient}
\end{equation}

Equation~\eqref{eq:app_relevance_positive_gradient} shows that the
relevance-discrimination term pushes all positive scores upward,
but its gradient magnitude is proportional to the current relative
share $r_m$. Optimizing this term alone therefore gives larger
upward score updates to positives that are already stronger.

Equation~\eqref{eq:app_balance_positive_gradient} provides a
complementary balancing effect.
When a view's share within the positive set is below $1/M$, this
term pushes its score upward; when its share exceeds $1/M$, the
term pushes its score downward.
Adding these two gradients recovers
Eq.~\eqref{eq:app_positive_gradient}.

For a negative document $n$, note that
\begin{equation}
    r_m
    =
    \frac{\exp(s_m/\tau)}
    {\sum_{a\in P}\exp(s_a/\tau)}
    \label{eq:app_positive_only_softmax}
\end{equation}
depends only on positive scores. Hence,
\begin{equation}
    \frac{\partial\mathcal{L}_{\mathrm{bal}}}{\partial s_n}=0,
    \qquad
    \frac{\partial\mathcal{L}_{\mathrm{rel}}}{\partial s_n}
    =
    \frac{p_n}{\tau}.
    \label{eq:app_component_negative_gradients}
\end{equation}

For the current query, the relevance-discrimination term thus
penalizes high-scoring irrelevant text, while the positive-view
balance term prevents relevant text from monopolizing the positive
probability mass and provides additional support for weaker
relevant image views.

\subsection{Constraining Cross-Modal Score Baselines}
\label{app:modality_offsets}

When each modality is trained with a separately normalized
InfoNCE objective, adding the same offset to all candidate scores
within a modality leaves its softmax probabilities unchanged.
Such objectives therefore cannot identify relative score offsets
between modalities.

Shared normalization removes this invariance to independent
modality-specific offsets.
Let $C_v$ denote the candidates belonging to modality $v$, and
consider adding an offset $c_v$ to all their scores.
Evaluated at the current scores, Eq.~\eqref{eq:app_score_gradient}
gives
\begin{equation}
\begin{aligned}
    \frac{\partial\mathcal{L}}{\partial c_v}
    &=
    \sum_{j\in C_v}
    \frac{\partial\mathcal{L}}{\partial s_j} \\
    &=
    \frac{1}{\tau}
    \left(
        \sum_{j\in C_v}p_j
        -
        \frac{|P\cap C_v|}{M}
    \right).
\end{aligned}
\label{eq:app_modality_offset_gradient}
\end{equation}

Trident includes one positive per modality, giving
\begin{equation}
    \frac{\partial\mathcal{L}}{\partial c_v}
    =
    \frac{1}{\tau}
    \left(
        \sum_{j\in C_v}p_j-\frac{1}{M}
    \right).
    \label{eq:app_trident_offset_gradient}
\end{equation}

A modality-specific shift now changes the shared probability
allocation and is constrained by the training objective.
Unlike separately normalized objectives, the shared objective is
not invariant to independently raising all scores of one modality.
It therefore constrains the relative score baselines that govern
competition between modalities.

\section{Model Performance on Natural Images}
\label{sec:appd:nature}


\begin{table*}[t]
\centering
\caption{Main results on four natural-image retrieval benchmarks under the \textbf{Origin} (each dataset's original, single-modality corpus) and \textbf{Mix} (mixed-modality corpora merged across all four datasets) settings. \textit{Avg} is the mean over the four benchmarks, computed separately for Origin and Mix. The best result in each column is in \textbf{bold}; the second best is \underline{underlined}. Our proposed models are highlighted in gray, with the corresponding ablation (w/o Multi-Positive InfoNCE) shown below.}
\label{tab:main_results_nature}
\renewcommand{\arraystretch}{1.08}  
\setlength{\tabcolsep}{5pt}          
\resizebox{\textwidth}{!}{%
\begin{tabular}{l *{10}{c}}
\toprule
\multirow{2.5}{*}{\textbf{Model}}
 & \multicolumn{2}{c}{Google-WIT} & \multicolumn{2}{c}{MSCOCO} & \multicolumn{2}{c}{VisualNews}
 & \multicolumn{2}{c}{OVEN} & \multicolumn{2}{c}{\textbf{Avg}} \\
\cmidrule(lr){2-3} \cmidrule(lr){4-5} \cmidrule(lr){6-7}
\cmidrule(lr){8-9} \cmidrule(lr){10-11}
 & Origin & Mix & Origin & Mix & Origin & Mix & Origin & Mix & Origin & Mix \\
\midrule
\multicolumn{11}{c}{\textit{CLIP-based Models}} \\
\midrule
SigLIP2-L/16-384
  & 22.51 & 6.29 & \textbf{81.83} & 40.94 & 50.21 & 17.58 & \underline{57.95} & 22.42 & 53.13 & 21.81 \\
\quad + GR-CLIP
  & 22.49 & 19.35 & \underline{81.60} & 62.02 & 52.38 & 43.04 & \textbf{60.49} & 44.66 & 54.24 & 42.27 \\
CLIP-ViT-L/14
  & \textbf{61.59} & 51.94 & 71.14 & 37.06 & \textbf{90.40} & 49.68 & 54.16 & 13.23 & 69.32 & 37.98 \\
\quad + GR-CLIP
  & \underline{60.66} & 66.65 & 72.00 & 60.89 & \underline{90.27} & 75.94 & 56.41 & 39.01 & 69.84 & 60.62 \\
AltCLIP
  & 60.30 & 49.17 & 76.26 & 52.28 & 88.66 & 57.03 & 54.47 & 18.44 & \textbf{69.92} & 44.23 \\
\quad + GR-CLIP
  & 59.63 & 68.08 & 74.95 & 62.96 & 89.30 & 75.69 & 55.68 & 36.61 & \underline{69.89} & 60.84 \\
Jina-CLIP-v2
  & 48.83 & \underline{75.59} & 73.81 & 60.31 & 64.82 & 63.91 & 40.37 & 25.90 & 56.96 & 56.43 \\
\quad + GR-CLIP
  & 50.78 & \textbf{82.28} & 73.11 & 71.56 & 67.48 & 76.10 & 40.71 & 26.31 & 58.02 & 64.06 \\
\rowcolor{gray!15}
\textbf{Trident-JinaCLIP} \textit{(Ours)}
  & 44.60 & 74.11 & 76.20 & \textbf{77.12} & 74.29 & \textbf{85.08} & 50.51 & \underline{45.70} & 61.40 & \textbf{70.50} \\
\rowcolor{gray!15}
\quad \textit{w/o Multi-Positive InfoNCE} 
  & 40.56 & 69.79 & 74.56 & \underline{74.41} & 71.75 & \underline{80.52} & 51.37 & \textbf{46.19} & 59.56 & \underline{67.73} \\
\midrule
\multicolumn{11}{c}{\textit{VLM-based Models}} \\
\midrule
UniME-Phi3.5-V-4.2B
  & 41.11 & 74.45 & 81.70 & 81.88 & 72.96 & 78.16 & 39.76 & 26.63 & 58.88 & 65.28 \\
VisRAG-Ret
  & 39.72 & 42.14 & 76.89 & 43.88 & 59.22 & 31.92 & 46.84 & 17.68 & 55.67 & 33.91 \\
Jina-embeddings-v5-omni-small
  & 42.46 & \underline{79.72} & 80.79 & 74.10 & 63.09 & 73.67 & 44.68 & 25.16 & 57.76 & 63.16 \\
Qwen3-VL-Embedding-2B
  & 45.92 & 70.66 & \underline{85.60} & 86.11 & 75.85 & 82.06 & \underline{59.03} & \textbf{55.85} & 66.60 & 73.67 \\
Qwen3-VL-Embedding-8B
  & \textbf{56.28} & 68.76 & \textbf{87.18} & \textbf{89.54} & \underline{79.20} & 84.25 & \textbf{59.41} & \underline{55.15} & \textbf{70.52} & \underline{74.43} \\
\rowcolor{gray!15}
\textbf{Trident-Qwen3VL-2B} \textit{(Ours)}
  & \underline{56.10} & \textbf{83.67} & 84.89 & \underline{86.86} & \textbf{80.57} & \textbf{88.10} & 57.67 & 48.49 & \underline{69.81} & \textbf{76.78} \\
\rowcolor{gray!15}
\quad \textit{w/o Multi-Positive InfoNCE} 
  & \underline{56.10} & 77.06 & 81.82 & 81.17 & 77.54 & \underline{86.81} & 57.79 & 51.03 & 68.31 & 74.02 \\
\bottomrule
\end{tabular}
}
\end{table*}

\textbf{Experimental Setup.} To verify that our findings extend to natural images, we train on 170K samples collected from the training sets of MMEB~\citep{jiang2024vlm2vec}, and evaluate on four natural-image retrieval benchmarks: Google-WIT~\citep{srinivasan2021wit}, MSCOCO~\citep{chen2015microsoft}, VisualNews~\citep{liu2021visual}, and OVEN~\citep{hu2023open} (Table~\ref{tab:main_results_nature}). \textit{Origin} denotes each dataset's original single-modality corpus, while \textit{Mix} directly adopts the mixed-modality corpus from MixBench~\citep{li2025closing}, which merges the four datasets.

\textbf{Existing models remain non-robust in mixed-modality retrieval on natural images.} Most baselines degrade substantially from Origin to Mix: average NDCG@10 drops from 53.13 to 21.81 for SigLIP2, from 69.32 to 37.98 for CLIP-ViT-L/14, and from 55.67 to 33.91 for VisRAG-Ret. Meanwhile, some models, including Trident, score higher in Mix than in Origin (e.g., Qwen3-VL-Embedding-2B rises from 66.60 to 73.67). This contrasts with document images: document images contain abundant text and thus align readily with textual semantics, whereas natural-image semantics are sparser and more abstract, making them harder to match with text queries, while the semantically equivalent text documents substituted in Mix are explicit and easier to retrieve. The Origin-to-Mix change on natural images therefore conflates dataset characteristics with robustness, and is better judged by comparing models on the same Mix corpus.

\textbf{By mitigating modality preference, Trident remains effective on natural images.} On the same Mix corpus, Trident, which trains toward modality balance, achieves the highest Mix average on both backbones (70.50 and 76.78), clearly surpassing the corresponding baselines Jina-CLIP-v2 (56.43) and Qwen3-VL-Embedding-2B (73.67), while its Origin average does not decrease. This shows that mitigating modality preference does not compromise single-modality retrieval. Consistent with the main text, removing Multi-Positive InfoNCE lowers the Mix average on both backbones, confirming that an objective that balances modalities during training is effective.

\section{Further Ablation Studies}
\label{sec:appd:ablation}

\begin{figure*}[t!]
  \centering
  \includegraphics[width=\textwidth]{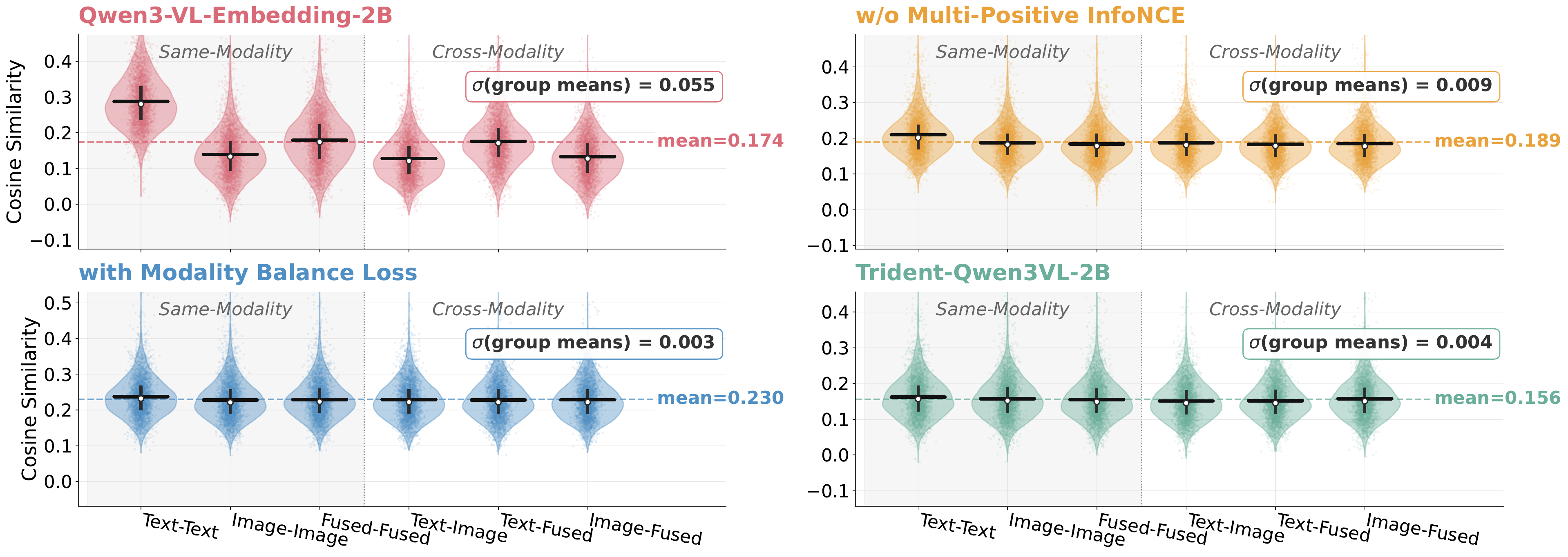}  
   \caption{Distribution of cosine similarities between same-modality and cross-modality positive pairs, ablation studies on Qwen3-VL-E-2B.}
  \label{fig:sim_change_appd}
\end{figure*}

We further examine how the two components of Trident contribute to mitigating modality preference. As shown in Figure~\ref{fig:sim_change_appd}, the original Qwen3-VL-Embedding-2B exhibits substantial disparity across same- and cross-modality positive pairs, with a standard deviation of $\sigma=0.055$ across the six group means. Training with our modality-balanced data construction alone (w/o Multi-Positive InfoNCE) reduces $\sigma$ to $0.009$, indicating that balanced positive sampling already alleviates modality-dependent disparity. Incorporating Multi-Positive InfoNCE further reduces $\sigma$ to $0.004$, consistent with the positive-view balancing effect characterized by Eq.~\ref{eq:loss_decomposition}: the shared softmax discourages the positive probability mass from being systematically concentrated in a single modality. Together with the retrieval gains in Table~\ref{tab:main_results}, these results show that data-level balance and objective-level coordination provide complementary contributions to mitigating modality preference.

\begin{figure*}[t!]
  \centering
  \includegraphics[width=\textwidth]{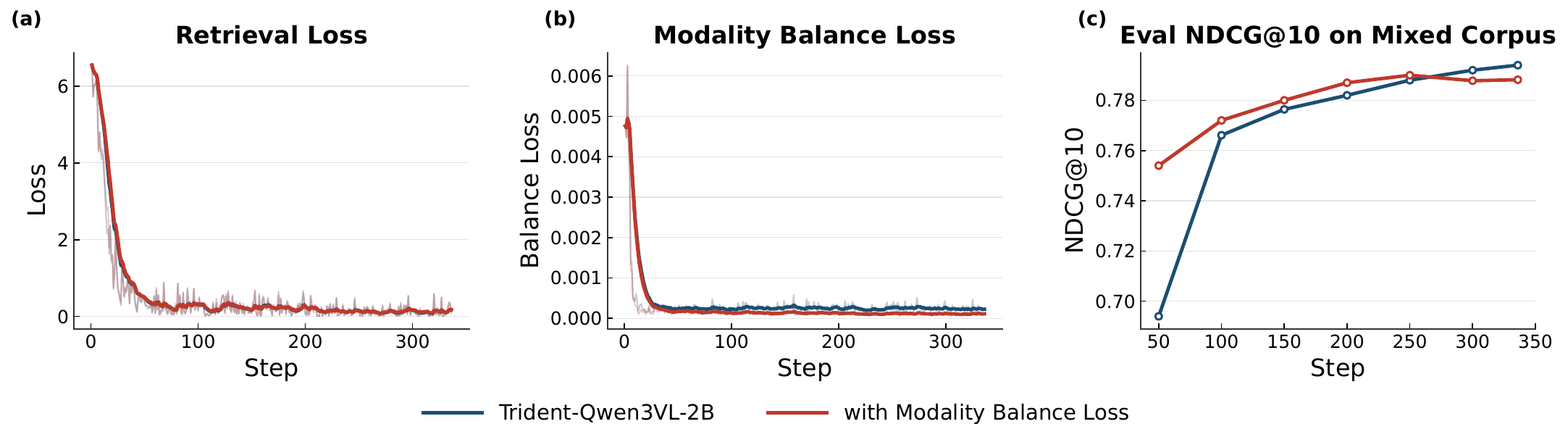}
    \caption{
    \textbf{Effect of explicit modality balancing.} We compare Multi-Positive InfoNCE with and without an additional Modality Balance Loss, corresponding to ``with Modality Balance Loss" and ``Trident-Qwen3VL-2B" respectively. Although explicit balancing further reduces modality discrepancy, it does not improve the final retrieval performance.}
  \label{fig:ablation_appd}
\end{figure*}

A natural question is whether stronger modality balancing would further improve retrieval. To test this, we augment Multi-Positive InfoNCE with an explicit Modality Balance Loss that directly penalizes similarity differences among the three positive views, and compare $w=200$ with the default $w=0$. As shown in Figure~\ref{fig:ablation_appd}(a), both settings exhibit similar convergence of the retrieval objective. More importantly, the balance loss decreases rapidly even when it is not explicitly optimized ($w=0$; Figure~\ref{fig:ablation_appd}(b)), providing additional empirical evidence that Multi-Positive InfoNCE already reduces modality disparity through its shared objective. Explicit balancing pushes this effect further, reducing $\sigma$ from $0.004$ to $0.003$ in Figure~\ref{fig:sim_change_appd}. However, this lower disparity does not yield better retrieval: although $w=200$ performs slightly better early in training, its advantage disappears and the final NDCG@10 is slightly lower than that of $w=0$ (Figure~\ref{fig:ablation_appd}(c)).

This result reveals an important distinction between \emph{mitigating modality preference} and \emph{enforcing similarity equality}. The positive-view balance induced by Multi-Positive InfoNCE operates on the relative probability mass within the shared softmax, discouraging systematic dominance by any single modality without directly requiring their cosine similarities to be identical. In contrast, the additional Modality Balance Loss explicitly pushes the similarities of Text, Image, and Fused views toward equality. Such a constraint can be overly restrictive because these views differ inherently in information content and granularity; for example, an image may contain visual evidence that is not fully captured by its textual description. Their optimal similarities to a query therefore need not coincide. The slightly better retrieval performance of $w=0$, despite its marginally larger similarity dispersion ($0.004$ vs.\ $0.003$), supports this distinction. \textbf{Modality balance does not require different modalities to have identical similarities; rather, it requires preventing any single modality from systematically dominating the shared representation space.}

\section{Full 3D Surface Comparisons Before and After Trident}
\label{sec:appd:comparison}

\begin{figure}[t]
    \centering
    \begin{subfigure}[b]{0.49\textwidth}
        \centering
        \includegraphics[width=\textwidth]{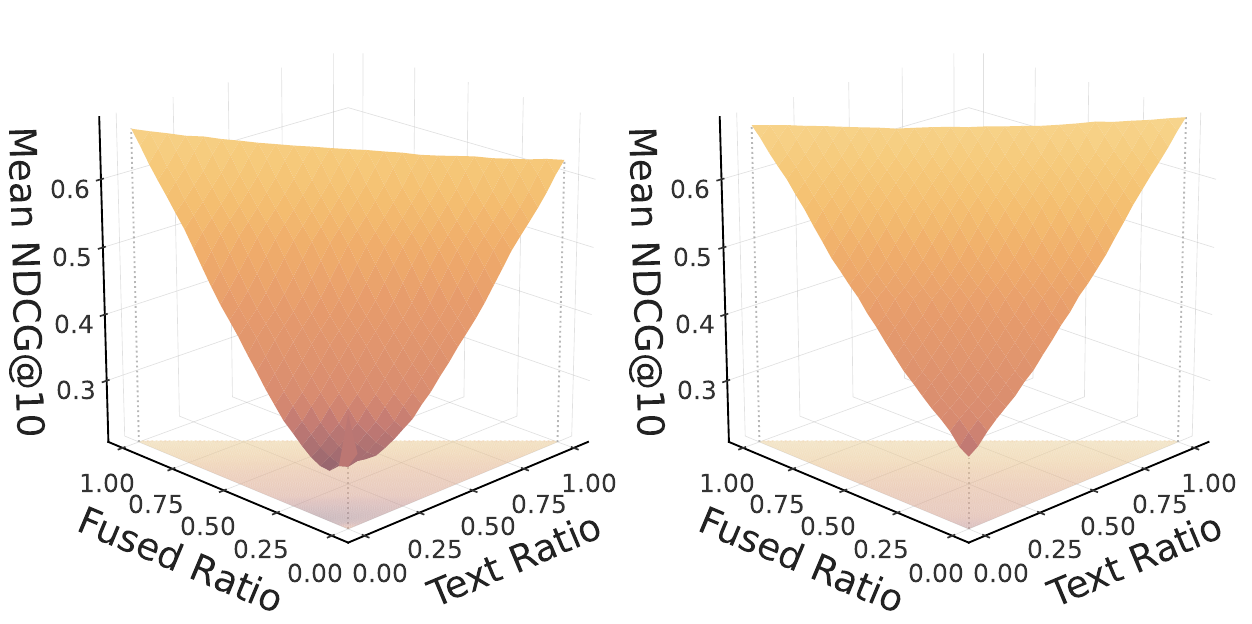}
        \caption{Jina-Clip-v2 on ChartQA}
        \label{fig:ndcg_compare_a}
    \end{subfigure}
    \hfill
    \begin{subfigure}[b]{0.49\textwidth}
        \centering
        \includegraphics[width=\textwidth]{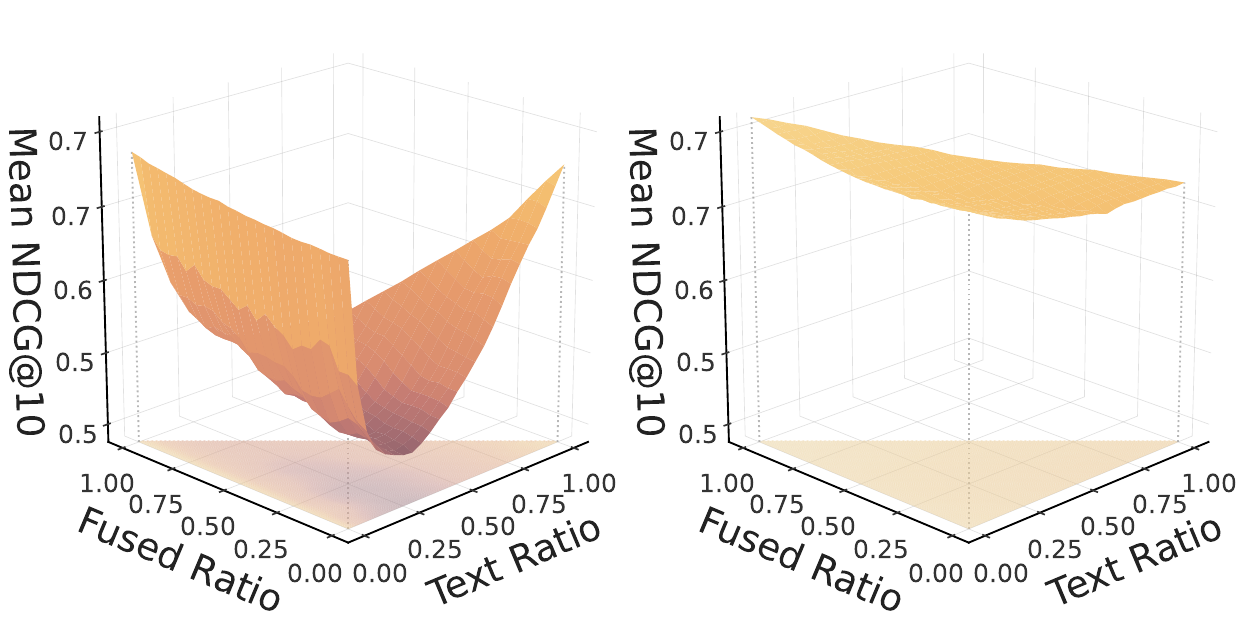}
        \caption{Qwen3-VL-E-2B on ChartQA}
        \label{fig:ndcg_compare_b}
    \end{subfigure}

    \begin{subfigure}[b]{0.49\textwidth}
        \centering
        \includegraphics[width=\textwidth]{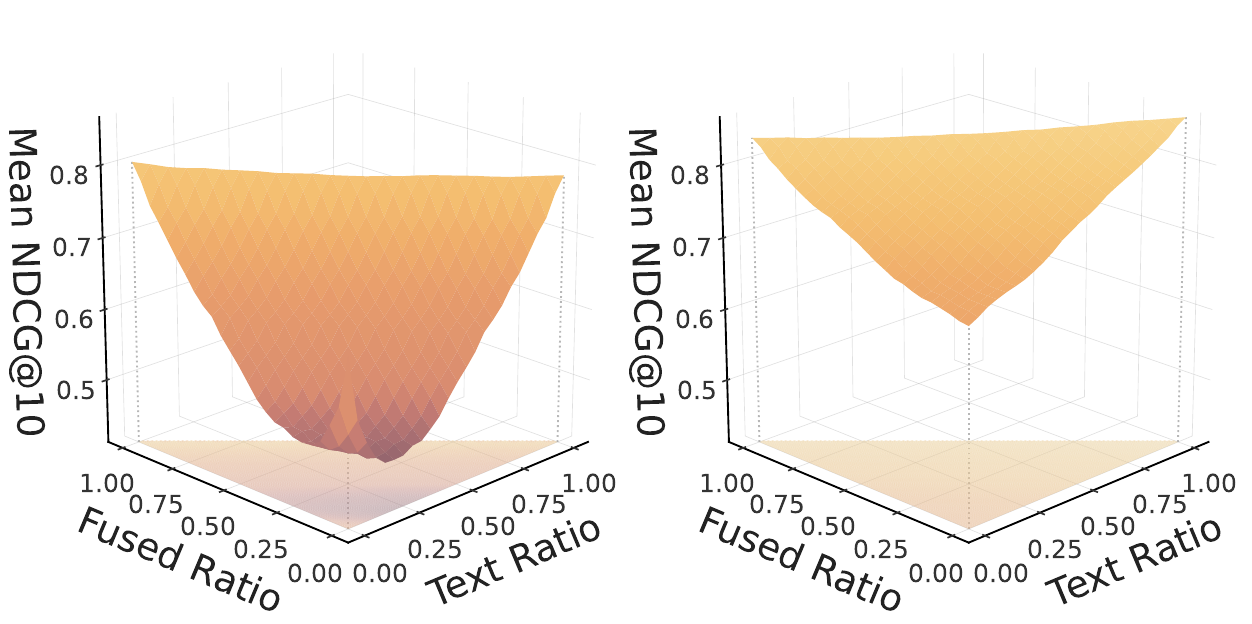}
        \caption{Jina-Clip-v2 on InfoVQA}
        \label{fig:ndcg_compare_a}
    \end{subfigure}
    \hfill
    \begin{subfigure}[b]{0.49\textwidth}
        \centering
        \includegraphics[width=\textwidth]{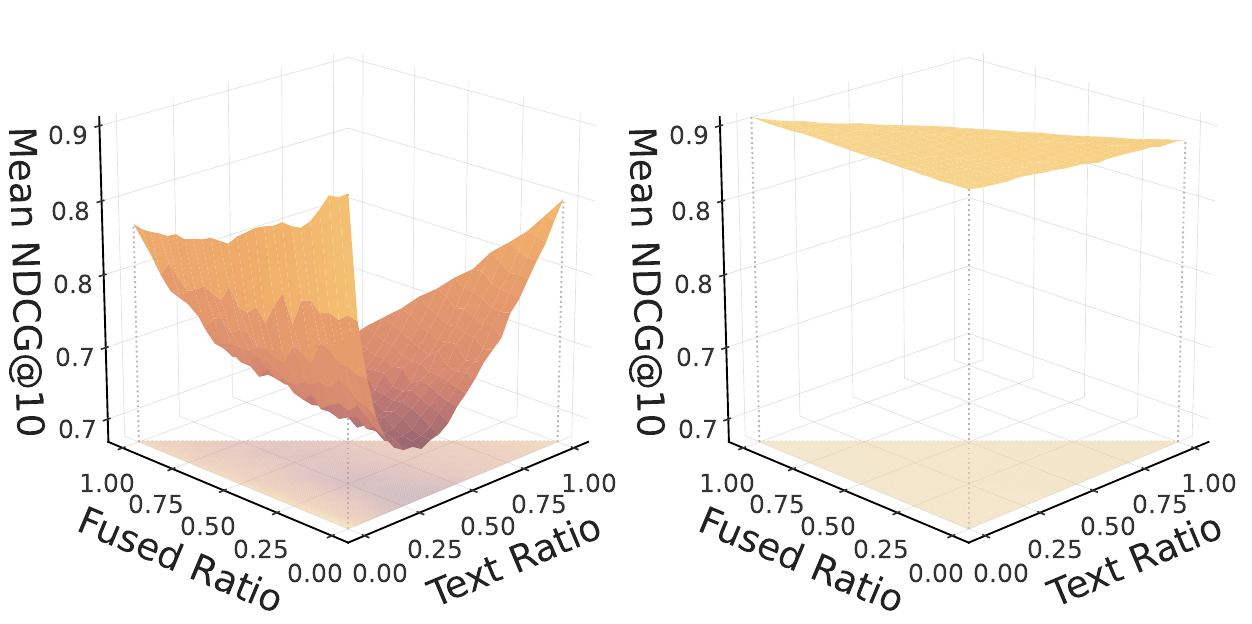}
        \caption{Qwen3-VL-E-2B on InfoVQA}
        \label{fig:ndcg_compare_b}
    \end{subfigure}

    \begin{subfigure}[b]{0.49\textwidth}
        \centering
        \includegraphics[width=\textwidth]{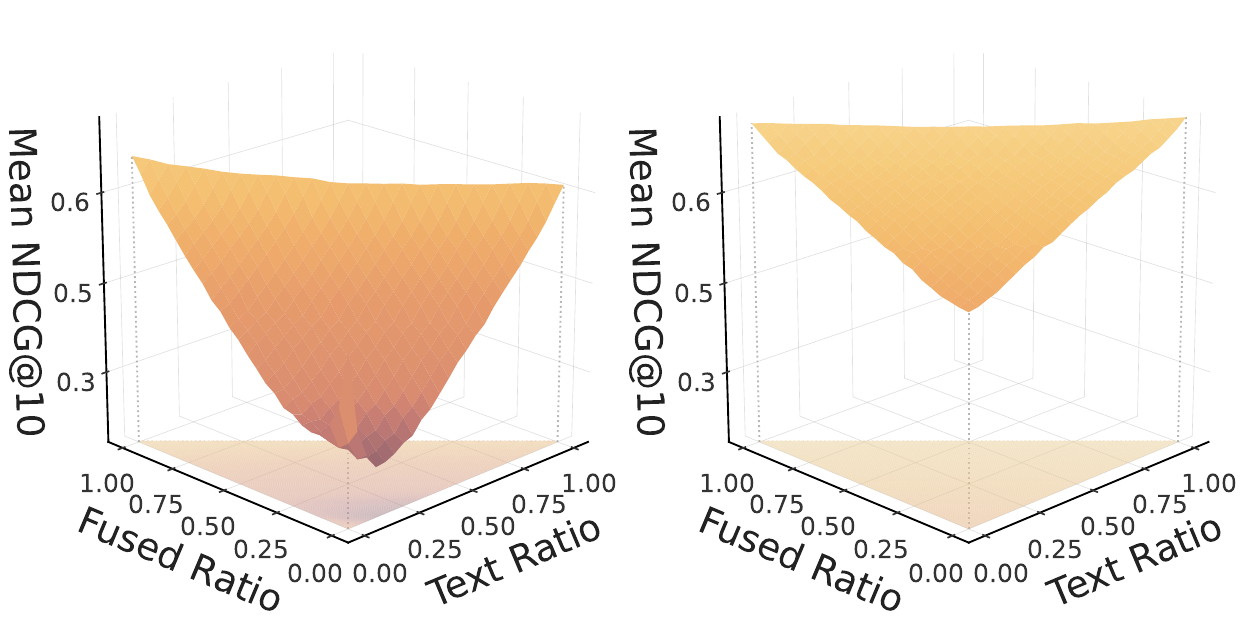}
        \caption{Jina-Clip-v2 on DocVQA}
        \label{fig:ndcg_compare_a}
    \end{subfigure}
    \hfill
    \begin{subfigure}[b]{0.49\textwidth}
        \centering
        \includegraphics[width=\textwidth]{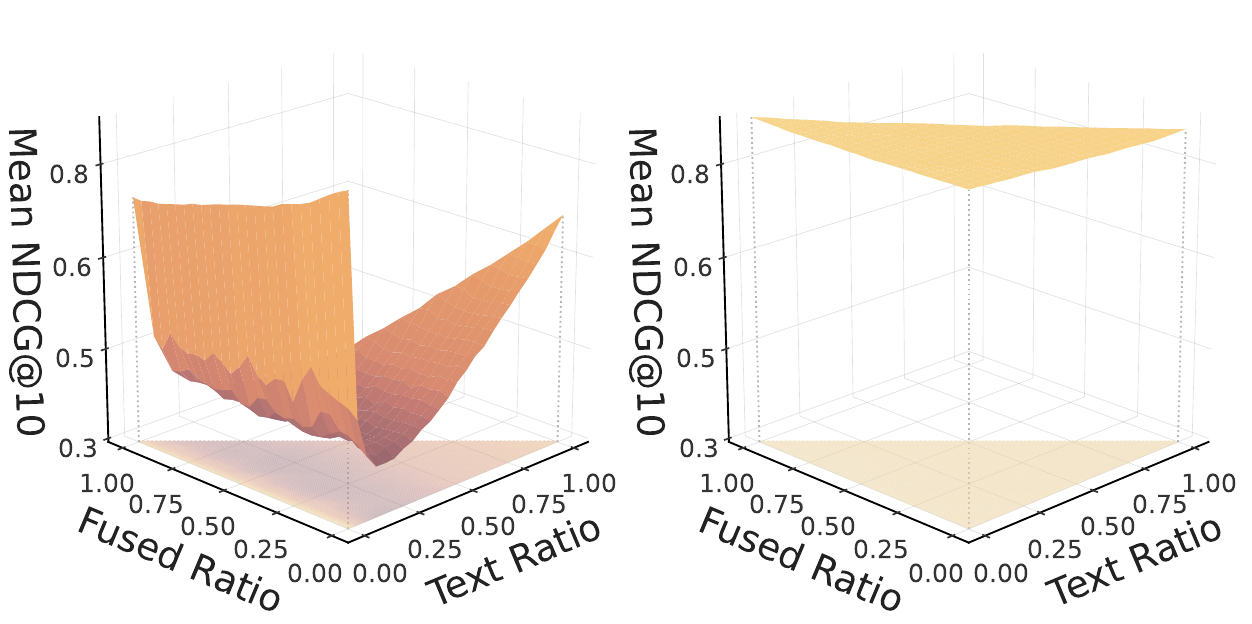}
        \caption{Qwen3-VL-E-2B on DocVQA}
        \label{fig:ndcg_compare_b}
    \end{subfigure}

    \begin{subfigure}[b]{0.49\textwidth}
        \centering
        \includegraphics[width=\textwidth]{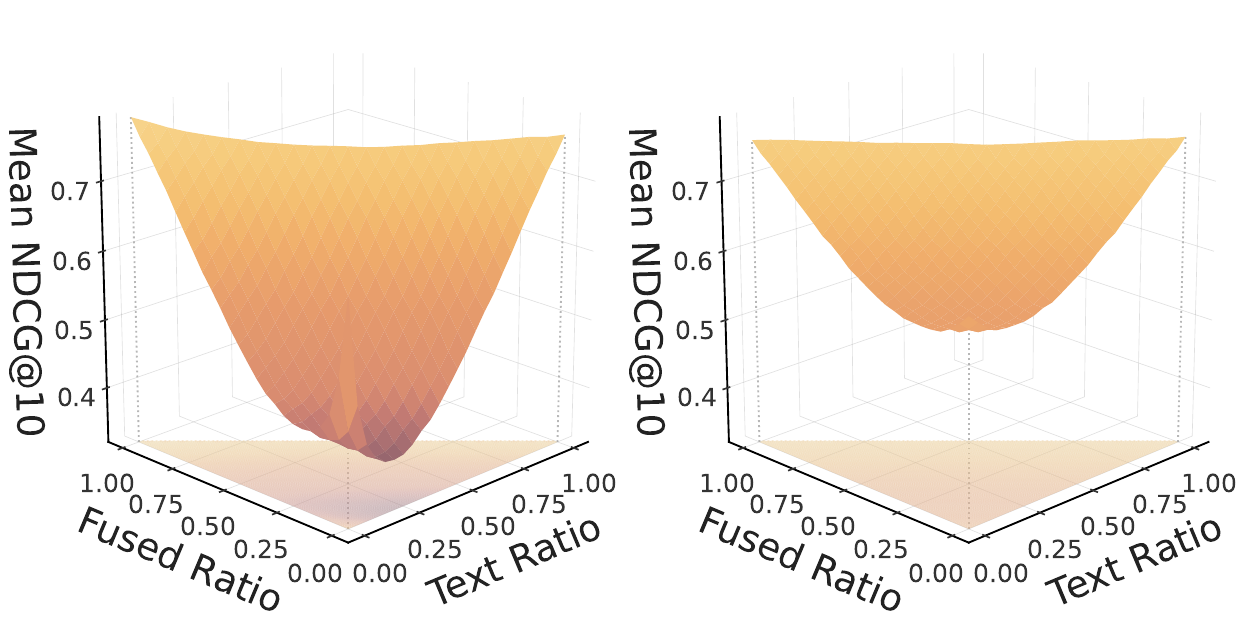}
        \caption{Jina-Clip-v2 on ViDoSeek}
        \label{fig:ndcg_compare_a}
    \end{subfigure}
    \hfill
    \begin{subfigure}[b]{0.49\textwidth}
        \centering
        \includegraphics[width=\textwidth]{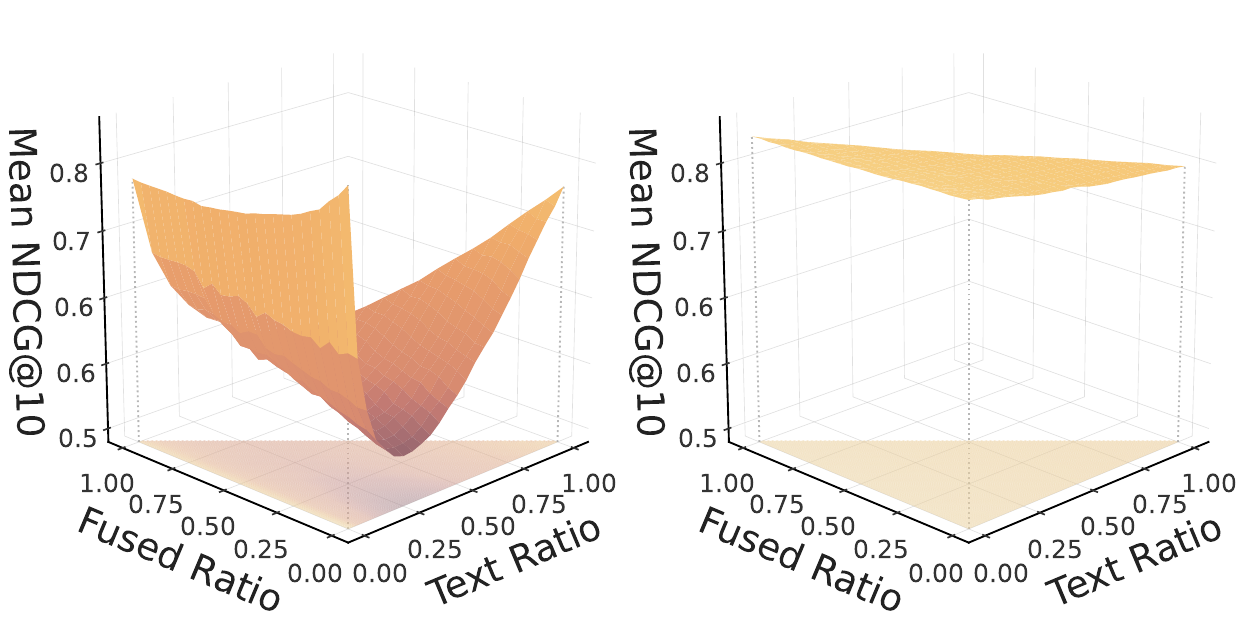}
        \caption{Qwen3-VL-E-2B on ViDoSeek}
        \label{fig:ndcg_compare_b}
    \end{subfigure}

    \caption{Mean NDCG@10 on four datasets as a function of corpus modality composition (Fused Ratio, Text Ratio), before (left) and after (right) Trident training.}
    \label{fig:ndcg_compare_appd}
\end{figure}

To further validate the robustness of Trident beyond the two datasets presented in the main text, we extend the 3D surface analysis to four additional benchmarks: ChartQA, InfoVQA, DocVQA, and ViDoSeek. Figure~\ref{fig:ndcg_compare_appd} reports Mean NDCG@10 as a function of corpus modality composition (Fused Ratio, Text Ratio) for both Jina-CLIP-v2 and Qwen3-VL-Embedding-2B, before (left) and after (right) Trident training.

Across all four datasets, the pattern observed in the main text is consistently reproduced. Before training, every surface exhibits a pronounced valley concentrated around balanced modality compositions, confirming that the modality preference of baseline retrievers is not an artifact specific to a single dataset but rather a systematic failure mode that generalizes across document types, ranging from natural images in ChartQA to dense text layouts in DocVQA and ViDoSeek. After Trident training, the valley is consistently eliminated, and Mean NDCG@10 remains stable across the full range of Fused Ratio and Text Ratio values. This consistency across datasets with markedly different visual and textual characteristics reinforces our central claim that Trident renders the model insensitive to corpus modality composition, rather than merely overfitting to the modality distribution of a particular training or evaluation set.

Beyond this overall consistency, a closer comparison across the two backbones reveals a subtler distinction. For Jina-CLIP-v2, although the pronounced valley is largely flattened after training, the resulting surface is not perfectly uniform: mild fluctuations remain visible, particularly on InfoVQA and DocVQA, where retrieval performance shows slight sensitivity to extreme modality ratios. We attribute this residual variation to the underlying CLIP-based architecture, in which text and image are encoded through separate towers and subsequently aligned only through a shared embedding space. Such a design inherently treats each modality as a distinct representational subspace, so that even after modality-balanced training, complete invariance to corpus composition is difficult to achieve.

In contrast, Qwen3-VL-Embedding-2B exhibits a markedly flatter surface after training, with performance variation across the entire Fused Ratio and Text Ratio grid becoming almost negligible on every dataset. We attribute this difference to the fact that Qwen3-VL-Embedding-2B is built upon a vision-language model backbone, wherein textual and visual tokens are processed within a single unified transformer rather than through separate modality-specific encoders. This architectural unification allows the model to represent text, image, and fused content within a shared token space from the outset, providing a more favorable inductive bias for eliminating modality preference once explicitly supervised with modality-balanced data. Taken together, these observations suggest that while Trident is effective regardless of backbone architecture, VLM-based embedding models offer a structurally more unified foundation for achieving genuinely modality-agnostic retrieval, whereas CLIP-based models, despite substantial improvement, retain a mild architectural bias toward treating modalities separately.

\section{Performance across Matryoshka Dimensions}

\begin{figure}[t]
    \centering
    \begin{subfigure}[b]{0.49\textwidth}
        \centering
        \includegraphics[width=\textwidth]{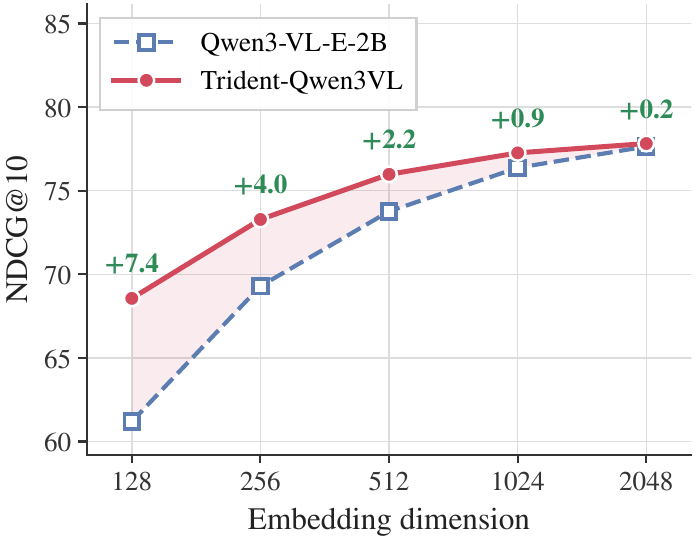}
        \caption{Performance on Origin Corpus}
        \label{fig:matryoshka:origin}
    \end{subfigure}
    \hfill
    \begin{subfigure}[b]{0.49\textwidth}
        \centering
        \includegraphics[width=\textwidth]{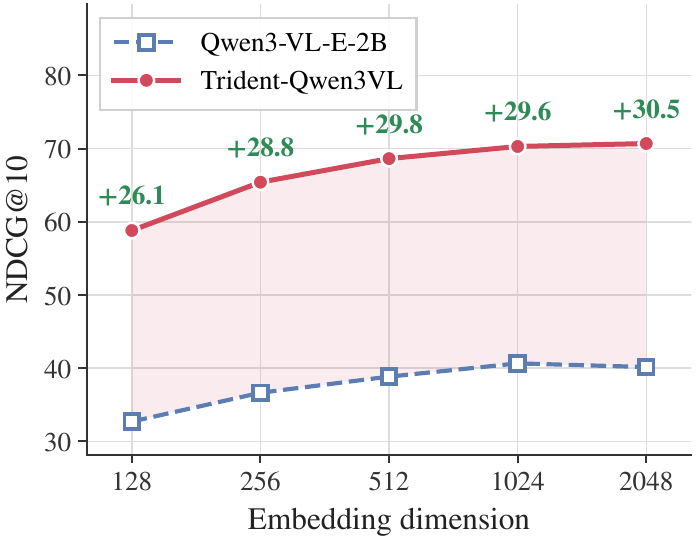}
        \caption{Performance on Mixed-Modality Corpus}
        \label{fig:matryoshka:mix}
    \end{subfigure}
    \caption{Performance across Matryoshka embedding dimensions.}
    \label{fig:matryoshka}
\end{figure}

The embedding dimension directly determines the storage and computation cost of a retrieval system, and Matryoshka representation learning~\citep{kusupati2022matryoshka} allows a single model to be used at multiple dimensions. To this end, we incorporate Matryoshka representation learning into the training of Trident (dimensions 128--2048) and compare Trident with Qwen3-VL-E-2B in terms of nDCG@10 across embedding dimensions (Figure~\ref{fig:matryoshka}).

\paragraph{Origin Corpus.}
As shown in Figure~\ref{fig:matryoshka:origin}, Trident performs at least on par with the baseline at every dimension, and its advantage grows as the dimension decreases: the two models are nearly tied at 2048 dimensions (+0.2), while Trident leads by +7.4 at 128 dimensions. This indicates that Trident does not sacrifice unimodal retrieval ability on the original corpus, and better preserves representation quality at low dimensions.

\paragraph{Mixed-Modality Corpus.}
As shown in Figure~\ref{fig:matryoshka:mix}, Trident outperforms the baseline by a large margin at all dimensions (+26.1 to +30.5), and this advantage does not shrink as the dimension increases. The baseline's nDCG@10 stays between 33 and 40 throughout, so a larger embedding dimension does little to alleviate its degradation on the mixed corpus, whereas Trident at only 128 dimensions already clearly surpasses the baseline at 2048 dimensions. This suggests that Trident's advantage in the mixed-modality setting stems not from greater representational capacity, but from effectively mitigating Modality Preference.

\end{document}